\documentclass[twocolumn, times, twocolappendix]{aastex631}

\newcommand\zcl{$z_{\rm cl}$}
\newcommand\Mcl{$M_{\rm cl}$}
\newcommand\Planck{\textit{Planck}}
\newcommand\SPHEREx{\textit{SPHEREx}}

\begin{document}

\title{
UPCluster-SZ: The Updated Catalog of Galaxy Clusters from the List of Planck Sunyaev-Zeldovich Sources
    }
    
\author{Hyeonguk Bahk}
\affiliation{Astronomy Program, Department of Physics and Astronomy, Seoul National University, 1 Gwanak-ro, Gwanak-gu, Seoul 08826, Republic of Korea}

\author{Ho Seong Hwang}
\affiliation{Astronomy Program, Department of Physics and Astronomy, Seoul National University, 1 Gwanak-ro, Gwanak-gu, Seoul 08826, Republic of Korea}
\affiliation{SNU Astronomy Research Center, Seoul National University, 1 Gwanak-ro, Gwanak-gu, Seoul 08826, Republic of Korea}

\correspondingauthor{Ho Seong Hwang}
\email{hhwang@astro.snu.ac.kr}

\begin{abstract}

We present the updated galaxy cluster catalog of the second \textit{Planck} catalog of Sunyaev-Zeldovich sources (PSZ2) through the compilation of the data for clusters and galaxies with  spectroscopically measured redshifts in the literature.
The original version of PSZ2 comprises 1653 SZ sources, of which 1203 have been validated as genuine galaxy clusters, while the remaining 450 sources are yet to be validated.
To increase the number of genuine clusters in PSZ2, we first update the validations of the cluster candidates and their redshift information using the data compiled for the confirmed clusters and the member galaxies in the literature.
We then use the galaxy redshift data in the fields of the remaining cluster candidates, by searching for possible member galaxies with measured spectroscopic redshifts around the Sunyaev-Zeldovich centroids.
In this search process, we classify clusters as strong candidates if they contain more than nine galaxies within a 4500 km s$^{-1}$ velocity range and within 15 arcmin
around the Sunyaev-Zeldovich centroids.
This process results in the validation of 139 new genuine clusters, the update of redshift information on 399 clusters, and the identification of 10 strong candidates,  which increases the number of validated clusters up to 1334 among the 1653 SZ sources. 
Our updated galaxy cluster catalog will be very useful for the studies of galaxy formation and cosmology through the combination with other all-sky surveys including \textit{WISE} and \textit{SPHEREx}.

\end{abstract}


\section{Introduction\label{sec:intro}}

Understanding our universe and the evolution of galaxies places significant emphasis on the statistical investigation of galaxy clusters, making it an indispensable element in astrophysical research.
In the current hierarchical formation scenario \citep{peebles80}, the physical properties of galaxy clusters, being the systems on top of the hierarchy, can provide strong constraints on cosmological parameters \citep{allen11}.
Moreover, galaxy clusters can serve as useful laboratories to investigate the formation and evolution of galaxies in a densely populated environment \citep[e.g.,][]{park09, kravtsov12}.
To facilitate these studies, well-defined cluster catalogs from homogeneous surveys are essential, containing key information such as celestial positions, redshifts, masses, and ideally, properties of member galaxies.

As part of such surveys, the upcoming \SPHEREx\ mission \citep[Spectro-Photometer for the History of the Universe and Ices Explorer;][]{dore14}, will provide the spectral energy distributions (SEDs) of $\sim 100,000$ cluster galaxies \citep{dore16, dore18}.
The \SPHEREx\ mission is the first all-sky near-infrared spectroscopic survey scheduled to launch in April 2025. It will provide the near-infrared ($0.75-4.8$ $\mu$m) spectra of $R=41-130$ at every $6.2"\times6.2"$ pixel on the sky.
To date, it has been only possible to analyze near-infrared window through either broad-band surveys such as \textit{IRAS} \citep{iras}, 2MASS \citep{2mass}, \textit{AKARI} \citep{akari}, and \textit{WISE} \citep{wise}, or pointing observations with near-infrared spectrograph of \textit{ISO} \citep{iso}, \textit{AKARI}, \textit{Spitzer} \citep{spitzer}, or \textit{JWST} \citep{jwst}.
However, broad-band photometry has limitations in accurately modeling the properties of galaxies \citep[e.g.,][]{nersesian23}, and the space observatories are too overloaded to survey near-infrared galaxy spectra for a statistical study dedicated to the galaxy clusters.
In this regard, \SPHEREx\ data will enable us to compare near-infrared SEDs of cluster members with those of field galaxies to better understand the physical mechanisms driving the evolution of cluster galaxies through the analysis of high-$z$ galaxies and the dust content of low-$z$ galaxies.

However, its angular resolution and spectral resolution are often too low to derive the physical parameters of individual cluster members without prior information about their precise locations, photometries, redshifts, etc. Thus, analyzing data from \SPHEREx\ requires a predefined cluster catalog as input. For cluster science with \SPHEREx, the cluster catalog may need to satisfy the following two characteristics: (\textit{i}) complete sample --- candidate clusters should be verified, and sample clusters should have measurements of cluster redshift \zcl\ and their mass \Mcl, (\textit{ii}) all-sky sample --- to benefit from the all-sky capability of \SPHEREx. Unfortunately, there are few such catalogs satisfying these conditions.

Among many cluster finding methods to construct a catalog, identification of clusters with intracluster medium (ICM) can satisfy these criteria more easily. 
Surveys based on X-ray (\citealp{rasscl, reflex, macs, mcxc}, see also \citealp{klein22}) or microwave \citep[i.e., Sunyaev-Zeldovich effect; SZ,][]{sz, psz1, psz1v2, psz2}, can homogeneously find clusters over the entire sky, making the optimal cluster sample for statistical analysis. Other cluster finding algorithms based on galaxy overdensity that are often accompanied with red sequence detection (e.g., MaDCoWS; \citealp{madcow}, redMaPPer; \citealp{redmapper}, see also \citealp{klein18,klein19}), might be prone to systematic effects due to the different depth of base imaging surveys. It is  often difficult to extend these catalogs to the all-sky sample. 
Still, for each candidate from the ICM-based cluster sample, we need to supplement the redshift information from other sources. Once the redshift is determined, the mass of a cluster can be estimated by proxies calibrated from the ICM quantities, which generally show only a small scatter.

The second Planck catalog of Sunyaev-Zeldovich sources (PSZ2; \citealp{psz2}, see also the first catalog, PSZ1; \citealp{psz1, psz1v2}) is a list of galaxy cluster candidates, which requires further updates regarding the validation and their redshift information. The PSZ2 catalog includes 1653 cluster candidates; among them, 1203 are validated and the remaining 450 are not still validated. In addition, it is important to note that not all validated clusters currently have their redshift measurements available. In this regard, significant efforts have been made to validate the remaining candidates and to measure their redshifts \citep[e.g.,][]{itp13a, burenin18, amodeo18}. However, this updating process is yet ongoing. Further details on the overall characteristics of the PSZ2 catalog and the follow-up studies will be discussed in Section \ref{ssec:sample}.

Here we present UPCluster-SZ, the updated PSZ2 catalog to provide the intermediate step of making a complete all-sky cluster catalog. We significantly increase the number of confirmed clusters in PSZ2 and update their redshifts by compiling follow-up studies. Additionally, our analysis of recent spectroscopic redshift measurements in the literature has enabled us to identify strong candidates. For the updated redshift measurements, we use the data from the NASA/IPAC Extragalactic Database (NED\footnote{\url{https://ned.ipac.caltech.edu}}), the Sloan Digital Sky Survey (SDSS) DR17 \citep{sdssdr17} spectroscopic redshifts, and the Dark Energy Spectroscopic Instrument (DESI) EDR \citep{desiedr} supplemented with numerous redshift surveys (see Section \ref{ssec:galcats}).  

We organize our contents as follows. We describe the cluster sample and the supplement data in Section \ref{sec:sample&data}. The methods for the validation, the compilation of the redshifts, and the identification of strong candidates are described in Section \ref{sec:methods}. Then we provide our updated PSZ2 catalog in Section \ref{sec:result} and discuss the results in Section \ref{sec:discussion}. A summary and conclusion are presented in Section \ref{sec:summary}. Throughout our analysis, we assume the standard $\Lambda$ Cold Dark Matter ($\Lambda$CDM) cosmology with $H_0 = 70 {\rm\ km\ s^{-1}\ Mpc^{-1}}$, $\Omega_{m,0} = 0.3$ and $\Omega_{\Lambda,0} = 0.7$.

\section{Sample and Data \label{sec:sample&data}}
\subsection{Sample \label{ssec:sample}}

To construct the all-sky cluster catalog, we adopt PSZ2 as our base catalog.
PSZ2 is the all-sky catalog of SZ sources, which were detected in a homogenous method throughout the all-sky measurements of seven different channels of \Planck\ \citep{psz1, psz1v2, psz2}. This homogeneity will provide us a good cluster sample for statistical studies with the least systematic effects, when compared with other cluster catalogs.

\begin{figure*}
\gridline{\fig{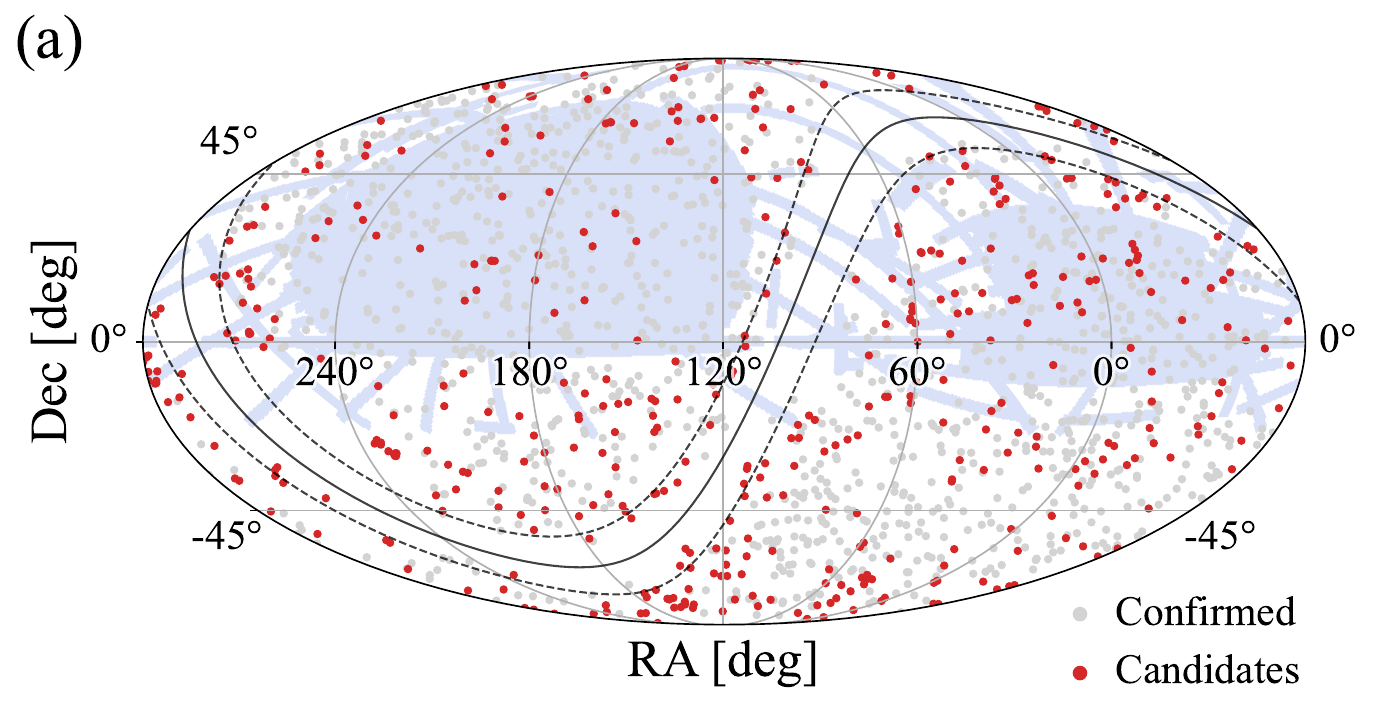}{0.48\textwidth}{}
          \fig{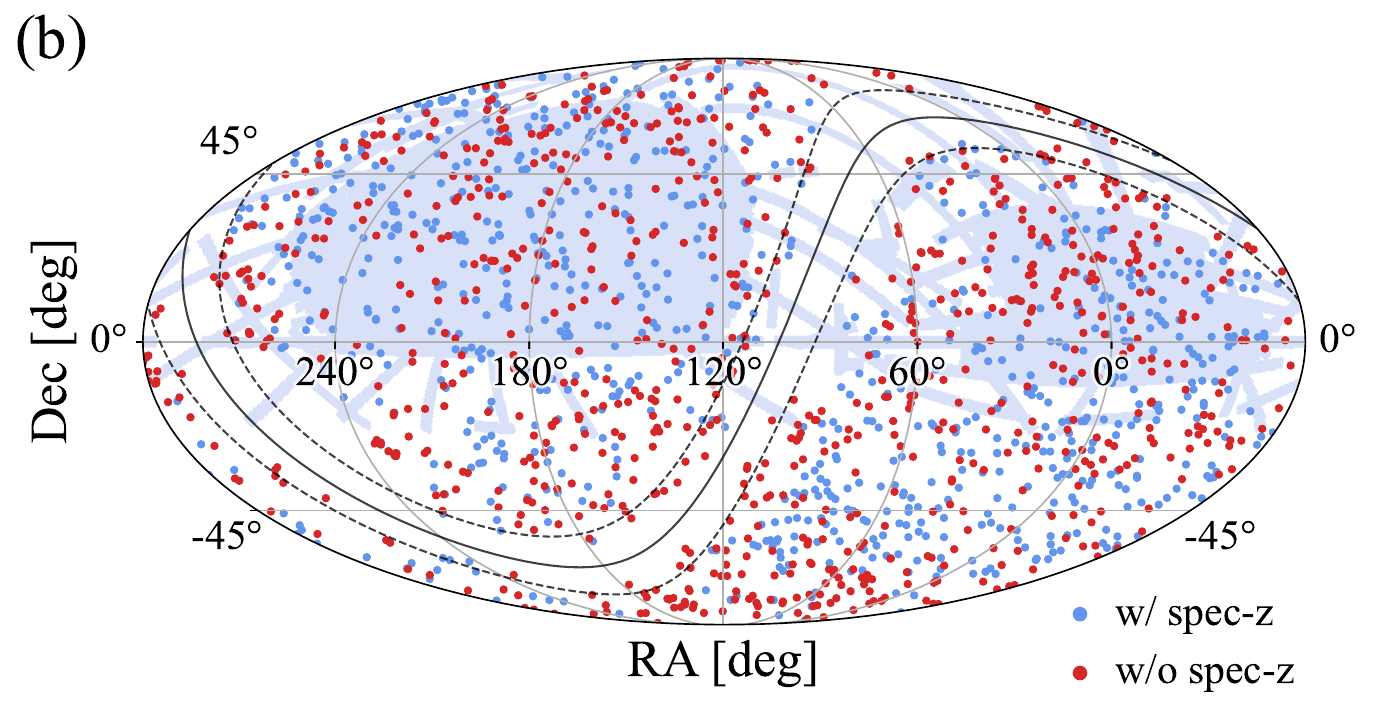}{0.48\textwidth}{}}
\caption{The sky distribution of the PSZ2 clusters and candidates. (a) Confirmed galaxy clusters (gray, 1203) and the candidate clusters (red, 450) in PSZ2. (b) Galaxy clusters whose spectroscopic redshifts are available (blue, 1094) and clusters (or candidates) whose spectroscopic redshifts are not available (red, 559) in PSZ2. The solid line shows the galactic plane, and the dashed lines correspond to the galactic latitude $b=\pm 10^{\circ}$. The blue-shaded region shows the SDSS footprint.\label{fig:skymap_psz}}
\end{figure*}

PSZ2 contains 1653 clusters and cluster candidates in total; among them, 1203 were validated as genuine clusters and 450 were candidates that need to be validated. The spatial distribution of these validated clusters and candidates is shown in the left panel of Figure \ref{fig:skymap_psz}. The SZ-detected cluster candidates cannot be validated without additional data from photometric and/or spectroscopic observations in X-ray/optical wavelengths. Note that there are fewer candidates in the SDSS footprint (e.g., $110^{\circ}\leq$ R.A. $\leq270^{\circ}$ and $0^{\circ}\leq$ Decl. $\leq70^{\circ}$) in Figure \ref{fig:skymap_psz}, because of its rich photometric and spectroscopic data targeting galaxies therein.

Out of the 1203 validated clusters, 1094 clusters have redshift information available, spanning a redshift range of $0<z<1$.
The right panel of Figure \ref{fig:skymap_psz} shows the clusters categorized based on the presence or absence of spectroscopic measurements for their redshift.
Specifically, PSZ2 consists of 829 clusters with spectroscopic redshifts and 258 with photometric redshifts.
Given the importance of accurate redshift information in estimating cluster mass and studying their evolution, we prioritize updating the redshifts of clusters with photometric redshifts and candidate clusters.
The information about the redshift type (spectroscopic or photometric redshift) is not provided by PSZ2 itself. Instead, we compile this information from the sources of measurements, as we will discuss in Section \ref{ssec:typeid}.

\subsection{Cluster Catalogs\label{ssec:clcats}}

\begin{deluxetable*}{cccccccc}
\tablecaption{Cluster Catalogs Used in This Work \label{tab:clcat}}
\tablehead{
\colhead{Number} & \colhead{Ref} & \colhead{Name} & \colhead{Inclusion} & \colhead{Telescope} & \colhead{$N_{\rm cl}$} & \colhead{$N_{\rm up}$} & \colhead{Validation}}
\colnumbers
\startdata
\multicolumn{8}{l}{PSZ2 Follow-up studies}\\
\cline{1-8}
1 & \citet{itp13a} & ITP13 & $\checkmark$ & INT, TNG, WHT, GTC & 115 & 55 & 50\\
2 & \citet{itp13b} & ITP13 & $\checkmark$ & INT, TNG, WHT, GTC & 75 & 22 & 50\\
3 & \citet{lp15a} & LP15 & $\checkmark$ & INT, TNG, GTC & 106 & 104 & 50\\
4 & \citet{lp15b} & LP15 & $\checkmark$ & INT, TNG, GTC & 120 & 105 & 50\\
5 & \citet{streblyanska18} & {} & $\checkmark$ & SDSS & 37 & 10 & 51\\
6 & \citet{burenin18} & RTT & $\checkmark$ & RTT150, Sayan, Calar Alto, BTA & 7 & 1 & 52\\
7 & \citet{zaznobin19} & RTT & $\checkmark$ & RTT150, Sayan, Calar Alto, BTA & 38 & 19 & 52\\
8 & \citet{zaznobin20} & RTT & $\checkmark$ & RTT150, Sayan, BTA & 67 & 9 & 52\\
9 & \citet{zaznobin21} & RTT & $\checkmark$ & RTT150, Sayan, Calar Alto & 23 & 3 & 52\\
10 & \citet{amodeo18} & {} & $\checkmark$ & Gemini, Keck, Palomar & 20 & 17 & 53\\
11 & \citet{boada19} & {} & $\checkmark$ & Mayall & 15 & 4 & 54\\
12 & \citet{zohren19} & {} & $\checkmark$ & WHT & 32 & 1 & 55\\
\tablebreak\\
\multicolumn{8}{l}{Non-PSZ2-based cluster catalogs}\\
\cline{1-8}
13 & \citet{burenin17} & {} & $\checkmark$ & SDSS, \textit{WISE} & 2964 & 130 & 60\\
14 & \citet{spt-sz19} & SPT-SZ 2500d & $\checkmark$ & SPT & 677 & 7 & 61\\
15 & \citet{spt-ecs20} & SPT-ECS & $\checkmark$ & SPT & 470 & 12 & 62\\
16 & \citet{sptpol19} & SPTpol 100d & $\times$ & SPT & 89 & 0 & {}\\
17 & \citet{buddendiek15} & {} & $\times$ & WHT, LBT & 47 & 0 & {}\\
18 & \citet{xclass} & X-CLASS & $\times$ & SDSS & 1646 & 0 & {}\\
19 & \citet{psz-mcmf} & PSZ-MCMF & $\checkmark$ & DES & 853 & 7 & 63\\
20 & \citet{rass-mcmf} & RASS-MCMF & $\checkmark$ & DESI LS DR10 & 8449 & 43 & 64\\
\enddata
\tablecomments{The descriptions of the columns are listed as follows: (1) Ordinal number for each catalog. (2) The reference. (3) The name of the program or catalog. (4) Inclusion status in our update. (5) Telescope used for the follow-up studies. (6) The number of clusters (candidates and non-detections) provided. (7) The number of clusters used for the update at this work. (8) The validation code for identifying the reference of validation for the catalog, as in PSZ2. The acronyms for the names of the listed telescopes are (i) INT (Isaac Newton Telescope, 2.5 m), (ii) TNG (Italian Telescopio Nazionale Galileo, 3.5 m), (iii) WHT (William Herschel Telescope, 4.2 m), (iv) GTC (Gran Telescopoi Canarias, 10.4 m), (v) RTT150 (Russian-Turkish Telescope, 1.5 m), (vi) Sayan (Sayan Observatory Telescope, 1.6 m), (vii) Calar Alto (Calar Alto Telescope, 3.5 m), (viii) BTA (Bolshoi Telescope Alt-azimutalnyi, 6 m), (ix) Gemini (Gemini Telescopes, 8.1 m), (x) Keck (Keck Telescopes, 10 m), (xi) Palomar (Palomar Telescope, 200 inch), (xii) Mayall (Mayall Telescope, 4 m), (xiii) SPT (South Pole Telescope, 10 m), (xiv) LBT (Large Binocular Telescope, 8.4 m), (xv) DES (Victor M. Blanco Telescope, 4 m), (xvi) DESI LS DR10 (Victor M. Blanco Telescope, 4 m; Mayall Telescope, 4 m; Bok Telescope, 2.3 m).}
\end{deluxetable*}

Because our goal is to validate the cluster candidates in PSZ2 and to update their redshift with spectroscopic data, we first need to compile the available data for our cluster sample from the literature. Therefore, we update PSZ2 by cross-matching PSZ2 clusters with recent cluster catalogs from follow-up studies and independent works.
In this section we describe the information of the cluster catalogs that we compile to update the PSZ2 clusters and candidates.
From each catalog we obtain various information on clusters including confirmation flag, redshift, and type of redshift measurement (photometric or spectroscopic). The following paragraphs summarize the relevant details of each catalog used in the update. We also distinguish the description of catalogs into two categories, the PSZ2 follow-up observations and non-PSZ2 based cluster catalogs. We list the basic information about these cluster catalogs in Table \ref{tab:clcat}.

\subsubsection{PSZ2 Follow-up Studies \label{sec:2.2.1}}

We find in the literature direct follow-up studies for PSZ (PSZ2 and PSZ1), whose purpose is to verify the PSZ candidates and to measure their redshifts from either imaging or spectroscopic survey: Roque de los Muchachos Observatory (ORM; or the European Northern observatory, ENO as denoted in PSZ2) follow-ups \citep{itp13a, itp13b, lp15a, lp15b, streblyanska18}, Russian-Turkish Telescope (RTT) follow-ups \citep{burenin18, zaznobin19, zaznobin20, zaznobin21}, Gemini and Keck follow-up \citep{amodeo18}, William Herschel Telescope (WHT) follow-up \citep{zohren19}, and Mayall Observatory follow-up \citep{boada19}. There are more follow-up observations other than these, but all of them have been already incorporated in PSZ2. We omit to describe them here and their information can be found in the PSZ papers \citep{psz1, psz1v2, psz2}.
Here we provide a brief description of each follow-up study:
\begin{itemize}
    \item \textit{ORM follow-ups}. After the two sequences of extensive photometric and spectroscopic observations, they use their confirmation criteria \citep[see Table 2. of both ][]{itp13a, lp15a} to validate the candidate clusters. We adopt their validations and redshift measurements, including both spectroscopic and photometric redshifts. The photometric redshifts are estimated from the cluster-red sequence method \citep{gladders00, lopes07} following \citet{eno}.
    \item \textit{RTT follow-ups}. These studies target the \Planck\ candidate clusters from PSZ2 and \citet{burenin17}, providing the spectroscopic measurements of the redshift for PSZ2 candidates. These redshifts are typically originated from the long-slit observation of the brightest cluster galaxies (BCGs) or a few member galaxies.
    \item \textit{Gemini+Keck follow-up}. \citet{amodeo18} provides reliable spectroscopic redshifts and velocity dispersions for 20 PSZ clusters in the SDSS-covered region, using the in-depth Gemini and Keck observations.
    \item \textit{Mayall follow-up}. \citet{boada19} conduct optical imaging observations to confirm PSZ2 candidates of high-significance detection, finding BCGs by visual inspection. The redshift information is all photometric, which is estimated using the Bayesian Photometric Redshifts \citep[BPZ;][]{bpz00, bpz06}.
    \item \textit{WHT follow-up}. \citet{zohren19} select PSZ2 candidates potentially located at $z>0.7$. They adopt the Multi-Component Matched Filter \citep[MCMF; ][]{klein18, klein19} and the red sequence finding technique used by \citet{vanderburg16} to estimate photometric redshift and richness. They also provide the spectroscopic redshifts for subset of their sample, which are obtained from long-slit observations.
    
\end{itemize}

\subsubsection{Non-PSZ2-based Cluster Catalog \label{sec:2.2.2}}

Alongside the direct follow-up observations, cluster catalogs selected from other surveys often contain the PSZ2 counterpart. We find the cluster catalogs based on extended \Planck\ SZ survey with lowered significance \citep{burenin17, madpsz}, independent SZ survey using the South Pole Telescope \citep{spt-sz19, sptpol19, spt-ecs20}, and X-ray surveys from ROSAT All Sky Survey \citep[RASS;][]{rasscat_all, buddendiek15} and \textit{XMM-Newton} observations \citep{xclass}. Here, we provide a brief description of each cluster catalog:
\begin{itemize}
    \item \citet{burenin17} presents an extension of the \Planck\ galaxy cluster catalog using an independent detection method of the SZ signal from the \Planck\ NILC Compton y-map \citep{tszmap}. They conduct cluster identification for the detected SZ sources inside the SDSS field, matching with redMapper and analyzing WISE and SDSS photometry to find the red sequence, together with the threshold in cluster infrared luminosity. For these clusters, the cluster redshifts are compiled from SDSS DR10, with the projected angular distance limit and deviation from the photometric redshift criteria, which is a process similar to the one in our cluster identification in Section \ref{ssec:val}.  It should be noted that the spectroscopic redshifts for the clusters with a small number of galaxy spectra $N_z$ could be wrong. In this catalog, we use only the clusters that have PSZ2 counterpart for our update.
    \item \citet{spt-sz19} provides the updated spectroscopic redshifts and improved calibration of photometric redshifts from the original 2500 deg$^2$ SPT-SZ survey \citep{spt-sz}. Similarly, the 2770 deg$^2$ SPTpol Extended Cluster Survey \citep[SPT-ECS;][]{spt-ecs20} contains the SZ-detected clusters (and candidates) in separate region, providing their photometric and spectroscopic redshifts. These catalogs have 174 PSZ2 counterparts when they are cross-matched in $5'$, while the SPTpol 100 deg$^2$ survey \citep{sptpol19} has no PSZ2 counterparts.
    \item \citet{buddendiek15} contains four PSZ clusters (PSZ2 G069.39+68.05, PSZ2 G073.31+67.52, PSZ2 G089.39+69.36, and PSZ2 G176.27+37.54). However, PSZ2 G069.39+68.05 and PSZ2 G073.31+67.52 are covered by ORM and RTT follow-ups. We decide not to include these spectroscopic redshifts from this catalog because their redshift estimates are consistent with those from the follow-up programs. PSZ2 G089.39+69.36 and PSZ2 G176.27+37.54 are also confirmed clusters in PSZ2, validated by SDSS high-z sample and redMapper, respectively (see Section \ref{ssec:typeid}). Because the redshift estimates for these two clusters are photometric ones in both PSZ2 and \citet{buddendiek15}, we again decide not to update the photometric redshift value with this catalog.
    \item The XMM Cluster Archive Super Survey \citep[X-CLASS;][]{xclass} catalog comprises the galaxy clusters of serendipitous detection from the \textit{XMM-Newton} X-ray observations. They provide the redshift validation and redshift measurement from the Multi Object Spectroscopy (MOS) follow-up program and additional compilation of galaxy redshift from the NED and SDSS DR16. However, all of the confirmed clusters in X-CLASS are covered with spectroscopic redshifts by the follow-up cluster catalogs (Section \ref{sec:2.2.1}). Their spectroscopic redshifts are consistent with each other, where the maximum difference $\Delta z/(1+z)=0.0027$. In the result, none of the clusters from X-CLASS is used for our update.
    \item The PSZ-MCMF catalog \citep{psz-mcmf} is the result of systematic follow-up of \Planck\ cluster candidate inside the DES \citep[Dark Energy Survey;][]{des} field. Using the MCMF method, they identified the optical counterparts of the \Planck\ SZ signal and estimated the richness and photometric redshift. Among the 853 MCMF confirmed clusters, we could identify 213 PSZ2 counterparts by cross-matching within a radius of $3'$, according to their matching process \citep[see Section 4.3.1 of ][]{psz-mcmf}.
    \item The RASS-MCMF catalog, an X-ray selected cluster catalog, is based on X-ray properties from RASS and optical cluster identifications from the DESI Legacy Survey DR10 \citep{desi-ls-overview-dey19} using the MCMF algorithm. They provide X-ray mass, optical richness and photometric redshift of the clusters. This catalog includes 842 PSZ2 counterparts identified through cross-matching within a $5'$ radius \citep[see Section 5.3 of][]{rass-mcmf}. 
\end{itemize}

\subsection{Galaxy Catalogs\label{ssec:galcats}}

To further identify strong signals of galaxy clusters from the candidates in PSZ2 and to estimate their redshifts, we compile the redshift data of galaxies from the literature (e.g., NED, SDSS DR17, and DESI EDR). We include SDSS DR17 \citep{sdssdr17} and DESI EDR \citep{desiedr} to extend NED, because of the incomplete inclusion of all galaxies in SDSS DR17 and DESI EDR within NED at the time of our analysis. In addition, to supplement the redshift of the faint galaxies we add the spectroscopic redshift information from the literature (e.g., \citealt{hecs13, hecs-sz, hecs-red, hectomap_dr2}, see \citealt{hwang10, hwang14} for other literature in detail).

From NED, we query the galaxies inside the searching radius of $15'$ near the PSZ2 positions. Here, we select only the object type of \texttt{G} and \texttt{QSO} to separate galaxies from the stars. This queried data is based on NED at the moment of Aug 2022. Then we merged this NED searched galaxy catalog with the supplemented SDSS DR17 catalog. We cross-match those two catalogs with matching threshold of $1''$ to resolve the duplications. We select the redshift of the supplemented SDSS DR17 over the redshift of the NED for the case of duplications. Our cluster identification and redshift evaluation processes based on this galaxy redshift catalog will be described at Section \ref{ssec:val}.


\section{Methods\label{sec:methods}}

\begin{figure}
\includegraphics[width=0.48\textwidth]{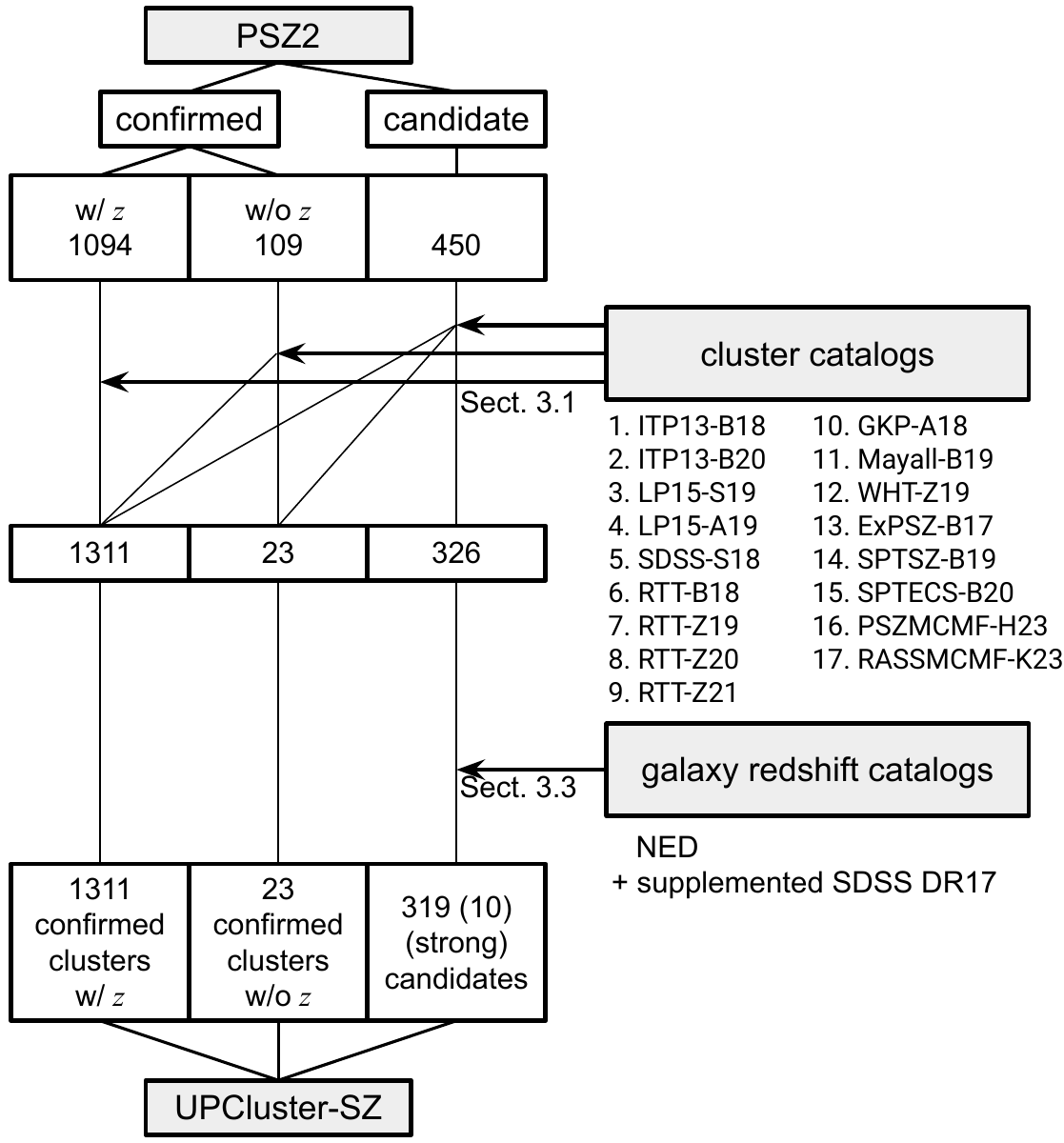}
\caption{The schematic flowchart of our compilation process. PSZ2 objects can be divided into 1203 confirmed clusters and 450 candidates. Among the confirmed clusters, 1094 clusters have their redshift information and 109 clusters do not have the redshift, before the inclusion of the cluster catalogs (see Section. \ref{ssec:clcatup} and \ref{ssec:clcats}) and validation using the galaxy catalogs (see Section. \ref{ssec:galcats} and \ref{ssec:val}). Cluster catalogs in the list are abbreviated with \texttt{Name-ReferenceYear}, which can be identified with the ordinal numbers that correspond to Column (1) in Table \ref{tab:clcat}.\label{fig:flowchart}}
\end{figure}

Here we explain the method we adopt to update the cluster validation and the redshift information, using the cluster and galaxy catalogs described in Section \ref{sec:sample&data}. The overall process of our update is illustrated in Figure \ref{fig:flowchart}. In short, we begin by merging the prioritized cluster catalogs. For remaining candidate clusters, we perform additional cluster identification process to provide strong candidates of clusters using the galaxy redshift catalog.
We count the number of galaxies with measured redshifts within a velocity range of $\Delta v < 4500$ km s$^{-1}$ as we move the reference redshift \zcl\ to compute the velocity range. Here, $\Delta v = c(z_{\rm gal}-z_{\rm cl})/(1+z_{\rm cl})$ represents the velocity difference between the galaxy redshift $z_{\rm gal}$ and the reference \zcl. We consider a cluster candidate as a strong candidate if their galaxy count exceeds nine at certain redshift, with the mean redshift recorded as the detection redshift.

In Section \ref{ssec:clcatup} we describe how we prioritize the cluster catalogs and which information is included for each cluster catalog. We explain the source of information that we used to determine the type of redshift measurement for each cluster in Section \ref{ssec:typeid}. We describe the parameters and criteria we used for  finding strong candidates using the galaxy catalog in Section \ref{ssec:val}. 

\subsection{Cluster Catalog Updates\label{ssec:clcatup}}

When we merge the validated clusters and their redshifts from the cluster catalogs into PSZ2, we assign the priority to each cluster catalog using three kinds of information: category of catalogs (follow-up and non-PSZ2 based), redshift type (uncertain, photometric, and spectroscopic), and observed number of galaxy spectra. This prioritization is necessary for duplicated entries across the cluster catalogs. We prefer the cluster catalogs from dedicated studies to validate the candidate clusters of PSZ2 and measure their redshifts. We select spectroscopic redshifts over photometric redshifts because photometric redshifts typically have significantly higher uncertainties than the spectroscopic redshifts. When the spectroscopic redshift is available from two or more cluster catalogs, we choose the cluster redshift derived from the highest number of galaxy spectra ($N_{\rm spec}$).

We apply this priority simply by manipulating the order of merging catalogs. We first update the clusters from follow-up catalogs. We begin by merging the most extensive ORM follow-up catalogs into ours. We update RTT follow-up clusters later because they typically have a small number of galaxy spectra than the ORM follow-up clusters. The merging sequence is in the order of cluster catalogs listed in Table \ref{tab:clcat} and Figure \ref{fig:flowchart}. We update a cluster if it is newly validated or if a higher-priority redshift becomes available compared to the previous step.

We mostly use the validated clusters from each cluster catalog. However, we also include the updated information on the cluster candidates from the ORM follow-ups. The ORM follow-ups provide the validation flag \texttt{Flag}=1, 2, 3, and ND, depending on their own validation criteria (see Table 2. of both \citealt{itp13a} and \citealt{lp15a}). They consider the cluster candidates with \texttt{Flag}=1, and 2 as validated clusters, whereas those with \texttt{Flag}=3 as ``weak associations'' with the SZ signal \citep{itp13a, lp15a}. These \texttt{Flag}=3 candidate clusters have redshift measurements while their validation status is unchanged. We update these redshifts as candidate redshifts, so users should be cautious when handling the redshift information for these candidates. 

We also update the validation flag as in PSZ2 \citep[\texttt{VALIDATION} column in PSZ2, see Table 9 and D.1. of][]{psz2}, with extended codes for the reference of the catalogs. The extended validation codes are listed in Table \ref{tab:clcat}. To encompass the non-detection (flagged as ``ND'') information of the ORM follow-ups, we make a new column for validation flag \texttt{VAL\_FLAG}. We denote validated clusters by ``V'', candidate clusters which still need validation as ``C'', and non-detections from ORM-follow-up as ``N'' in \texttt{VAL\_FLAG}.

\subsection{Identification of Redshift Type\label{ssec:typeid}}

We compile the type of redshift information and classify it into three cases: \texttt{spec}, \texttt{phot}, and \texttt{unct}. Each type corresponds to the spectroscopic redshift, the photometric redshift, and the uncertain type of redshift; if no redshift is available for a cluster, we omit this information from our records.
Here, the photometric redshift measurement includes the case from the scaling relation of the red sequence \citep[see for the concept,][]{gladders98}.
We first identify the type of redshift provided by PSZ2 for each validation source. Then we add the type information according to each cluster catalog when we merge them into PSZ2. These clarified redshift types are recorded in \texttt{Z\_FLAG} column of our updated catalog. When available, we also record the number of observed galaxies used to determine the cluster redshift.

To identify the redshift type, we divide the confirmed PSZ2 clusters using the validation code (\texttt{VALIDATION}). For example, the ENO follow-up validation sample, the redMaPPer non-blind sample within the SDSS footprint, the MCXC sample, and the ACT sample (\texttt{VALIDATION}=10, 13, 21, 23) are based on the spectroscopic redshift measurement. On the other hand, the PanSTARRS and the redMapper sample (\texttt{VALIDATION}=12, 24) are based on the photometric redshift measurement.

When necessary, we consider the redshift type of the individual clusters in some validation samples. Most of the RTT follow-up sample (\texttt{VALIDATION}=11) is based on spectroscopic observations, with the exception of four clusters (PSZ2 G048.21-65.00, PSZ2 G157.82+21.28, PSZ2 G223.04-20.27, and PSZ2 G224.01-11.14) which are estimated from the photometric data of RTT150 \citep[Table 1,][]{rttfollowup} hence marked as \texttt{phot}. The SDSS high-z sample \citep[\texttt{VALIDATION}=14, see Appendix A in][]{psz2} has two clusters with spectroscopic redshifts (PSZ2 G076.18-47.30 and PSZ2 G087.39+50.92) and two clusters with photometric redshifts (PSZ2 G089.39+69.36 and PSZ2 G097.52+51.70). For both the legacy validation sample from PSZ1 (\texttt{VALIDATION}=20) and the updated sample of the \Planck\ catalog of Sunyaev-Zeldovich sources \citep[PSZ1v2;][\texttt{VALIDATION}=25]{psz1v2}, we inherit the redshift flag (\texttt{Q\_Z}) from PSZ1v2. In this case, we changed the \texttt{estim} flag from PSZ1v2 to \texttt{phot}, because the distinction between the estimated redshift and the photometric redshift is unclear.

We identify the type of redshift for the SPT-SZ sample (\texttt{VALIDATION}=22) by matching the target IDs (\texttt{REDSHIFT\_ID} column in PSZ2) with those from the SPT-SZ catalog \citep{spt-sz}. The SPT-SZ catalog provides the \texttt{n\_z} column to mark the spectroscopic measurement. Three clusters (PSZ2 G347.58-35.35, PSZ2 G278.33-41.53, and PSZ2 G348.90-67.37) from the SPT-SZ sample do not follow the naming convention of SPT-SZ (\texttt{REDSHIFT\_ID}=ACO S 1121, ACO 3216, and ACO S 871, respectively), so we manually find their redshift type, which corresponds to \texttt{spec}, \texttt{phot}, and \texttt{spec}, respectively.

For the NED validation sample (\texttt{VALIDATION}=30), we search the redshift source at NED, using the NED identifier from \texttt{REDSHIFT\_ID}. We check the individual reference of the redshift measurement provided by the NED, and identify which type of redshift is measured. However, we fail to confirm the type for four clusters (PSZ2 G132.29+44.51, PSZ2 G227.63-52.28, PSZ2 G264.92+44.70 and PSZ2 G326.01+17.21). We mark the redshift type of these validated clusters with unidentified redshift as \texttt{unct}.

\subsection{Classification of Strong Candidates with Galaxy Redshift Catalogs\label{ssec:val}}

We use the catalog of galaxies with spectroscopic redshifts (Section \ref{ssec:galcats}), to classify the strong candidates and estimate their redshifts.  Previous studies \citep{burenin17, xclass} have determined the cluster redshift using the archival spectroscopic sample of galaxies. However, caution is required when validating and estimating redshifts using this method, particularly when only a small number of galaxy spectra are available \citep{burenin17}. Nevertheless, the high accuracy of the spectroscopic redshifts allows us to confidently identify a reliable clustering signal.

We try to adopt the detection process as simple as possible because our spectroscopic sample of galaxies is very heterogeneous in terms of its sample selection and completeness. Unlike the analysis of homogeneous photometric data, statistical approaches such as overdensity estimation \citep[e.g., ][]{madcow} and evaluating contamination \citep[e.g., ][]{madpsz} are very challenging to apply on the validation using the inhomogeneous spectroscopic sample. Many studies have already conducted such analysis using photometric data, constructing the cluster catalogs described in Section \ref{ssec:clcats}. Hence we focus on the identification of clusters using the spectroscopic redshift measurement of individual galaxies.

\begin{figure}
\gridline{\fig{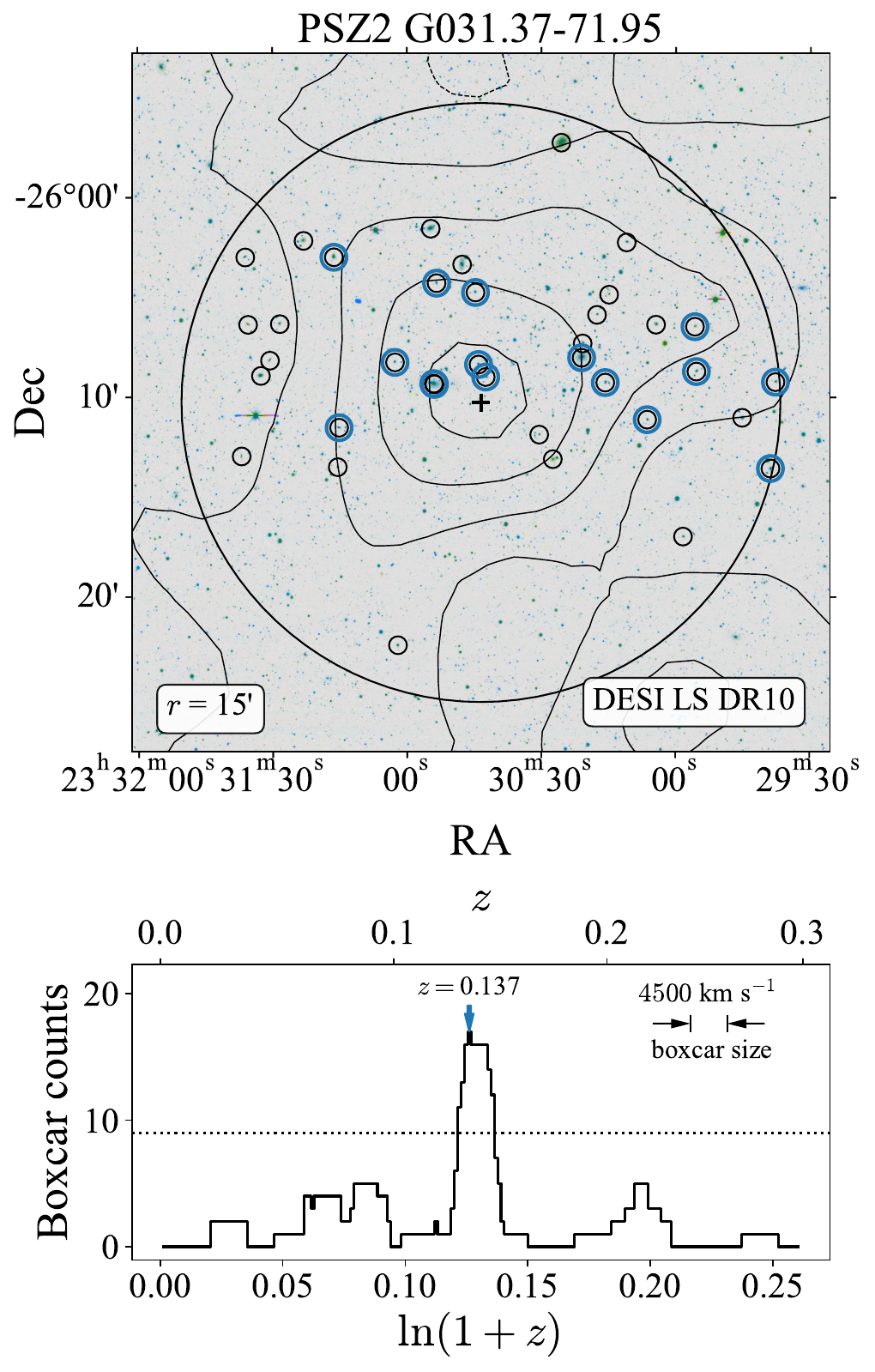}{0.48\textwidth}{}}
\caption{\textit{Top}: Cluster detection of PSZ2 G031.37-71.95 using the catalog of galaxies with spectroscopic redshifts. The galaxies that comprise the detection peak are shown by blue circles, while all the available spectroscopic redshift galaxies inside the radius $15'$ from its SZ center are represented by black circles. The searching radius is shown by the thick black circle, where its center is located at the PSZ2 centroid marked as a black cross. The contour of MILCA Compton y-map provided by \Planck\ is overlaid on the DESI Legacy Survey DR10 image, with thin solid (positive) and dashed (negative) lines. \textit{Bottom}: The number count of spectroscopic redshift galaxies in the range of $\Delta v < 4500$ km s$^{-1}$ with respect to the trailing redshift. The bin size is smaller than the velocity range by factor of 11, hence its name ``boxcar count''. Our detection threshold of nine galaxies is shown by the dotted line. The corresponding detection peak for the blue-circled galaxies in the velocity range is annotated by blue arrow with its corresponding redshift ($z=0.137$). \label{fig:method}}
\end{figure}

To identify a clustering signal in a given field of PSZ2 source, we detect an excess on the number of galaxies through the redshift bins. We use the spectroscopic catalog of galaxies within $15'$ from the center of PSZ2 objects (black open circles in Figure \ref{fig:method}, see Section \ref{ssec:galcats}). We first construct a redshift grid in the logarithmic space, $\ln(1+z)$, to keep the velocity scale uniform. Considering the typical velocity dispersion $\sim 1000$ km s$^{-1}$ of the galaxy members in a cluster, we count the number of galaxies within a boxcar that has a size of $\Delta \ln (1+z)=0.015$, which corresponds to $\Delta v \sim 4500$ km s$^{-1}$ on the velocity scale. We set this size of $\Delta v$ sufficiently large not to lose any possible galaxy members. To avoid losing galaxy members at the fixed boundary, we set the sampling of the trailing redshift to be 11 times smaller than the boxcar size. The bottom panel of Figure \ref{fig:method} shows an example of this process for PSZ2 G031.37-71.95. 

To find a local maximum of the boxcar counts that corresponds to the cluster location, we use \texttt{find\_peaks} function from the \textsc{Scipy} library. We select the local maxima of the count through the redshift, ensuring that the distance between the neighboring peaks is larger than the boxcar size.
Then, we consider a peak with a height greater than nine as a detection of a cluster signal, if present.
We set the detection threshold to nine by assuming a simple Poisson error, resulting in an expected signal-to-noise ratio of three.

We only adopt this method to the PSZ2 objects of $Q_{\rm neural}>0.4$ to avoid the false signal caused by foreground contamination from infrared sources. PSZ2 provides the quality flag $Q_{\rm neural}$, which is determined using the supervised artificial neural network \citep{aghanim15}. This flag indicates the contribution of sources to \Planck\ SED other than the SZ effect, with particular sensitivity to the infrared-induced spurious detections \citep[Section 4.7, ][]{psz2}. Because low-quality detections are typically indicated by $Q_{\rm neural}<0.4$, we exclude these low-quality SZ detections ($\sim 27$\%) from our spectroscopic detection process.

In addition, to minimize the contamination by nearby clusters and the misidentification of wrong counterparts, we discard the case where the projected physical distance ($d$) between the PSZ2 center and the mean center of galaxies contributing to the peak is greater than 1.5 Mpc. We select the highest peak for the case of multiple peaks passing this distance criterion. Once the detection peak is determined, we record the mean redshift of the member galaxies of the peak as the redshift of the cluster. We update the redshift of these strong candidates to our updated catalog. We also record the number of galaxies in \texttt{N\_SPEC} column and the type of redshift measurement \texttt{Z\_FLAG} as \texttt{spec}. We separate this sample of strong candidates from the other candidates, marking them as ``S'' in \texttt{VAL\_FLAG}.

\subsection{Mass Updates Using SZ Mass Proxy\label{ssec:massup}}

We add the information on the mass of clusters that are updated with new validation status and new redshift information. We calculate the SZ mass, $M_{\rm SZ}$ (and its upper and lower errors) with the new redshift estimates using the $M_{\rm SZ}(z)$ table of PSZ2 \citep[see Appendix D.2. of][]{psz2}. We update the \texttt{MSZ} (\texttt{MSZ\_ERR\_UP} and \texttt{MSZ\_ERR\_LOW}) column with the $M_{\rm SZ}$ values (and their upper and lower errors) that we obtained.


\section{Results\label{sec:result}}

Here, we summarize the results from the validation process and the classification of strong candidates (see Sections \ref{ssec:clcatup} and \ref{ssec:val}) in Section \ref{ssec:valresult}, and present the resulting catalog of UPCluster-SZ in Section \ref{ssec:catresult}. 

\subsection{Updates with Cluster and Galaxy Catalogs\label{ssec:valresult}}

From the cluster catalogs compiled in Section \ref{ssec:clcats}, 404 clusters are validated or redshift-updated through the merging process in Section \ref{ssec:clcatup}. The number of validated clusters using the cluster catalogs is 139; among them, 77 and 62 have spectroscopic and photometric redshift measurements, respectively. The number of confirmed PSZ2 clusters with updated redshift information is 265; among them, 225 and 40 are updated with spectroscopic and photometric redshifts, respectively.

Using the galaxy catalogs compiled in Section \ref{ssec:galcats}, we could find the cluster signals for 18 cluster candidates. To quantify the reliability of our selection criteria for clusters, we apply the same method to the confirmed clusters of known redshift. The top panel of Figure \ref{fig:red1-1} shows the comparison of the reference redshift and the redshift derived from our detection method. The redshifts from our method match well with the reference redshift for most of the clusters.
However, there are some cases where we underestimate (or overestimate) the cluster redshifts provided by previous studies. Such cases can happen if we adopt lower thresholds as in Figure  \ref{fig:red1-1}. We identify outliers as cases where the relative redshift difference $|\Delta z| /(1+z)$ exceeds 0.015 (equivalent to $|\Delta v|>$ 4500 km s$^{-1}$). We can reduce the outlier fraction using higher thresholds, with approximately 10\% observed for our threshold of nine.
When examining the outlier fraction across different redshift bins, as illustrated in the bottom and side panels of the upper part of Figure \ref{fig:red1-1}, we observe an increasing trend. Both the true (reference) and estimated redshifts from our method show that the outlier fraction exceeds 20\% at high-z bins.

\begin{figure}
\plotone{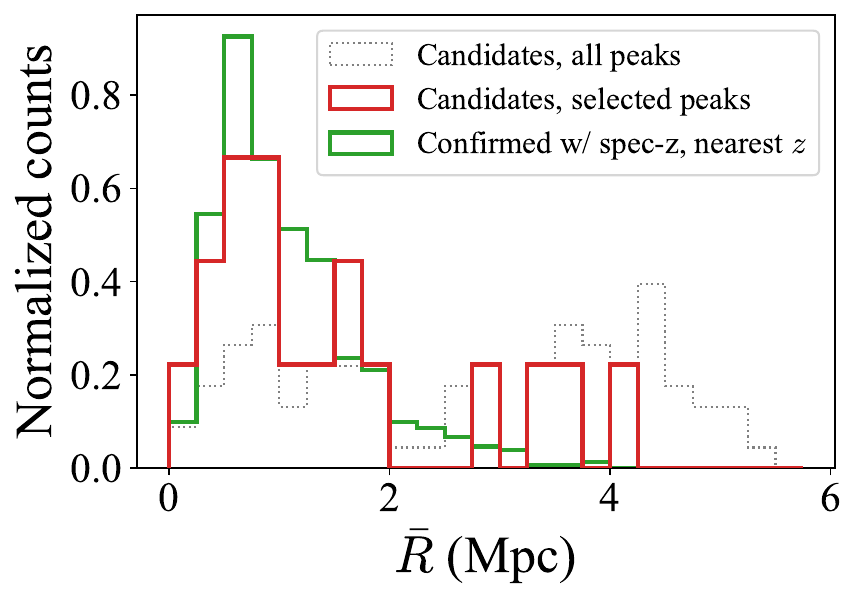}
\caption{The normalized histogram of detected cluster signal using our method for confirmed clusters versus $\bar{R}$, which represents the mean distance of member galaxies from the PSZ2 center. The dotted gray line denotes all cluster detection signal without any selection process, and the red line shows the selected peak by the number of galaxies and the galaxy center offsets relative to the PSZ2 center. The green line shows the detections of each confirmed cluster, whose redshifts are the closest to their reference redshifts. The detections of candidate clusters show additional bump on $\bar{R}>2$, which is not significant for the confirmed clusters. \label{fig:Rhist}}
\end{figure}

To decrease the chance of false detection further, we adopt an additional selection criterion of $\bar{R}\leq2$ Mpc, where $\bar{R}$ is the mean projected physical distance of member galaxies from the PSZ centroid. This criterion helps to prioritize clusters with concentrated member galaxies in a small region. Figure \ref{fig:Rhist} shows the distribution of the mean projected distance $\bar{R}$. Here we select the cluster detections (green line) that are closest to the reference redshifts of the confirmed clusters, with a peak around $\bar{R}\sim0.6$ Mpc. Among all cluster detections in the field of the candidate clusters (gray dotted line), many detections have the mean distance of member galaxies greater than 2 Mpc, indicating a distribution that is very different from that of the confirmed clusters. The selected detections of candidate clusters, by richness and mean offset ($d\leq1.5$ Mpc), also show bimodal distribution separated at $\bar{R}=2$ Mpc. Hence we adopt the threshold of $\bar{R}\leq2$ Mpc, classifying detection signals with concentrated member galaxies as strong candidates, where their distribution is similar to the distribution of the confirmed clusters.

As a result of applying the cluster detection and selection criteria using the galaxy catalog, we classify 10 candidates as strong candidates. During this process, we have also conducted visual inspection of each candidate, and have revised the classification if necessary (see Section \ref{ssec:indv}). 

Furthermore, we assess the possibility of updating redshift information for validated clusters without spectroscopic redshifts, following the same procedure as done for the candidate clusters. For clusters with photometric redshifts, we only consider the peaks within a redshift difference of $|z-z_{\rm phot}|/(1+z)<0.06$, to utilize the photometric redshift information as in \citet{burenin17}. This threshold is based on the largest photometric redshift uncertainty reported in the cluster catalogs \citep[$\sigma_{z_{\rm phot}}=0.062$ from][]{zohren19}. Despite our efforts, the spectroscopic data did not provide any new updates for the redshifts of these validated clusters.

\subsection{The Resulting UPCluster-SZ Catalog\label{ssec:catresult}}

Using the process described in the previous section, we have obtained the updated PSZ2 catalog, named UPCluster-SZ, which is an all-sky galaxy cluster catalog (Table \ref{tab:resultcat}). In the UPCluster-SZ, we provide information on the original PSZ2 detection \citep[PSZ2 index, name, SNR, sky position, positional uncertainty, detection pipeline, $Y_{\rm 5R500}$, $M_{\rm SZ}$;][]{psz2} along with the redshift and validation code. We include new columns indicating redshift type (\texttt{Z\_FLAG}), number of galaxy spectra (\texttt{N\_SPEC}), validation flag (\texttt{VAL\_FLAG}), and additional notes (\texttt{NOTES}) to track the updating process and benefit user applications such as sample selection.

Table \ref{tab:resultstat} summarizes the numbers for our UPCluster-SZ in terms of validation status and the availability of redshift information. Our update affects a total of 547 PSZ2 clusters. Specifically, we update the redshift information of 265 confirmed clusters in PSZ2, along with the new validations of 139 PSZ2 cluster candidates. In total, UPCluster-SZ contains 1334 confirmed clusters, 10 strong candidates among 197 candidates, and 122 non-detections. The non-detections should be considered as candidate clusters even though some of them could be false SZ signals, because they are not the confirmed false signals but the candidate clusters whose optical counterparts are not visible in deep optical images \citep{itp13a, lp15a}. Exceptions to the updating process outlined in Section \ref{sec:methods} are addressed in Section \ref{ssec:indv}.

\movetabledown=1.5in
\begin{rotatetable*}
\begin{deluxetable*}{ccccccccccccc}
\tablecaption{UPCluster-SZ\label{tab:resultcat}}
\tablehead{
\colhead{Index} & \colhead{Name} & \colhead{SNR} & \colhead{RA} & \colhead{DEC} & \colhead{POS\_ERR} & \colhead{$z$} & \colhead{$z$ Flag} & \colhead{$N_{\rm spec}$} & \colhead{$Y_{5R500}$} & \colhead{Validation} & \colhead{Validation Flag} & \colhead{$M_{\rm SZ}$}\\
\colhead{} & \colhead{} & \colhead{} & \colhead{deg} & \colhead{deg} & \colhead{arcmin} & \colhead{} & \colhead{} & \colhead{} & \colhead{0.001 arcmin$^2$} & \colhead{} & 
\colhead{} & \colhead{$10^{14}M_{\odot}$}}
\colnumbers
\startdata
1 & PSZ2 G000.04+45.13 & 6.75 & 229.1905 & -1.0172 & 4.11 & 0.1198 & spec & -- & $5.48 \pm 1.90$ & 20 & V & $3.96^{+0.39}_{-0.37}$  \\
2 & PSZ2 G000.13+78.04 & 9.26 & 203.5587 & 20.2560 & 2.06 & 0.1710 & spec & -- & $4.36 \pm 1.84$ & 20 & V & $5.12^{+0.35}_{-0.32}$  \\
3 & PSZ2 G000.40-41.86 & 9.70 & 316.0845 & -41.3542 & 2.43 & 0.1651 & spec & -- & $4.51 \pm 1.05$ & 21 & V & $5.30^{+0.32}_{-0.34}$  \\
4 & PSZ2 G000.77-35.69 & 6.58 & 307.9728 & -40.5987 & 2.34 & 0.3416 & spec & -- & $1.61 \pm 0.52$ & 21 & V & $6.33^{+0.59}_{-0.61}$  \\
5 & PSZ2 G002.04-22.15 & 5.13 & 291.3596 & -36.5179 & 5.02 & -1.0000 & -- & -- & $1.93 \pm 0.64$ & -1 & C & --  \\
6 & PSZ2 G002.08-68.28 & 4.75 & 349.6324 & -36.3326 & 5.43 & 0.1400 & spec & -- & $3.13 \pm 1.59$ & 20 & V & $2.84^{+0.37}_{-0.46}$  \\
7 & PSZ2 G002.42+69.64 & 4.62 & 210.9933 & 15.6884 & 2.43 & 0.1802 & spec & 3 & $2.69 \pm 1.31$ & 60 & V & $3.51^{+3.97}_{-3.02}$  \\
8 & PSZ2 G002.77-56.16 & 9.20 & 334.6595 & -38.8794 & 2.28 & 0.1411 & spec & -- & $4.05 \pm 1.03$ & 21 & V & $4.41^{+0.30}_{-0.30}$  \\
9 & PSZ2 G002.82+39.23 & 8.07 & 235.0152 & -3.2851 & 2.43 & 0.1533 & spec & -- & $9.70 \pm 2.64$ & 21 & V & $5.74^{+0.48}_{-0.48}$  \\
10 & PSZ2 G002.88-22.45 & 4.80 & 292.0148 & -35.8761 & 7.24 & -1.0000 & -- & -- & $1.94 \pm 0.71$ & -1 & C & --  \\
11 & PSZ2 G003.21-76.04 & 4.79 & 358.3512 & -33.2932 & 5.27 & 0.6576 & phot & -- & $1.04 \pm 0.38$ & 64 & V & $5.88^{+6.54}_{-5.12}$  \\
12 & PSZ2 G003.91-42.03 & 9.33 & 316.4675 & -38.7532 & 2.43 & 0.1521 & spec & -- & $5.36 \pm 1.11$ & 21 & V & $4.72^{+0.32}_{-0.37}$  \\
13 & PSZ2 G003.93-59.41 & 17.36 & 338.6081 & -37.7413 & 2.43 & 0.1510 & spec & -- & $5.66 \pm 0.72$ & 21 & V & $7.19^{+0.26}_{-0.26}$  \\
14 & PSZ2 G004.04+42.23 & 4.76 & 233.2733 & -0.6754 & 5.28 & 0.1548 & spec & 3 & $3.21 \pm 1.79$ & 60 & V & $3.64^{+4.11}_{-3.14}$  \\
15 & PSZ2 G004.13+56.81 & 6.76 & 221.8651 & 8.4517 & 3.40 & 0.3800 & spec & -- & $3.09 \pm 0.93$ & 20 & V & $7.12^{+0.64}_{-0.68}$  \\
16 & PSZ2 G004.45-19.55 & 9.07 & 289.2667 & -33.5416 & 2.43 & 0.5400 & spec & -- & $1.96 \pm 0.36$ & 20 & V & $10.36^{+0.68}_{-0.72}$  \\
17 & PSZ2 G005.91-28.26 & 5.98 & 299.8069 & -34.8918 & 2.06 & 0.1728 & spec & -- & $2.29 \pm 0.89$ & 21 & V & $4.29^{+0.39}_{-0.42}$  \\
18 & PSZ2 G006.05+29.43 & 5.09 & 244.4061 & -7.2625 & 3.46 & 0.2523 & spec & -- & $1.79 \pm 0.74$ & 20 & V & $5.52^{+0.65}_{-0.71}$  \\
19 & PSZ2 G006.16-69.49 & 4.55 & 350.5375 & -34.5752 & 4.46 & -1.0000 & -- & -- & $1.33 \pm 0.68$ & -1 & C & --  \\
20 & PSZ2 G006.38+62.03 & 5.65 & 218.3080 & 12.4863 & 2.43 & 0.2370 & spec & 2 & $1.36 \pm 0.42$ & 60 & V & $4.83^{+5.38}_{-4.26}$  \\
\enddata
\tablecomments{First 20 rows of UPCluster-SZ. (1) PSZ2 Index (ID). (2) PSZ2 Name (NAME). (3) PSZ2 signal-to-noise ratio (SNR). (4) Right ascension in degree (RA; J2000). (5) Declination in degree (DEC; J2000). (6) Positional uncertainty in arcmin (POS\_ERR). (7) Redshift (Z). (8) Redshift flag (Z\_FLAG; Section \ref{ssec:typeid}). (9) Number of galaxy spectra (N\_SPEC; Section \ref{ssec:clcatup}). (10) $Y_{\rm 5R500}$ (Y5R500). (11) Validation code (VALIDATION; Section \ref{ssec:clcatup}). (12) Validation flag (VAL\_FLAG; Section \ref{ssec:clcatup}). (13) $M_{\rm SZ}$ (MSZ). We omit here the column of the notes on individual objects (NOTES).}
\tablerefs{Citations from the NOTES column: \citet{lp15b}, \citet{itp13a}, \citet{boada19}, \cite{burenin17}, 
\citet{burenin18}, \citet{streblyanska18}, \citet{lp15a}, \citet{zaznobin19}, and \citet{zohren19}.}
\end{deluxetable*}
\end{rotatetable*}

\begin{figure}
\plotone{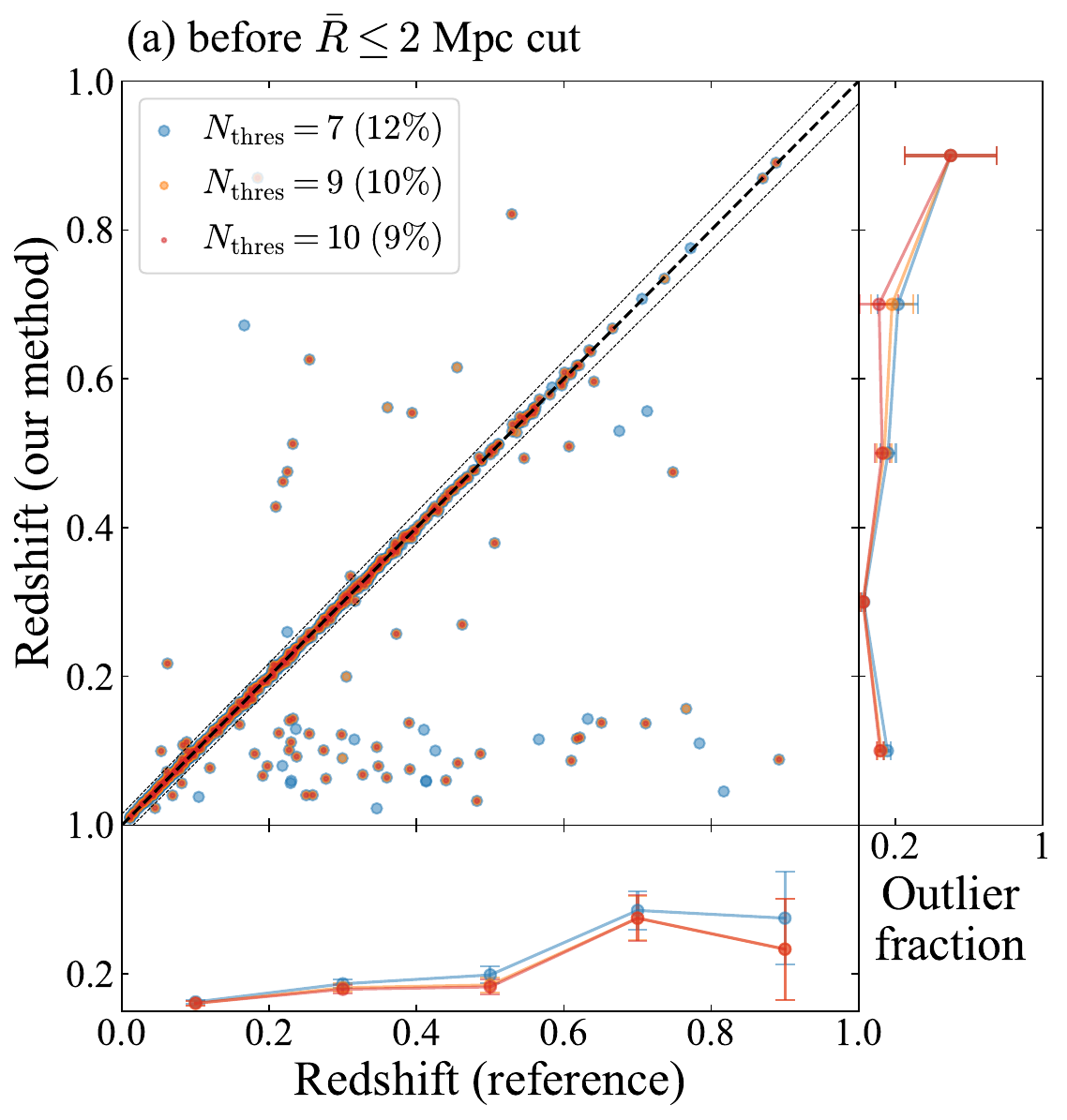}
\plotone{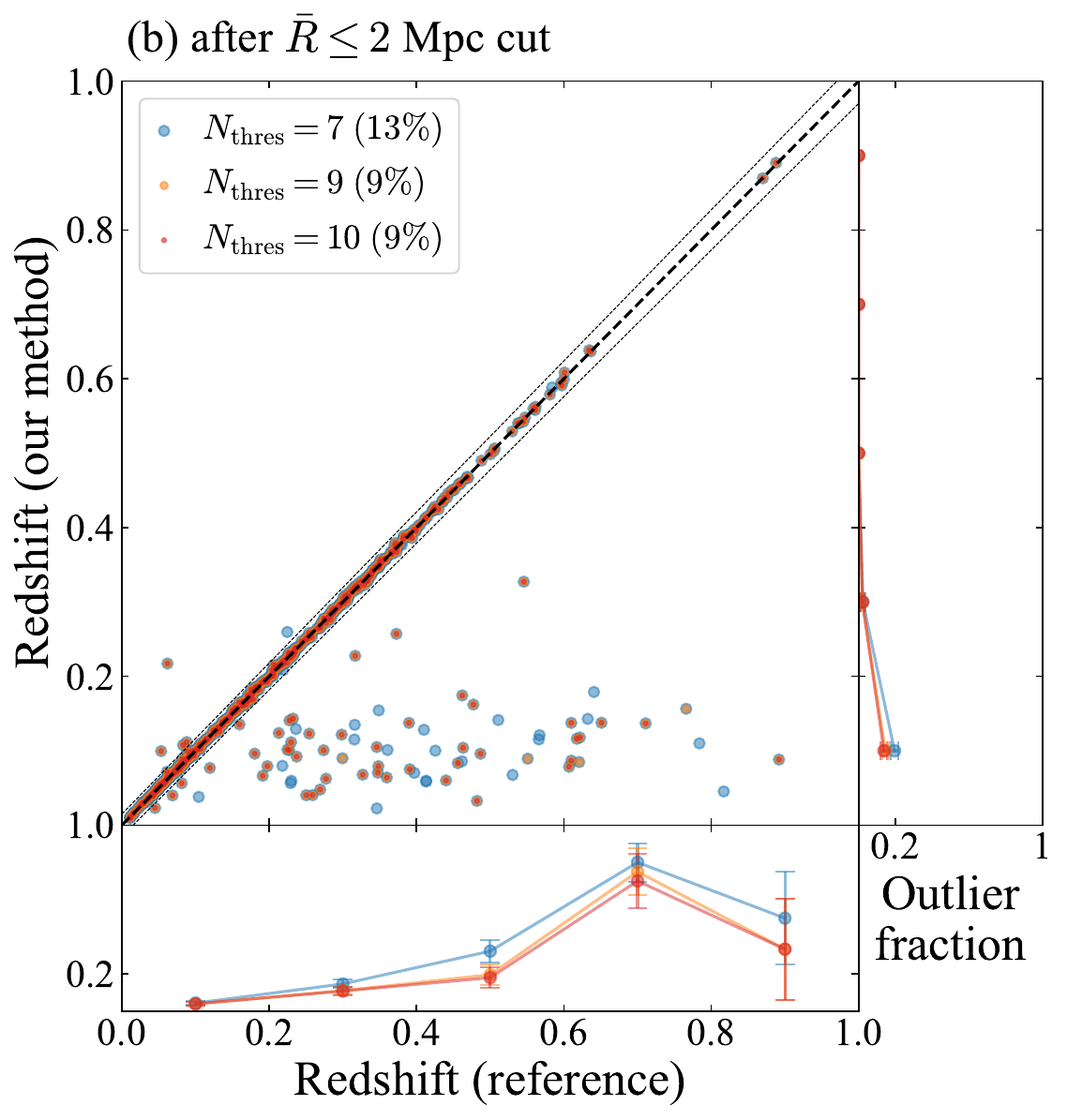}
\caption{
Comparison between redshift estimations of the confirmed clusters using our galaxy catalog detection method and their reference redshift, shown before (a) and after (b) applying the $\bar{R} \leq 2$ cut. The symbol color and size represent different detection thresholds: blue, orange, and red correspond to thresholds of 7, 9, and 10, respectively. We define outliers as cases where the relative redshift difference $|\Delta z| /(1+z)$ exceeds 0.015, which corresponds to a velocity limit of 4500 km s$^{-1}$. The dotted lines in the figure represent this velocity limit. The legend provides the total outlier fraction for each threshold, enclosed in parentheses. The bottom and right panels show the outlier fraction as a function of redshift, sharing the color scheme with the main panels.
\label{fig:red1-1}}
\end{figure}

\begin{figure}
\plotone{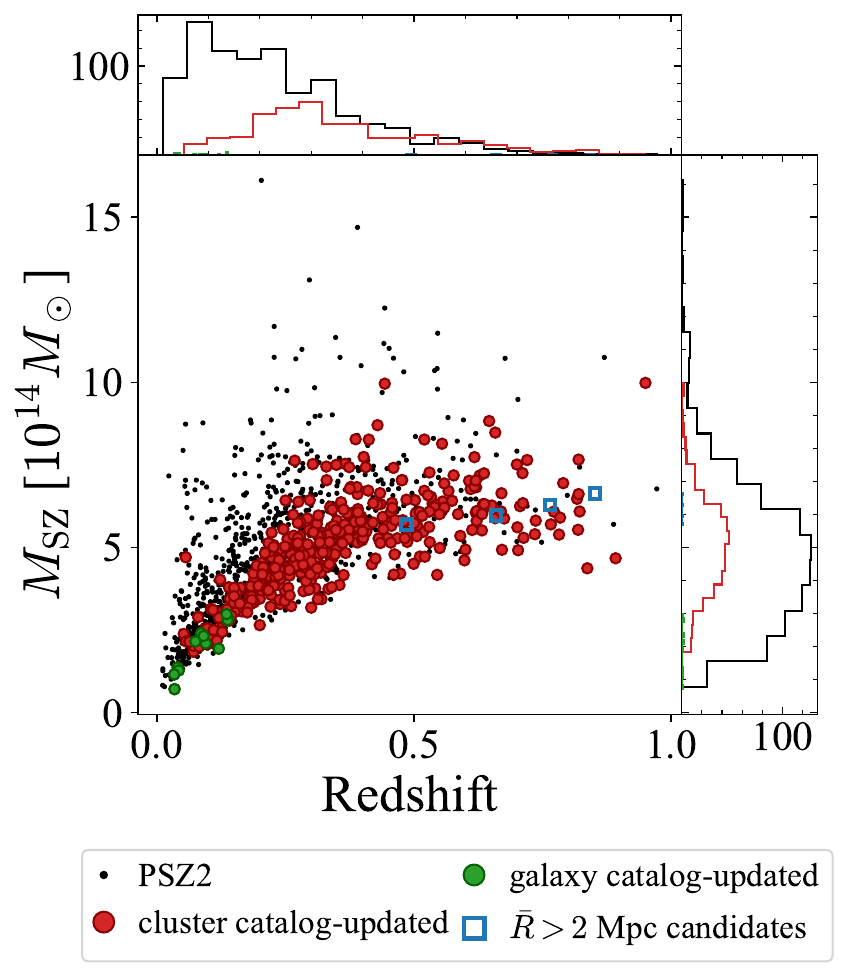}
\caption{Mass-redshift distribution of galaxy clusters in our UPCluster-SZ. The black dots represent the confirmed cluster of the original PSZ2 whose redshift information is available. The red dots represent the confirmed clusters updated from the cluster catalogs, and the green dots show the strong candidates classified based on our method using the galaxy catalog. The blue squares indicate the candidates of cluster detection with only $\bar{R}>2$ Mpc, which is excluded for the strong candidates. Histograms for the redshift and mass are displayed along each axis, with the same colors for the main panel. Candidate clusters and confirmed clusters without any redshift measurements are not shown in this figure. \label{fig:result}}
\end{figure}

Figure \ref{fig:result} shows the mass-redshift ($M_{\rm SZ}$-$z$) distribution of the subsample of UPCluster-SZ whose redshifts are available. Newly confirmed galaxy clusters from the cluster catalog (red symbols) and the strong candidates identified with the spectroscopic redshift data (green symbols) fall in the regime where both cluster mass and redshift are low; the mass distribution in the right panel generally follows that of the previously confirmed clusters at $z>0.2$. This may be attributed to observational bias, as the higher-mass sample is likely to be detected already, especially in the local universe. On the other hand, this could be partly because of underestimation of cluster redshifts as in Figure \ref{fig:red1-1}.

There are still 319 candidate clusters in UPCluster-SZ that require further confirmation. In addition, further update is required for 204 confirmed clusters only having photometric redshifts and 23 confirmed clusters without any redshift information. For 122 candidates that have been categorized as non-detections in the previous follow-up studies, further observations are necessary to examine whether they are false detections or not.

\begin{deluxetable}{lccccr}
\tablecaption{Result Statistics of UPCluster-SZ\label{tab:resultstat}}
\tablehead{
\colhead{Redshift} & \colhead{Type} & \colhead{Confirmed} & \colhead{Candidate} & \colhead{Non-detection} & \colhead{Total}}
\startdata
{}        & spec & 1103 & 18  & {}  & 1121 \\
Available & phot & 204  & 2   & {}  & 206  \\
{}        & unct & 4    & {}  & {}  & 4    \\
Unknown   & {}   & 23   & 177 & 122 & 322  \\
\cline{1-6}
Total     & {}   & 1334 & 197 & 122 & 1653 \\
\enddata
\tablecomments{The types of the redshift are the spectroscopic redshift, photometric redshift, and uncertain redshift (see Section \ref{ssec:typeid}). The redshift information for non-detections is not displayed because the optical counterpart of the SZ sources must be identified \textit{a priori} to the measurement of their redshift. The redshift information for the candidate clusters is obtained from the ORM follow-ups, with \texttt{Flag}=3.}
\end{deluxetable}


\section{Discussion\label{sec:discussion}}

\subsection{Reliability of Our Sample\label{ssec:relval}}
The PSZ2 galaxy cluster catalog is constructed from the all-sky survey data with the SZ effect, which offers the homogeneous selection of galaxy clusters. This makes the \Planck\ SZ catalog very valuable for testing cosmological models, as understanding the selection function is crucial to ensure unbiased cosmological analyses \citep{allen11} in the era of precision cosmology.
Therefore, it is essential to identify genuine clusters from the list of SZ-detected sources through follow-up observations \citep{vanderburg16, zohren19}. This helps to obtain a deeper understanding of the selection function and to examine the purity of the SZ cluster signal detections from \Planck\ ; the expected purity of PSZ2 is 83\%-87\% \citep{psz2}.

One of the challenges for a SZ-based cluster survey is to understand the effect of contamination \citep{allen11}. \citet{psz2} points out that the major contaminant of SZ sources is the infrared emission from dust near the Galactic plane, which were actually reported in some studies \citep[e.g., ][]{itp13a,streblyanska18,lp15a,burenin17,khatri16}. We have also found such cases; for example, PSZ2 G136.31+54.67 is located at an excess in the SFD dust map \citep{sfd98}, which shows no clustering feature on the optical image of the DESI Legacy Survey image.

We test the false-positive rate of our detection method using the galaxy catalog. The bottom panel of Figure \ref{fig:red1-1} shows the selected redshift of the validated sample compared to the reference redshift. After we apply the additional criterion of $\bar{R}\leq2$ Mpc, we reduce the number of cases overestimating the cluster redshift.  On the other hand, the number of cases underestimating the cluster redshift increases with this criterion, resulting in only a small improvement ($\sim 1$\%) in the total outlier fraction.
Regarding the estimated redshift bins, the outlier fraction remains below 2\%, with the exception of the lowest redshift bin (see the bottom side panel of Figure \ref{fig:red1-1}). Given that our entire sample of strong candidates falls within this bin, we anticipate the false detection rate to be approximately 15\% for these candidates.

\begin{figure}
\plotone{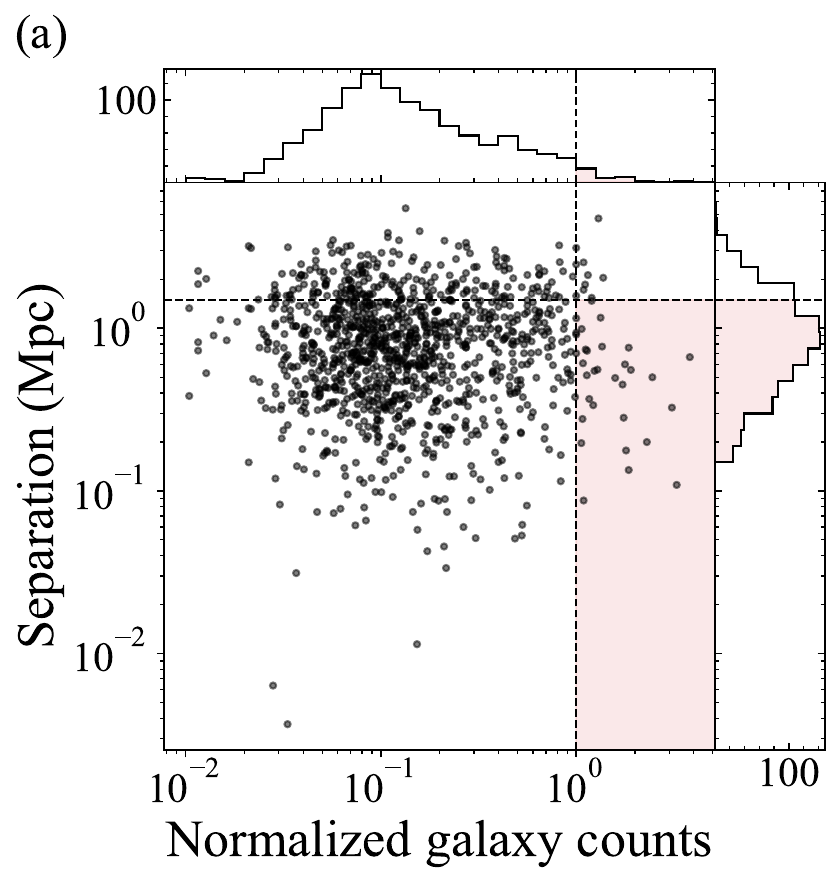}
\plotone{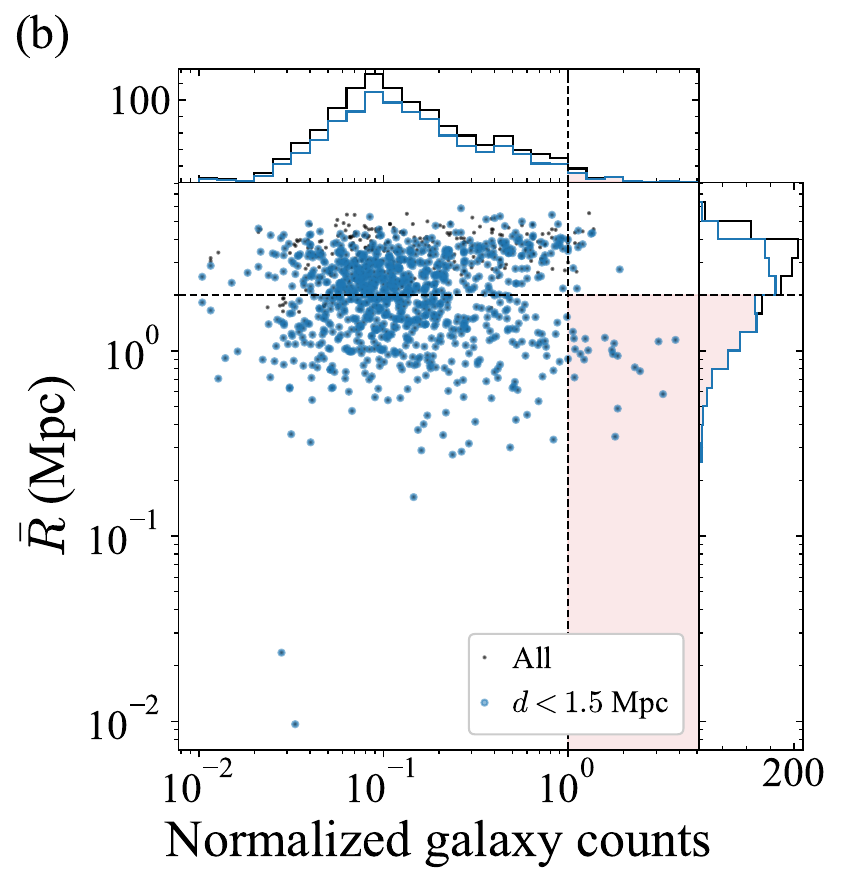}
\caption{The distribution of cluster detections obtained from our method (Section \ref{ssec:val}) for the confirmed PSZ2 clusters, with respect to the normalized galaxy counts and (a) separation and (b) $\bar{R}$. The normalized galaxy count is the galaxy count normalized by the number of member galaxies of the detections of their closest redshift. The separation $d$ is the offset of the optical galaxy center from the PSZ2 center, and $\bar{R}$ is the mean distance of the galaxies from the PSZ2 center.  In panel (b), blue represents detections that satisfy the separation criterion ($d<1.5$ Mpc), while black dots represent all detections prior to the separation cut. The red-shaded area indicates the region where detections may result in false-positive selections.
 \label{fig:normNz}}
\end{figure}

We expect that the false detection rate is sufficiently low under the assumption that our candidate clusters (and clusters confirmed without redshift) have spectroscopic measurements similar to those of the validated sample.
The distribution of all detected clustering signals for the validated sample is shown in the top panel of Figure \ref{fig:normNz}, illustrating the separation (projected distance between the PSZ center and the mean center of the galaxies) and the normalized galaxy count (number of member galaxies of the cluster signal normalized by the galaxy number of the signal selected by the closest redshift from the reference redshift) of the signals.
The false detections with our criteria fall in the red-shaded area of Figure \ref{fig:normNz}, which are only $\sim 2.4$ \% of the whole detections. Similarly, the bottom panel of Figure \ref{fig:normNz} shows the distribution after the separation cut ($d<1.5$ Mpc), with respect to the mean distance of the member galaxies from the PSZ center ($\bar{R}$) and the normalized galaxy count. In this case, the false detection rate slightly reduced to $\sim 2.1$ \%. In summary, we expect the contamination rate to be 2\%-15\%, which is based on the two distinct statistics; the false positive rate discussed here (Figure \ref{fig:normNz}) and the outlier fraction discussed in Section \ref{sec:result} (Figure \ref{fig:red1-1}).

We believe that this process can detect relatively unrelaxed clusters, which corresponds to the Bautz-Morgan type III \citep{bmtype}. We can take advantage of the spectroscopic sample of galaxies to find galaxy overdensity in a given field that does not show obvious concentration of galaxies by eye. There are many previous studies that have identified the optical counterparts of PSZ cluster candidates by detecting the red sequence \citep[e.g., ][]{redmapper, vanderburg16, zohren19}; this is because the high significance (S/N$>$4.5) of \Planck\ SZ detection tend to select massive clusters (e.g., $M_{500}>5\times10^{14}M_{\odot}$ at $z\sim0.5$, with 80\% completeness), which contain many evolved galaxies that form a prominent red sequence in the color-magnitude diagram. On the other hand, our method allows us to detect the clusters even with relatively smaller masses at low redshifts.

\begin{figure}
\includegraphics[width=0.48\textwidth]{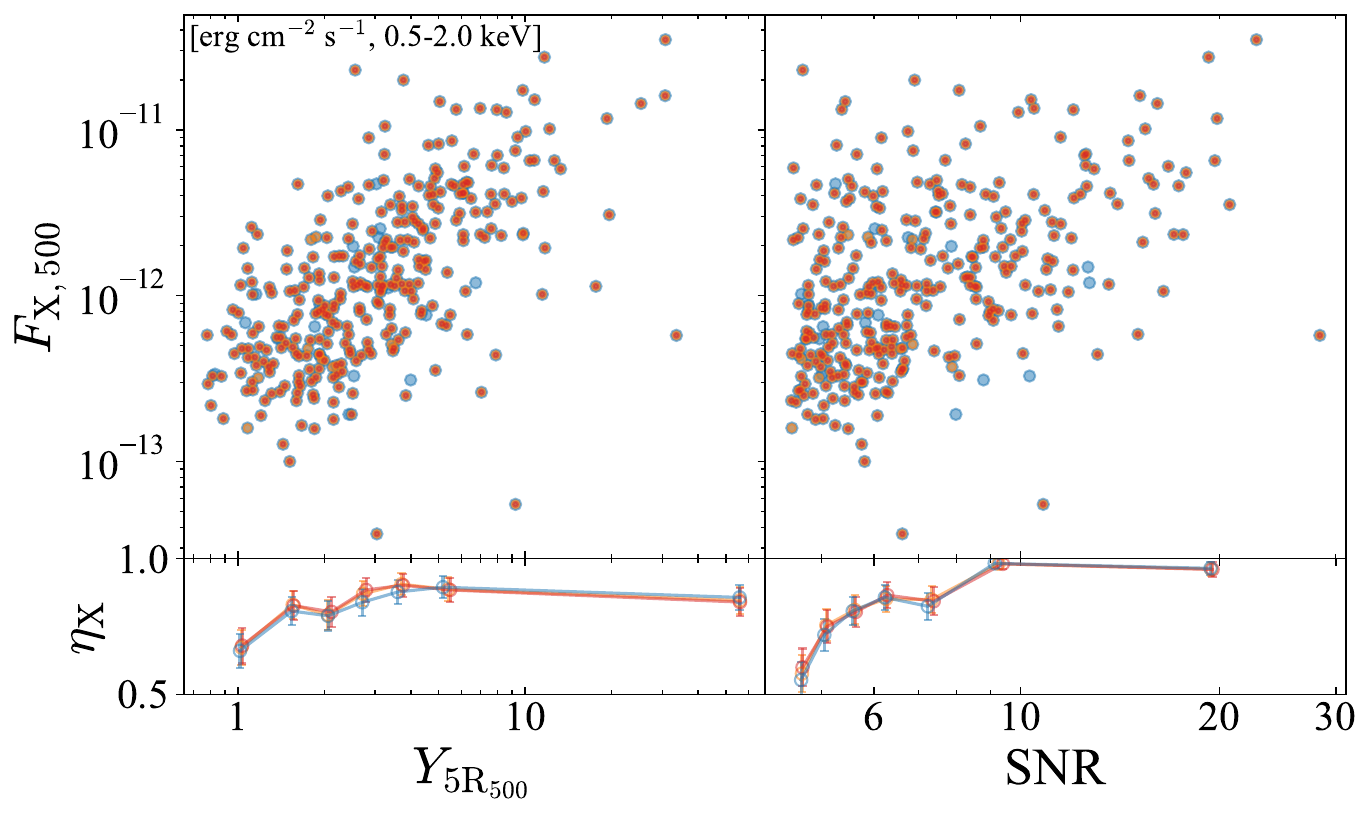}
\caption{The X-ray flux at 0.5-2.0 keV and the fraction of clusters $\eta_X$ with X-ray counterparts from the RASS-MCMF catalog, plotted against the integrated Compton y-parameter ($Y_{5R_{500}}$) and the SNR of the SZ signal. The colored markers represent the confirmed clusters blindly detected using our galaxy catalog method, following the same color scheme as in Figure \ref{fig:red1-1}, corresponding to thresholds of 7 (blue), 9 (orange), and 10 (red). The SZ parameters ($Y_{5R_{500}}$ and SNR) are binned to ensure an equal number of elements within each bin.
 \label{fig:xcomp}}
\end{figure}

In addition, we can compare our catalog with the one based on X-ray detection (e.g. RASS-MCMF) to examine the reliability of our strong candidates.
This comparison is useful because X-ray emission from hot gas of a galaxy cluster can confirm the presence of a genuine cluster when combined with the SZ detection.
The top left panel of Figure \ref{fig:xcomp} shows this correlation between the X-ray flux and the SZ Compton y-parameter.
As RASS-MCMF clusters associated with PSZ2 are already incorporated into our catalog as validated ones, none of our strong candidates have corresponding X-ray counterparts in RASS-MCMF.

To assess the reliability of the candidates without RASS X-ray detection, we calculate the fraction of X-ray-detected clusters among those already validated in our UPCluster-SZ using RASS-MCMF clusters. To avoid the confusion from the complicated masked area of RASS-MCMF, we focus only on the objects in a clean region with galactic latitudes of $|b|>45^{\circ}$. We then determine $\eta_X$, the fraction of clusters with RASS-MCMF counterparts within a $5'$ matching radius. We calculate $\eta_X$ for various $Y_{5R_{500}}$ and SNR bins, where $Y_{5R_{500}}$ represents the integrated Compton y-parameter provided by \Planck\ \citep{psz2}. The $\eta_X$ increases with $Y_{5R_{500}}$ and SNR as expected (see bottom panels of Figure \ref{fig:xcomp}), indicating that $\eta_X$ does not reach 100\% for the clusters with low $Y_{5R_{500}}$ and SNR. This suggests the presence of genuine clusters without X-ray detection (potentially due to X-ray detection limits) even though they are already validated through various methods (see Section \ref{ssec:clcats}). Given that our strong candidates have $Y_{5R_{500}}$ values ranging from 0.7 to 11.8 (median 2.3) and SNR values from 4.5 to 7.2 (median 5.0), and considering the relatively low $\eta_X$ in these ranges, we can conclude that the lack of X-ray detection for the strong candidates does not necessarily mean that they are not reliable. We anticipate future observational efforts will help confirm their nature.

\begin{figure}
\plotone{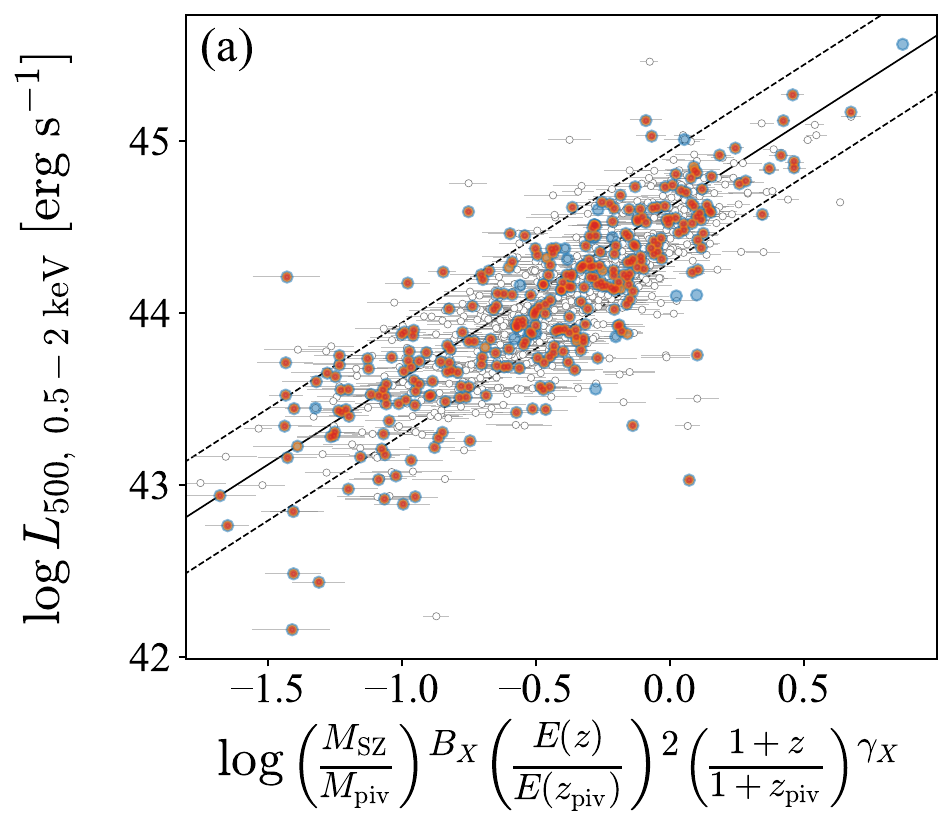}
\plotone{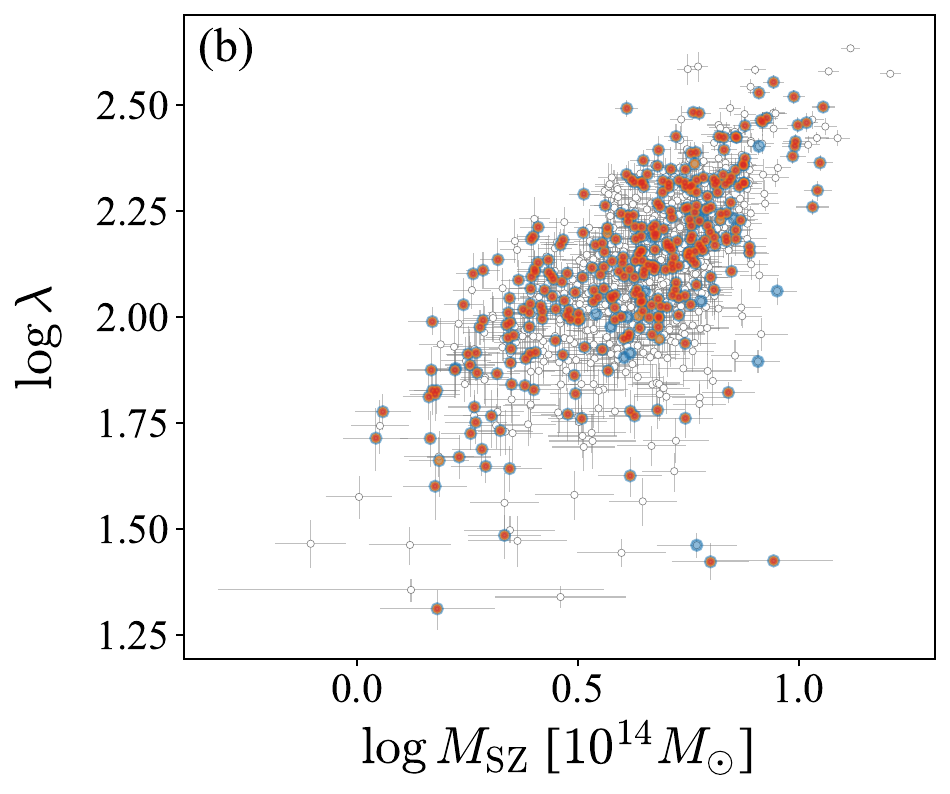}
\caption{The scaling relations between (a) X-ray luminosity in the 0.5-2.0 keV band ($L_{\rm 500}$), $M_{\rm SZ}$, and redshift; and (b) the MCMF richness parameter ($\lambda$) and $M_{\rm SZ}$. Both $L_{\rm 500}$ and $\lambda$ values are sourced from the RASS-MCMF catalog. Colored markers represent confirmed clusters whose redshifts were determined through analysis of the galaxy redshift catalog, with colors corresponding to those used in Figures \ref{fig:red1-1} and \ref{fig:xcomp}. Open circles show the total set of PSZ2 counterparts in the RASS-MCMF clusters, where we show their 1$\sigma$ error bars except for $L_{\rm 500}$. The power-law form and its parameters for the $L_{\rm 500}$-$M_{\rm SZ}$-$z$ scaling relation are derived from \citet{bulbul19}. Solid and dashed lines represent the best-fit relation and the 3$\sigma$ intrinsic scatter range, respectively.
 \label{fig:scale}}
\end{figure}

We also analyze the alignment of clusters identified by our method with key scaling relations, particularly the X-ray luminosity-mass and the richness-mass relations. We adopt the scaling relation between X-ray luminosity in the 0.5-2.0 keV band ($L_{500}$) and $M_{500}$ from \citet{bulbul19}:
\begin{equation}
    L_{500, {\rm 0.5-2 keV}} = A_X \left(\frac{M_{500}}{M_{\rm piv}}\right)^{B_X}\left(\frac{E(z)}{E(z_{\rm piv})}\right)^2\left(\frac{1+z}{1+z_{\rm piv}}\right)^{\gamma_X}
\end{equation}
with parameters $A_X=4.15\times10^{44}$ erg s$^{-1}$, $B_X = 1.91$, $\gamma_X = 0.20$, $M_{\rm piv}=6.35\times10^{14}M_\odot$, and $z_{\rm piv}=0.45$. Figure \ref{fig:scale} shows this relation as a solid line and the range of three times the intrinsic scatter ($3\sigma_{\rm int}$) as dashed lines. The richness-mass scaling is based on the MCMF richness $\lambda$ from the RASS-MCMF catalog.

Upon reidentifying PSZ2 clusters with RASS-MCMF counterparts through galaxy redshift catalog analysis, we recalculated the values of $M_{\rm SZ}$ for these newly determined redshifts, as outlined in Section \ref{ssec:massup}. In Figure \ref{fig:scale}, colored circles represent our identified clusters, while white circles denote all the RASS-MCMF counterparts. The recalculated $M_{\rm SZ}$ values align well with the overall RASS-MCMF dataset, indicating our selection of strong candidates does not introduce systematic biases relative to these scaling relations. Notably, we find that most outliers in these relations are multiple systems, potentially leading to overestimated $L_{500}$ or $M_{\rm SZ}$ due to projection effects of multiple systems contributing to the X-ray or SZ signals. This suggests limitations arising from observational constraints rather than our methodology. We document such multiple systems in the \texttt{NOTES} column of our catalog for further reference.

\subsection{Notes on Some Individual Objects\label{ssec:indv}}
Here we discuss some individual PSZ2 objects that require special attention during the updating process, which could be possible contaminants, multiple clusters, etc.
We include this additional information in our catalog at the \texttt{NOTES} column.
First of all, for the updates based on the cluster catalogs, eight confirmed clusters (PSZ2 G052.93+10.42, PSZ2 G098.67-07.04, PSZ2 G098.75-28.63, PSZ2 G108.13-09.21, PSZ2 G143.70-08.59, PSZ2 G163.04-27.80, PSZ2 G198.73+13.34 and PSZ2 G199.73+36.98) are invalidated as the ORM follow-ups fail to detect any galaxy overdensity \citep[flagged as ``ND'',][]{itp13a, itp13b, lp15b}.
Similarly, RASS-MCMF provides redshift corrections for five confirmed PSZ2 clusters \citep[PSZ2 G091.40-51.01, PSZ2 G109.86+27.94, PSZ2 G181.71-68.65, PSZ2 G281.09-42.51, and PSZ2 G287.00-35.24; see Table A1 of ][]{rass-mcmf}. We have updated these redshifts accordingly.

According to our updating sequence, PSZ2 G136.02-47.15 would have been validated by Burenin et al. (2017). Yet, we maintain its candidate status because of concerns raised by earlier studies (Streblyanska et al., 2018; Zohren et al., 2019). These studies argue that the optical counterparts near $z_{\rm spec}=0.465$ and $z_{\rm phot}=0.61$ lack sufficient richness and are too distantly separated to account for the SZ signal, suggesting that it may predominantly result from noise. We find a clustering signal at $z=0.465$ using the galaxy redshift catalog (Section \ref{ssec:val}), but this signal is also rejected by the distance criterion $\bar{R}\leq2$, where $\bar{R}$ has a value of 4.0 in this case.

We also find the close proximity between PSZ2 G280.76-52.30 and PSZ2 G280.78-52.22, with a separation of only $5.2'$. While PSZ2 G280.76-52.30 is initially identified as SPT-CL J0240-5952 in PSZ2, consideration of their positions and comparison of their $M_{\rm SZ}$ with SPT mass estimates lead us to reassign their counterparts: PSZ2 G280.76-52.30 to SPT-CL J0240-5946 and PSZ2 G280.78-52.22 to SPT-CL J0240-5952.

Additionally, we have included the clusters from PSZ-MCMF and RASS-MCMF in our catalog. These clusters may be contaminated by overdensities caused by foreground and background galaxies. Nevertheless, their measure of the random superposition probability, $f_{\rm cont}$ \citep[see Section 3.2.2 of]{psz-mcmf}, indicates that the majority of the PSZ2 counterparts are regarded as confirmed clusters. However, a few clusters, namely PSZ2 G018.92-33.64, PSZ2 G195.03-79.43, and PSZ2 G224.37-47.33, exhibit $f_{\rm cont} > 0.1$, indicating a higher likelihood of contamination.

Below we list the specifics of individual PSZ2 objects validated using the galaxy redshift catalog as in Section \ref{ssec:val}:

\textit{PSZ2 G027.99-69.85}. We find a clustering signal at $z=0.137$ for this object.
However, upon examining the optical image from DESI LS DR10, we find a rich galaxy cluster (or supercluster) with an unknown redshift, possibly higher than 0.137 (H. Bahk et al. in preparation). As a result, we were unable to determine the redshift for this object and only updated its validation status.

\textit{PSZ2 G044.21+52.13}.
Our method validates this candidate as a system at $z=0.075$. However, according to the ORM follow-up study \citep{lp15a}, they find a cluster at a higher redshift of $z=0.377$. The velocity dispersion of this system falls below their confirmation criteria, resulting in it being classified as a cluster candidate with \texttt{Flag}=3. This raises the possibility that the SZ signal may correspond to the overlap of two systems.

\textit{PSZ2 G051.48-30.87}. Clustering signals at $z=0.056$ and $z=0.135$ are detected, where more prominent signal at $z=0.135$ is selected. This suggests the presence of overlapping multiple clusters along the line of sight, each contributing to the SZ signal. However, it is worth noting that the ORM follow-up \citep{lp15a} classifies this as a non-detection. Additionally, some hints of Galactic cirrus can be observed in the optical image, raising the possibility of contamination by Galactic dust emission.  This candidate has $Q_{\rm neural}=0.97>0.4$, which suggests a high-quality detection. Further investigation is needed to accurately determine the nature of this source and its contribution to the SZ signal.

\textit{PSZ2 G110.69-46.25}.
We have observed that the MILCA Compton y-map in this field exhibits a double-peaked structure, indicating the possibility of multiple systems overlapping along the line of sight. Our analysis, thanks to the rich spectroscopic data from SDSS, has allowed us to detect multiple clustering signals at redshifts $z=0.086, 0.165, 0.304, 0.539$. However, based on our validation criterion $\bar{R}$, the clustering signals at $z=0.304$ and $z=0.539$ have been rejected. Further deep observations are necessary to thoroughly investigate these multiple clustering signals.

\textit{PSZ2 G332.11-23.63}. We initially detected a cluster signal for this candidate. However, upon visual inspection, it became evident that there was a clear misidentification of another nearby cluster, PSZ2 G332.29-23.57. As a result, we exclude PSZ2 G332.11-23.63 from the strong candidate sample.

\section{Summary\label{sec:summary}}

In this work, we have updated the \Planck -selected all-sky galaxy cluster catalog, which becomes UPCluster-SZ. We compile the cluster catalogs from follow-up studies for validation and updating redshift information, and use the catalog of galaxies with spectroscopic redshifts to provide an additional sample of strong candidates. The UPCluster-SZ contains 1334 validated clusters and 319 cluster candidates, of which 122 candidates are flagged as non-detections. The redshift information is available for 1311 validated clusters, of which 1103 have spectroscopic redshifts. We also provide the sample of 10 strong candidates identified by the spectroscopic galaxy catalog.

Despite the update, there are still 319 candidate clusters in UPCluster-SZ that require validation. Additionally, we should further update the redshifts of 227 confirmed clusters, which are either obtained photometrically or not obtained at all. Further observations are needed to complete the validation process and the redshift measurement, for more comprehensive and accurate understanding of the \Planck\ cluster population.

In conclusion, the UPCluster-SZ catalog in this study will be a valuable resource for future galaxy cluster studies, such as the legacy science of \SPHEREx\ mission. The updates and validations of the PSZ2 clusters highlight the challenges in identifying and characterizing galaxy clusters, particularly in the presence of contamination. The UPCluster-SZ catalog will enable further studies regarding the formation and evolution of galaxy clusters. It can also be an important basis for planning further follow-up observations for various scientific purposes. Moreover, the combination of the data from large spectroscopic surveys including DESI \citep{desi}, 4MOST \citep{4most}, and A-SPEC (H. Hwang et al. in preparation), and from imaging surveys including the Rubin Observatory Legacy Survey of Space and Time \citep[LSST, ][]{lsst} will be very useful for further completing this all-sky galaxy cluster catalog from \Planck\ SZ observations.

\section*{acknowledgments}
We thank the referee for constructive comments that improved the manuscript. HSH acknowledges the support by the National Research Foundation of Korea (NRF) grant funded by the Korea government (MSIT) (No. 2021R1A2C1094577).
Funding for the Sloan Digital Sky 
Survey IV has been provided by the 
Alfred P. Sloan Foundation, the U.S. 
Department of Energy Office of 
Science, and the Participating 
Institutions. 
SDSS-IV acknowledges support and 
resources from the Center for High 
Performance Computing  at the 
University of Utah. The SDSS 
website is www.sdss4.org.
SDSS-IV is managed by the 
Astrophysical Research Consortium 
for the Participating Institutions 
of the SDSS Collaboration including 
the Brazilian Participation Group, 
the Carnegie Institution for Science, 
Carnegie Mellon University, Center for 
Astrophysics | Harvard \& 
Smithsonian, the Chilean Participation 
Group, the French Participation Group, 
Instituto de Astrof\'isica de 
Canarias, The Johns Hopkins 
University, Kavli Institute for the 
Physics and Mathematics of the 
Universe (IPMU) / University of 
Tokyo, the Korean Participation Group, 
Lawrence Berkeley National Laboratory, 
Leibniz Institut f\"ur Astrophysik 
Potsdam (AIP),  Max-Planck-Institut 
f\"ur Astronomie (MPIA Heidelberg), 
Max-Planck-Institut f\"ur 
Astrophysik (MPA Garching), 
Max-Planck-Institut f\"ur 
Extraterrestrische Physik (MPE), 
National Astronomical Observatories of 
China, New Mexico State University, 
New York University, University of 
Notre Dame, Observat\'ario 
Nacional / MCTI, The Ohio State 
University, Pennsylvania State 
University, Shanghai 
Astronomical Observatory, United 
Kingdom Participation Group, 
Universidad Nacional Aut\'onoma 
de M\'exico, University of Arizona, 
University of Colorado Boulder, 
University of Oxford, University of 
Portsmouth, University of Utah, 
University of Virginia, University 
of Washington, University of 
Wisconsin, Vanderbilt University, 
and Yale University.
This research used data obtained with the Dark Energy Spectroscopic Instrument (DESI). DESI construction and operations is managed by the Lawrence Berkeley National Laboratory. This material is based upon work supported by the U.S. Department of Energy, Office of Science, Office of High-Energy Physics, under Contract No. DE–AC02–05CH11231, and by the National Energy Research Scientific Computing Center, a DOE Office of Science User Facility under the same contract. Additional support for DESI was provided by the U.S. National Science Foundation (NSF), Division of Astronomical Sciences under Contract No. AST-0950945 to the NSF’s National Optical-Infrared Astronomy Research Laboratory; the Science and Technology Facilities Council of the United Kingdom; the Gordon and Betty Moore Foundation; the Heising-Simons Foundation; the French Alternative Energies and Atomic Energy Commission (CEA); the National Council of Science and Technology of Mexico (CONACYT); the Ministry of Science and Innovation of Spain (MICINN), and by the DESI Member Institutions: www.desi.lbl.gov/collaborating-institutions. The DESI collaboration is honored to be permitted to conduct scientific research on Iolkam Du’ag (Kitt Peak), a mountain with particular significance to the Tohono O’odham Nation. Any opinions, findings, and conclusions or recommendations expressed in this material are those of the author(s) and do not necessarily reflect the views of the U.S. National Science Foundation, the U.S. Department of Energy, or any of the listed funding agencies.
This research uses services or data provided by the SPectra Analysis and Retrievable Catalog Lab (SPARCL) and the Astro Data Lab, which are both part of the Community Science and Data Center (CSDC) program at NSF's National Optical-Infrared Astronomy Research Laboratory. NOIRLab is operated by the Association of Universities for Research in Astronomy (AURA), Inc. under a cooperative agreement with the National Science Foundation.
This research has made use of the NASA/IPAC Extragalactic Database (NED), which is funded by the National Aeronautics and Space Administration and operated by the California Institute of Technology.

\facility{\Planck, Astro Data Lab}
\software{Astropy \citep{astropy:2013, astropy:2018, astropy:2022}, Astroquery \citep{astroquery}, Scipy \citep{2020SciPy-NMeth}, Numpy \citep{numpy}, Matplotlib \citep{matplotlib}}

\bibliography{ms}

\begin{thebibliography}{}
\expandafter\ifx\csname natexlab\endcsname\relax\def\natexlab#1{#1}\fi
\providecommand{\url}[1]{\href{#1}{#1}}
\providecommand{\dodoi}[1]{doi:~\href{http://doi.org/#1}{\nolinkurl{#1}}}
\providecommand{\doeprint}[1]{\href{http://ascl.net/#1}{\nolinkurl{http://ascl.net/#1}}}
\providecommand{\doarXiv}[1]{\href{https://arxiv.org/abs/#1}{\nolinkurl{https://arxiv.org/abs/#1}}}

\bibitem[{{Abdurro'uf} {et~al.}(2022){Abdurro'uf}, {Accetta}, {Aerts}, {Silva Aguirre}, {Ahumada}, {Ajgaonkar}, {Filiz Ak}, {Alam}, {Allende Prieto}, {Almeida}, {Anders}, {Anderson}, {Andrews}, {Anguiano}, {Aquino-Ort{\'\i}z}, {Arag{\'o}n-Salamanca}, {Argudo-Fern{\'a}ndez}, {Ata}, {Aubert}, {Avila-Reese}, {Badenes}, {Barb{\'a}}, {Barger}, {Barrera-Ballesteros}, {Beaton}, {Beers}, {Belfiore}, {Bender}, {Bernardi}, {Bershady}, {Beutler}, {Bidin}, {Bird}, {Bizyaev}, {Blanc}, {Blanton}, {Boardman}, {Bolton}, {Boquien}, {Borissova}, {Bovy}, {Brandt}, {Brown}, {Brownstein}, {Brusa}, {Buchner}, {Bundy}, {Burchett}, {Bureau}, {Burgasser}, {Cabang}, {Campbell}, {Cappellari}, {Carlberg}, {Wanderley}, {Carrera}, {Cash}, {Chen}, {Chen}, {Cherinka}, {Chiappini}, {Choi}, {Chojnowski}, {Chung}, {Clerc}, {Cohen}, {Comerford}, {Comparat}, {da Costa}, {Covey}, {Crane}, {Cruz-Gonzalez}, {Culhane}, {Cunha}, {Dai}, {Damke}, {Darling}, {Davidson}, {Davies}, {Dawson}, {De Lee}, {Diamond-Stanic}, {Cano-D{\'\i}az}, {S{\'a}nchez},
  {Donor}, {Duckworth}, {Dwelly}, {Eisenstein}, {Elsworth}, {Emsellem}, {Eracleous}, {Escoffier}, {Fan}, {Farr}, {Feng}, {Fern{\'a}ndez-Trincado}, {Feuillet}, {Filipp}, {Fillingham}, {Frinchaboy}, {Fromenteau}, {Galbany}, {Garc{\'\i}a}, {Garc{\'\i}a-Hern{\'a}ndez}, {Ge}, {Geisler}, {Gelfand}, {G{\'e}ron}, {Gibson}, {Goddy}, {Godoy-Rivera}, {Grabowski}, {Green}, {Greener}, {Grier}, {Griffith}, {Guo}, {Guy}, {Hadjara}, {Harding}, {Hasselquist}, {Hayes}, {Hearty}, {Hern{\'a}ndez}, {Hill}, {Hogg}, {Holtzman}, {Horta}, {Hsieh}, {Hsu}, {Hsu}, {Huber}, {Huertas-Company}, {Hutchinson}, {Hwang}, {Ibarra-Medel}, {Chitham}, {Ilha}, {Imig}, {Jaekle}, {Jayasinghe}, {Ji}, {Johnson}, {Jones}, {J{\"o}nsson}, {Katkov}, {Khalatyan}, {Kinemuchi}, {Kisku}, {Knapen}, {Kneib}, {Kollmeier}, {Kong}, {Kounkel}, {Kreckel}, {Krishnarao}, {Lacerna}, {Lane}, {Langgin}, {Lavender}, {Law}, {Lazarz}, {Leung}, {Leung}, {Lewis}, {Li}, {Li}, {Lian}, {Liang}, {Lin}, {Lin}, {Lin}, {Lintott}, {Long}, {Longa-Pe{\~n}a}, {L{\'o}pez-Cob{\'a}}, {Lu},
  {Lundgren}, {Luo}, {Mackereth}, {de la Macorra}, {Mahadevan}, {Majewski}, {Manchado}, {Mandeville}, {Maraston}, {Margalef-Bentabol}, {Masseron}, {Masters}, {Mathur}, {McDermid}, {Mckay}, {Merloni}, {Merrifield}, {Meszaros}, {Miglio}, {Di Mille}, {Minniti}, {Minsley}, {Monachesi}, {Moon}, {Mosser}, {Mulchaey}, {Muna}, {Mu{\~n}oz}, {Myers}, {Myers}, {Nadathur}, {Nair}, {Nandra}, {Neumann}, {Newman}, {Nidever}, {Nikakhtar}, {Nitschelm}, {O'Connell}, {Garma-Oehmichen}, {Luan Souza de Oliveira}, {Olney}, {Oravetz}, {Ortigoza-Urdaneta}, {Osorio}, {Otter}, {Pace}, {Padilla}, {Pan}, {Pan}, {Parikh}, {Parker}, {Peirani}, {Pe{\~n}a Ram{\'\i}rez}, {Penny}, {Percival}, {Perez-Fournon}, {Pinsonneault}, {Poidevin}, {Poovelil}, {Price-Whelan}, {B{\'a}rbara de Andrade Queiroz}, {Raddick}, {Ray}, {Rembold}, {Riddle}, {Riffel}, {Riffel}, {Rix}, {Robin}, {Rodr{\'\i}guez-Puebla}, {Roman-Lopes}, {Rom{\'a}n-Z{\'u}{\~n}iga}, {Rose}, {Ross}, {Rossi}, {Rubin}, {Salvato}, {S{\'a}nchez}, {S{\'a}nchez-Gallego}, {Sanderson}, {Santana
  Rojas}, {Sarceno}, {Sarmiento}, {Sayres}, {Sazonova}, {Schaefer}, {Schiavon}, {Schlegel}, {Schneider}, {Schultheis}, {Schwope}, {Serenelli}, {Serna}, {Shao}, {Shapiro}, {Sharma}, {Shen}, {Shetrone}, {Shu}, {Simon}, {Skrutskie}, {Smethurst}, {Smith}, {Sobeck}, {Spoo}, {Sprague}, {Stark}, {Stassun}, {Steinmetz}, {Stello}, {Stone-Martinez}, {Storchi-Bergmann}, {Stringfellow}, {Stutz}, {Su}, {Taghizadeh-Popp}, {Talbot}, {Tayar}, {Telles}, {Teske}, {Thakar}, {Theissen}, {Tkachenko}, {Thomas}, {Tojeiro}, {Hernandez Toledo}, {Troup}, {Trump}, {Trussler}, {Turner}, {Tuttle}, {Unda-Sanzana}, {V{\'a}zquez-Mata}, {Valentini}, {Valenzuela}, {Vargas-Gonz{\'a}lez}, {Vargas-Maga{\~n}a}, {Alfaro}, {Villanova}, {Vincenzo}, {Wake}, {Warfield}, {Washington}, {Weaver}, {Weijmans}, {Weinberg}, {Weiss}, {Westfall}, {Wild}, {Wilde}, {Wilson}, {Wilson}, {Wilson}, {Wolf}, {Wood-Vasey}, {Yan}, {Zamora}, {Zasowski}, {Zhang}, {Zhao}, {Zheng}, {Zheng}, \& {Zhu}}]{sdssdr17}
{Abdurro'uf}, {Accetta}, K., {Aerts}, C., {et~al.} 2022, \apjs, 259, 35, \dodoi{10.3847/1538-4365/ac4414}

\bibitem[{{Aghanim} {et~al.}(2015){Aghanim}, {Hurier}, {Diego}, {Douspis}, {Macias-Perez}, {Pointecouteau}, {Comis}, {Arnaud}, \& {Montier}}]{aghanim15}
{Aghanim}, N., {Hurier}, G., {Diego}, J.~M., {et~al.} 2015, \aap, 580, A138, \dodoi{10.1051/0004-6361/201424963}

\bibitem[{{Aguado-Barahona} {et~al.}(2019){Aguado-Barahona}, {Barrena}, {Streblyanska}, {Ferragamo}, {Rubi{\~n}o-Mart{\'\i}n}, {Tramonte}, \& {Lietzen}}]{lp15b}
{Aguado-Barahona}, A., {Barrena}, R., {Streblyanska}, A., {et~al.} 2019, \aap, 631, A148, \dodoi{10.1051/0004-6361/201936034}

\bibitem[{{Allen} {et~al.}(2011){Allen}, {Evrard}, \& {Mantz}}]{allen11}
{Allen}, S.~W., {Evrard}, A.~E., \& {Mantz}, A.~B. 2011, \araa, 49, 409, \dodoi{10.1146/annurev-astro-081710-102514}

\bibitem[{{Amodeo} {et~al.}(2018){Amodeo}, {Mei}, {Stanford}, {Lawrence}, {Bartlett}, {Stern}, {Chary}, {Shim}, {Marleau}, {Melin}, \& {Rodr{\'\i}guez-Gonz{\'a}lvez}}]{amodeo18}
{Amodeo}, S., {Mei}, S., {Stanford}, S.~A., {et~al.} 2018, \apj, 853, 36, \dodoi{10.3847/1538-4357/aa98dd}

\bibitem[{{Astropy Collaboration} {et~al.}(2013){Astropy Collaboration}, {Robitaille}, {Tollerud}, {Greenfield}, {Droettboom}, {Bray}, {Aldcroft}, {Davis}, {Ginsburg}, {Price-Whelan}, {Kerzendorf}, {Conley}, {Crighton}, {Barbary}, {Muna}, {Ferguson}, {Grollier}, {Parikh}, {Nair}, {Unther}, {Deil}, {Woillez}, {Conseil}, {Kramer}, {Turner}, {Singer}, {Fox}, {Weaver}, {Zabalza}, {Edwards}, {Azalee Bostroem}, {Burke}, {Casey}, {Crawford}, {Dencheva}, {Ely}, {Jenness}, {Labrie}, {Lim}, {Pierfederici}, {Pontzen}, {Ptak}, {Refsdal}, {Servillat}, \& {Streicher}}]{astropy:2013}
{Astropy Collaboration}, {Robitaille}, T.~P., {Tollerud}, E.~J., {et~al.} 2013, \aap, 558, A33, \dodoi{10.1051/0004-6361/201322068}

\bibitem[{{Astropy Collaboration} {et~al.}(2018){Astropy Collaboration}, {Price-Whelan}, {Sip{\H{o}}cz}, {G{\"u}nther}, {Lim}, {Crawford}, {Conseil}, {Shupe}, {Craig}, {Dencheva}, {Ginsburg}, {Vand erPlas}, {Bradley}, {P{\'e}rez-Su{\'a}rez}, {de Val-Borro}, {Aldcroft}, {Cruz}, {Robitaille}, {Tollerud}, {Ardelean}, {Babej}, {Bach}, {Bachetti}, {Bakanov}, {Bamford}, {Barentsen}, {Barmby}, {Baumbach}, {Berry}, {Biscani}, {Boquien}, {Bostroem}, {Bouma}, {Brammer}, {Bray}, {Breytenbach}, {Buddelmeijer}, {Burke}, {Calderone}, {Cano Rodr{\'\i}guez}, {Cara}, {Cardoso}, {Cheedella}, {Copin}, {Corrales}, {Crichton}, {D'Avella}, {Deil}, {Depagne}, {Dietrich}, {Donath}, {Droettboom}, {Earl}, {Erben}, {Fabbro}, {Ferreira}, {Finethy}, {Fox}, {Garrison}, {Gibbons}, {Goldstein}, {Gommers}, {Greco}, {Greenfield}, {Groener}, {Grollier}, {Hagen}, {Hirst}, {Homeier}, {Horton}, {Hosseinzadeh}, {Hu}, {Hunkeler}, {Ivezi{\'c}}, {Jain}, {Jenness}, {Kanarek}, {Kendrew}, {Kern}, {Kerzendorf}, {Khvalko}, {King}, {Kirkby}, {Kulkarni},
  {Kumar}, {Lee}, {Lenz}, {Littlefair}, {Ma}, {Macleod}, {Mastropietro}, {McCully}, {Montagnac}, {Morris}, {Mueller}, {Mumford}, {Muna}, {Murphy}, {Nelson}, {Nguyen}, {Ninan}, {N{\"o}the}, {Ogaz}, {Oh}, {Parejko}, {Parley}, {Pascual}, {Patil}, {Patil}, {Plunkett}, {Prochaska}, {Rastogi}, {Reddy Janga}, {Sabater}, {Sakurikar}, {Seifert}, {Sherbert}, {Sherwood-Taylor}, {Shih}, {Sick}, {Silbiger}, {Singanamalla}, {Singer}, {Sladen}, {Sooley}, {Sornarajah}, {Streicher}, {Teuben}, {Thomas}, {Tremblay}, {Turner}, {Terr{\'o}n}, {van Kerkwijk}, {de la Vega}, {Watkins}, {Weaver}, {Whitmore}, {Woillez}, {Zabalza}, \& {Astropy Contributors}}]{astropy:2018}
{Astropy Collaboration}, {Price-Whelan}, A.~M., {Sip{\H{o}}cz}, B.~M., {et~al.} 2018, \aj, 156, 123, \dodoi{10.3847/1538-3881/aabc4f}

\bibitem[{{Astropy Collaboration} {et~al.}(2022){Astropy Collaboration}, {Price-Whelan}, {Lim}, {Earl}, {Starkman}, {Bradley}, {Shupe}, {Patil}, {Corrales}, {Brasseur}, {N{"o}the}, {Donath}, {Tollerud}, {Morris}, {Ginsburg}, {Vaher}, {Weaver}, {Tocknell}, {Jamieson}, {van Kerkwijk}, {Robitaille}, {Merry}, {Bachetti}, {G{"u}nther}, {Aldcroft}, {Alvarado-Montes}, {Archibald}, {B{'o}di}, {Bapat}, {Barentsen}, {Baz{'a}n}, {Biswas}, {Boquien}, {Burke}, {Cara}, {Cara}, {Conroy}, {Conseil}, {Craig}, {Cross}, {Cruz}, {D'Eugenio}, {Dencheva}, {Devillepoix}, {Dietrich}, {Eigenbrot}, {Erben}, {Ferreira}, {Foreman-Mackey}, {Fox}, {Freij}, {Garg}, {Geda}, {Glattly}, {Gondhalekar}, {Gordon}, {Grant}, {Greenfield}, {Groener}, {Guest}, {Gurovich}, {Handberg}, {Hart}, {Hatfield-Dodds}, {Homeier}, {Hosseinzadeh}, {Jenness}, {Jones}, {Joseph}, {Kalmbach}, {Karamehmetoglu}, {Ka{l}uszy{'n}ski}, {Kelley}, {Kern}, {Kerzendorf}, {Koch}, {Kulumani}, {Lee}, {Ly}, {Ma}, {MacBride}, {Maljaars}, {Muna}, {Murphy}, {Norman}, {O'Steen},
  {Oman}, {Pacifici}, {Pascual}, {Pascual-Granado}, {Patil}, {Perren}, {Pickering}, {Rastogi}, {Roulston}, {Ryan}, {Rykoff}, {Sabater}, {Sakurikar}, {Salgado}, {Sanghi}, {Saunders}, {Savchenko}, {Schwardt}, {Seifert-Eckert}, {Shih}, {Jain}, {Shukla}, {Sick}, {Simpson}, {Singanamalla}, {Singer}, {Singhal}, {Sinha}, {Sip{H{o}}cz}, {Spitler}, {Stansby}, {Streicher}, {{{S}}umak}, {Swinbank}, {Taranu}, {Tewary}, {Tremblay}, {Val-Borro}, {Van Kooten}, {Vasovi{'c}}, {Verma}, {de Miranda Cardoso}, {Williams}, {Wilson}, {Winkel}, {Wood-Vasey}, {Xue}, {Yoachim}, {Zhang}, {Zonca}, \& {Astropy Project Contributors}}]{astropy:2022}
{Astropy Collaboration}, {Price-Whelan}, A.~M., {Lim}, P.~L., {et~al.} 2022, apj, 935, 167, \dodoi{10.3847/1538-4357/ac7c74}

\bibitem[{{Barrena} {et~al.}(2018){Barrena}, {Streblyanska}, {Ferragamo}, {Rubi{\~n}o-Mart{\'\i}n}, {Aguado-Barahona}, {Tramonte}, {G{\'e}nova-Santos}, {Hempel}, {Lietzen}, {Aghanim}, {Arnaud}, {B{\"o}hringer}, {Chon}, {Democles}, {Dahle}, {Douspis}, {Lasenby}, {Mazzotta}, {Melin}, {Pointecouteau}, {Pratt}, {Rossetti}, \& {van der Burg}}]{itp13a}
{Barrena}, R., {Streblyanska}, A., {Ferragamo}, A., {et~al.} 2018, \aap, 616, A42, \dodoi{10.1051/0004-6361/201732315}

\bibitem[{{Barrena} {et~al.}(2020){Barrena}, {Ferragamo}, {Rubi{\~n}o-Mart{\'\i}n}, {Streblyanska}, {Aguado-Barahona}, {Tramonte}, {G{\'e}nova-Santos}, {Hempel}, {Lietzen}, {Aghanim}, {Arnaud}, {B{\"o}hringer}, {Chon}, {Dahle}, {Douspis}, {Lasenby}, {Mazzotta}, {Melin}, {Pointecouteau}, {Pratt}, \& {Rossetti}}]{itp13b}
{Barrena}, R., {Ferragamo}, A., {Rubi{\~n}o-Mart{\'\i}n}, J.~A., {et~al.} 2020, \aap, 638, A146, \dodoi{10.1051/0004-6361/202037552}

\bibitem[{{Bautz} \& {Morgan}(1970)}]{bmtype}
{Bautz}, L.~P., \& {Morgan}, W.~W. 1970, \apjl, 162, L149, \dodoi{10.1086/180643}

\bibitem[{{Ben{\'\i}tez}(2000)}]{bpz00}
{Ben{\'\i}tez}, N. 2000, \apj, 536, 571, \dodoi{10.1086/308947}

\bibitem[{{Bleem} {et~al.}(2015){Bleem}, {Stalder}, {de Haan}, {Aird}, {Allen}, {Applegate}, {Ashby}, {Bautz}, {Bayliss}, {Benson}, {Bocquet}, {Brodwin}, {Carlstrom}, {Chang}, {Chiu}, {Cho}, {Clocchiatti}, {Crawford}, {Crites}, {Desai}, {Dietrich}, {Dobbs}, {Foley}, {Forman}, {George}, {Gladders}, {Gonzalez}, {Halverson}, {Hennig}, {Hoekstra}, {Holder}, {Holzapfel}, {Hrubes}, {Jones}, {Keisler}, {Knox}, {Lee}, {Leitch}, {Liu}, {Lueker}, {Luong-Van}, {Mantz}, {Marrone}, {McDonald}, {McMahon}, {Meyer}, {Mocanu}, {Mohr}, {Murray}, {Padin}, {Pryke}, {Reichardt}, {Rest}, {Ruel}, {Ruhl}, {Saliwanchik}, {Saro}, {Sayre}, {Schaffer}, {Schrabback}, {Shirokoff}, {Song}, {Spieler}, {Stanford}, {Staniszewski}, {Stark}, {Story}, {Stubbs}, {Vanderlinde}, {Vieira}, {Vikhlinin}, {Williamson}, {Zahn}, \& {Zenteno}}]{spt-sz}
{Bleem}, L.~E., {Stalder}, B., {de Haan}, T., {et~al.} 2015, \apjs, 216, 27, \dodoi{10.1088/0067-0049/216/2/27}

\bibitem[{{Bleem} {et~al.}(2020){Bleem}, {Bocquet}, {Stalder}, {Gladders}, {Ade}, {Allen}, {Anderson}, {Annis}, {Ashby}, {Austermann}, {Avila}, {Avva}, {Bayliss}, {Beall}, {Bechtol}, {Bender}, {Benson}, {Bertin}, {Bianchini}, {Blake}, {Brodwin}, {Brooks}, {Buckley-Geer}, {Burke}, {Carlstrom}, {Rosell}, {Carrasco Kind}, {Carretero}, {Chang}, {Chiang}, {Citron}, {Moran}, {Costanzi}, {Crawford}, {Crites}, {da Costa}, {de Haan}, {De Vicente}, {Desai}, {Diehl}, {Dietrich}, {Dobbs}, {Eifler}, {Everett}, {Flaugher}, {Floyd}, {Frieman}, {Gallicchio}, {Garc{\'\i}a-Bellido}, {George}, {Gerdes}, {Gilbert}, {Gruen}, {Gruendl}, {Gschwend}, {Gupta}, {Gutierrez}, {Halverson}, {Harrington}, {Henning}, {Heymans}, {Holder}, {Hollowood}, {Holzapfel}, {Honscheid}, {Hrubes}, {Huang}, {Hubmayr}, {Irwin}, {James}, {Jeltema}, {Joudaki}, {Khullar}, {Klein}, {Knox}, {Kuropatkin}, {Lee}, {Li}, {Lidman}, {Lowitz}, {MacCrann}, {Mahler}, {Maia}, {Marshall}, {McDonald}, {McMahon}, {Melchior}, {Menanteau}, {Meyer}, {Miquel}, {Mocanu},
  {Mohr}, {Montgomery}, {Nadolski}, {Natoli}, {Nibarger}, {Noble}, {Novosad}, {Padin}, {Palmese}, {Parkinson}, {Patil}, {Paz-Chinch{\'o}n}, {Plazas}, {Pryke}, {Ramachandra}, {Reichardt}, {Remolina Gonz{\'a}lez}, {Romer}, {Roodman}, {Ruhl}, {Rykoff}, {Saliwanchik}, {Sanchez}, {Saro}, {Sayre}, {Schaffer}, {Schrabback}, {Serrano}, {Sharon}, {Sievers}, {Smecher}, {Smith}, {Soares-Santos}, {Stark}, {Story}, {Suchyta}, {Tarle}, {Tucker}, {Vanderlinde}, {Veach}, {Vieira}, {Wang}, {Weller}, {Whitehorn}, {Wu}, {Yefremenko}, \& {Zhang}}]{spt-ecs20}
{Bleem}, L.~E., {Bocquet}, S., {Stalder}, B., {et~al.} 2020, \apjs, 247, 25, \dodoi{10.3847/1538-4365/ab6993}

\bibitem[{{Boada} {et~al.}(2019){Boada}, {Hughes}, {Menanteau}, {Doze}, {Barrientos}, \& {Infante}}]{boada19}
{Boada}, S., {Hughes}, J.~P., {Menanteau}, F., {et~al.} 2019, \apj, 871, 188, \dodoi{10.3847/1538-4357/aaf3a0}

\bibitem[{{Bocquet} {et~al.}(2019){Bocquet}, {Dietrich}, {Schrabback}, {Bleem}, {Klein}, {Allen}, {Applegate}, {Ashby}, {Bautz}, {Bayliss}, {Benson}, {Brodwin}, {Bulbul}, {Canning}, {Capasso}, {Carlstrom}, {Chang}, {Chiu}, {Cho}, {Clocchiatti}, {Crawford}, {Crites}, {de Haan}, {Desai}, {Dobbs}, {Foley}, {Forman}, {Garmire}, {George}, {Gladders}, {Gonzalez}, {Grandis}, {Gupta}, {Halverson}, {Hlavacek-Larrondo}, {Hoekstra}, {Holder}, {Holzapfel}, {Hou}, {Hrubes}, {Huang}, {Jones}, {Khullar}, {Knox}, {Kraft}, {Lee}, {von der Linden}, {Luong-Van}, {Mantz}, {Marrone}, {McDonald}, {McMahon}, {Meyer}, {Mocanu}, {Mohr}, {Morris}, {Padin}, {Patil}, {Pryke}, {Rapetti}, {Reichardt}, {Rest}, {Ruhl}, {Saliwanchik}, {Saro}, {Sayre}, {Schaffer}, {Shirokoff}, {Stalder}, {Stanford}, {Staniszewski}, {Stark}, {Story}, {Strazzullo}, {Stubbs}, {Vanderlinde}, {Vieira}, {Vikhlinin}, {Williamson}, \& {Zenteno}}]{spt-sz19}
{Bocquet}, S., {Dietrich}, J.~P., {Schrabback}, T., {et~al.} 2019, \apj, 878, 55, \dodoi{10.3847/1538-4357/ab1f10}

\bibitem[{{B{\"o}hringer} {et~al.}(2004){B{\"o}hringer}, {Schuecker}, {Guzzo}, {Collins}, {Voges}, {Cruddace}, {Ortiz-Gil}, {Chincarini}, {De Grandi}, {Edge}, {MacGillivray}, {Neumann}, {Schindler}, \& {Shaver}}]{reflex}
{B{\"o}hringer}, H., {Schuecker}, P., {Guzzo}, L., {et~al.} 2004, \aap, 425, 367, \dodoi{10.1051/0004-6361:20034484}

\bibitem[{{Buddendiek} {et~al.}(2015){Buddendiek}, {Schrabback}, {Greer}, {Hoekstra}, {Sommer}, {Eifler}, {Erben}, {Erler}, {Hicks}, {High}, {Hildebrandt}, {Marrone}, {Morris}, {Muzzin}, {Reiprich}, {Schirmer}, {Schneider}, \& {von der Linden}}]{buddendiek15}
{Buddendiek}, A., {Schrabback}, T., {Greer}, C.~H., {et~al.} 2015, \mnras, 450, 4248, \dodoi{10.1093/mnras/stv783}

\bibitem[{{Bulbul} {et~al.}(2019){Bulbul}, {Chiu}, {Mohr}, {McDonald}, {Benson}, {Bautz}, {Bayliss}, {Bleem}, {Brodwin}, {Bocquet}, {Capasso}, {Dietrich}, {Forman}, {Hlavacek-Larrondo}, {Holzapfel}, {Khullar}, {Klein}, {Kraft}, {Miller}, {Reichardt}, {Saro}, {Sharon}, {Stalder}, {Schrabback}, \& {Stanford}}]{bulbul19}
{Bulbul}, E., {Chiu}, I.~N., {Mohr}, J.~J., {et~al.} 2019, \apj, 871, 50, \dodoi{10.3847/1538-4357/aaf230}

\bibitem[{{Burenin}(2017)}]{burenin17}
{Burenin}, R.~A. 2017, Astronomy Letters, 43, 507, \dodoi{10.1134/S1063773717080035}

\bibitem[{{Burenin} {et~al.}(2018){Burenin}, {Bikmaev}, {Khamitov}, {Zaznobin}, {Khorunzhev}, {Eselevich}, {Afanasiev}, {Dodonov}, {Rubi{\~n}o-Mart{\'\i}n}, {Aghanim}, \& {Sunyaev}}]{burenin18}
{Burenin}, R.~A., {Bikmaev}, I.~F., {Khamitov}, I.~M., {et~al.} 2018, Astronomy Letters, 44, 297, \dodoi{10.1134/S1063773718050018}

\bibitem[{{Coe} {et~al.}(2006){Coe}, {Ben{\'\i}tez}, {S{\'a}nchez}, {Jee}, {Bouwens}, \& {Ford}}]{bpz06}
{Coe}, D., {Ben{\'\i}tez}, N., {S{\'a}nchez}, S.~F., {et~al.} 2006, \aj, 132, 926, \dodoi{10.1086/505530}

\bibitem[{{Dark Energy Survey Collaboration} {et~al.}(2016){Dark Energy Survey Collaboration}, {Abbott}, {Abdalla}, {Aleksi{\'c}}, {Allam}, {Amara}, {Bacon}, {Balbinot}, {Banerji}, {Bechtol}, {Benoit-L{\'e}vy}, {Bernstein}, {Bertin}, {Blazek}, {Bonnett}, {Bridle}, {Brooks}, {Brunner}, {Buckley-Geer}, {Burke}, {Caminha}, {Capozzi}, {Carlsen}, {Carnero-Rosell}, {Carollo}, {Carrasco-Kind}, {Carretero}, {Castander}, {Clerkin}, {Collett}, {Conselice}, {Crocce}, {Cunha}, {D'Andrea}, {da Costa}, {Davis}, {Desai}, {Diehl}, {Dietrich}, {Dodelson}, {Doel}, {Drlica-Wagner}, {Estrada}, {Etherington}, {Evrard}, {Fabbri}, {Finley}, {Flaugher}, {Foley}, {Fosalba}, {Frieman}, {Garc{\'\i}a-Bellido}, {Gaztanaga}, {Gerdes}, {Giannantonio}, {Goldstein}, {Gruen}, {Gruendl}, {Guarnieri}, {Gutierrez}, {Hartley}, {Honscheid}, {Jain}, {James}, {Jeltema}, {Jouvel}, {Kessler}, {King}, {Kirk}, {Kron}, {Kuehn}, {Kuropatkin}, {Lahav}, {Li}, {Lima}, {Lin}, {Maia}, {Makler}, {Manera}, {Maraston}, {Marshall}, {Martini}, {McMahon},
  {Melchior}, {Merson}, {Miller}, {Miquel}, {Mohr}, {Morice-Atkinson}, {Naidoo}, {Neilsen}, {Nichol}, {Nord}, {Ogando}, {Ostrovski}, {Palmese}, {Papadopoulos}, {Peiris}, {Peoples}, {Percival}, {Plazas}, {Reed}, {Refregier}, {Romer}, {Roodman}, {Ross}, {Rozo}, {Rykoff}, {Sadeh}, {Sako}, {S{\'a}nchez}, {Sanchez}, {Santiago}, {Scarpine}, {Schubnell}, {Sevilla-Noarbe}, {Sheldon}, {Smith}, {Smith}, {Soares-Santos}, {Sobreira}, {Soumagnac}, {Suchyta}, {Sullivan}, {Swanson}, {Tarle}, {Thaler}, {Thomas}, {Thomas}, {Tucker}, {Vieira}, {Vikram}, {Walker}, {Wechsler}, {Weller}, {Wester}, {Whiteway}, {Wilcox}, {Yanny}, {Zhang}, \& {Zuntz}}]{des}
{Dark Energy Survey Collaboration}, {Abbott}, T., {Abdalla}, F.~B., {et~al.} 2016, \mnras, 460, 1270, \dodoi{10.1093/mnras/stw641}

\bibitem[{{de Jong} {et~al.}(2019){de Jong}, {Agertz}, {Berbel}, {Aird}, {Alexander}, {Amarsi}, {Anders}, {Andrae}, {Ansarinejad}, {Ansorge}, {Antilogus}, {Anwand-Heerwart}, {Arentsen}, {Arnadottir}, {Asplund}, {Auger}, {Azais}, {Baade}, {Baker}, {Baker}, {Balbinot}, {Baldry}, {Banerji}, {Barden}, {Barklem}, {Barth{\'e}l{\'e}my-Mazot}, {Battistini}, {Bauer}, {Bell}, {Bellido-Tirado}, {Bellstedt}, {Belokurov}, {Bensby}, {Bergemann}, {Bestenlehner}, {Bielby}, {Bilicki}, {Blake}, {Bland-Hawthorn}, {Boeche}, {Boland}, {Boller}, {Bongard}, {Bongiorno}, {Bonifacio}, {Boudon}, {Brooks}, {Brown}, {Brown}, {Br{\"u}ggen}, {Brynnel}, {Brzeski}, {Buchert}, {Buschkamp}, {Caffau}, {Caillier}, {Carrick}, {Casagrande}, {Case}, {Casey}, {Cesarini}, {Cescutti}, {Chapuis}, {Chiappini}, {Childress}, {Christlieb}, {Church}, {Cioni}, {Cluver}, {Colless}, {Collett}, {Comparat}, {Cooper}, {Couch}, {Courbin}, {Croom}, {Croton}, {Daguis{\'e}}, {Dalton}, {Davies}, {Davis}, {de Laverny}, {Deason}, {Dionies}, {Disseau}, {Doel},
  {D{\"o}scher}, {Driver}, {Dwelly}, {Eckert}, {Edge}, {Edvardsson}, {Youssoufi}, {Elhaddad}, {Enke}, {Erfanianfar}, {Farrell}, {Fechner}, {Feiz}, {Feltzing}, {Ferreras}, {Feuerstein}, {Feuillet}, {Finoguenov}, {Ford}, {Fotopoulou}, {Fouesneau}, {Frenk}, {Frey}, {Gaessler}, {Geier}, {Gentile Fusillo}, {Gerhard}, {Giannantonio}, {Giannone}, {Gibson}, {Gillingham}, {Gonz{\'a}lez-Fern{\'a}ndez}, {Gonzalez-Solares}, {Gottloeber}, {Gould}, {Grebel}, {Gueguen}, {Guiglion}, {Haehnelt}, {Hahn}, {Hansen}, {Hartman}, {Hauptner}, {Hawkins}, {Haynes}, {Haynes}, {Heiter}, {Helmi}, {Aguayo}, {Hewett}, {Hinton}, {Hobbs}, {Hoenig}, {Hofman}, {Hook}, {Hopgood}, {Hopkins}, {Hourihane}, {Howes}, {Howlett}, {Huet}, {Irwin}, {Iwert}, {Jablonka}, {Jahn}, {Jahnke}, {Jarno}, {Jin}, {Jofre}, {Johl}, {Jones}, {J{\"o}nsson}, {Jordan}, {Karovicova}, {Khalatyan}, {Kelz}, {Kennicutt}, {King}, {Kitaura}, {Klar}, {Klauser}, {Kneib}, {Koch}, {Koposov}, {Kordopatis}, {Korn}, {Kosmalski}, {Kotak}, {Kovalev}, {Kreckel}, {Kripak}, {Krumpe},
  {Kuijken}, {Kunder}, {Kushniruk}, {Lam}, {Lamer}, {Laurent}, {Lawrence}, {Lehmitz}, {Lemasle}, {Lewis}, {Li}, {Lidman}, {Lind}, {Liske}, {Lizon}, {Loveday}, {Ludwig}, {McDermid}, {Maguire}, {Mainieri}, {Mali}, {Mandel}, {Mandel}, {Mannering}, {Martell}, {Martinez Delgado}, {Matijevic}, {McGregor}, {McMahon}, {McMillan}, {Mena}, {Merloni}, {Meyer}, {Michel}, {Micheva}, {Migniau}, {Minchev}, {Monari}, {Muller}, {Murphy}, {Muthukrishna}, {Nandra}, {Navarro}, {Ness}, {Nichani}, {Nichol}, {Nicklas}, {Niederhofer}, {Norberg}, {Obreschkow}, {Oliver}, {Owers}, {Pai}, {Pankratow}, {Parkinson}, {Paschke}, {Paterson}, {Pecontal}, {Parry}, {Phillips}, {Pillepich}, {Pinard}, {Pirard}, {Piskunov}, {Plank}, {Pl{\"u}schke}, {Pons}, {Popesso}, {Power}, {Pragt}, {Pramskiy}, {Pryer}, {Quattri}, {Queiroz}, {Quirrenbach}, {Rahurkar}, {Raichoor}, {Ramstedt}, {Rau}, {Recio-Blanco}, {Reiss}, {Renaud}, {Revaz}, {Rhode}, {Richard}, {Richter}, {Rix}, {Robotham}, {Roelfsema}, {Romaniello}, {Rosario}, {Rothmaier}, {Roukema}, {Ruchti},
  {Rupprecht}, {Rybizki}, {Ryde}, {Saar}, {Sadler}, {Sahl{\'e}n}, {Salvato}, {Sassolas}, {Saunders}, {Saviauk}, {Sbordone}, {Schmidt}, {Schnurr}, {Scholz}, {Schwope}, {Seifert}, {Shanks}, {Sheinis}, {Sivov}, {Sk{\'u}lad{\'o}ttir}, {Smartt}, {Smedley}, {Smith}, {Smith}, {Sorce}, {Spitler}, {Starkenburg}, {Steinmetz}, {Stilz}, {Storm}, {Sullivan}, {Sutherland}, {Swann}, {Tamone}, {Taylor}, {Teillon}, {Tempel}, {ter Horst}, {Thi}, {Tolstoy}, {Trager}, {Traven}, {Tremblay}, {Tresse}, {Valentini}, {van de Weygaert}, {van den Ancker}, {Veljanoski}, {Venkatesan}, {Wagner}, {Wagner}, {Walcher}, {Waller}, {Walton}, {Wang}, {Winkler}, {Wisotzki}, {Worley}, {Worseck}, {Xiang}, {Xu}, {Yong}, {Zhao}, {Zheng}, {Zscheyge}, \& {Zucker}}]{4most}
{de Jong}, R.~S., {Agertz}, O., {Berbel}, A.~A., {et~al.} 2019, The Messenger, 175, 3, \dodoi{10.18727/0722-6691/5117}

\bibitem[{{DESI Collaboration} {et~al.}(2016){DESI Collaboration}, {Aghamousa}, {Aguilar}, {Ahlen}, {Alam}, {Allen}, {Allende Prieto}, {Annis}, {Bailey}, {Balland}, {Ballester}, {Baltay}, {Beaufore}, {Bebek}, {Beers}, {Bell}, {Bernal}, {Besuner}, {Beutler}, {Blake}, {Bleuler}, {Blomqvist}, {Blum}, {Bolton}, {Briceno}, {Brooks}, {Brownstein}, {Buckley-Geer}, {Burden}, {Burtin}, {Busca}, {Cahn}, {Cai}, {Cardiel-Sas}, {Carlberg}, {Carton}, {Casas}, {Castander}, {Cervantes-Cota}, {Claybaugh}, {Close}, {Coker}, {Cole}, {Comparat}, {Cooper}, {Cousinou}, {Crocce}, {Cuby}, {Cunningham}, {Davis}, {Dawson}, {de la Macorra}, {De Vicente}, {Delubac}, {Derwent}, {Dey}, {Dhungana}, {Ding}, {Doel}, {Duan}, {Ealet}, {Edelstein}, {Eftekharzadeh}, {Eisenstein}, {Elliott}, {Escoffier}, {Evatt}, {Fagrelius}, {Fan}, {Fanning}, {Farahi}, {Farihi}, {Favole}, {Feng}, {Fernandez}, {Findlay}, {Finkbeiner}, {Fitzpatrick}, {Flaugher}, {Flender}, {Font-Ribera}, {Forero-Romero}, {Fosalba}, {Frenk}, {Fumagalli}, {Gaensicke}, {Gallo},
  {Garcia-Bellido}, {Gaztanaga}, {Pietro Gentile Fusillo}, {Gerard}, {Gershkovich}, {Giannantonio}, {Gillet}, {Gonzalez-de-Rivera}, {Gonzalez-Perez}, {Gott}, {Graur}, {Gutierrez}, {Guy}, {Habib}, {Heetderks}, {Heetderks}, {Heitmann}, {Hellwing}, {Herrera}, {Ho}, {Holland}, {Honscheid}, {Huff}, {Hutchinson}, {Huterer}, {Hwang}, {Illa Laguna}, {Ishikawa}, {Jacobs}, {Jeffrey}, {Jelinsky}, {Jennings}, {Jiang}, {Jimenez}, {Johnson}, {Joyce}, {Jullo}, {Juneau}, {Kama}, {Karcher}, {Karkar}, {Kehoe}, {Kennamer}, {Kent}, {Kilbinger}, {Kim}, {Kirkby}, {Kisner}, {Kitanidis}, {Kneib}, {Koposov}, {Kovacs}, {Koyama}, {Kremin}, {Kron}, {Kronig}, {Kueter-Young}, {Lacey}, {Lafever}, {Lahav}, {Lambert}, {Lampton}, {Landriau}, {Lang}, {Lauer}, {Le Goff}, {Le Guillou}, {Le Van Suu}, {Lee}, {Lee}, {Leitner}, {Lesser}, {Levi}, {L'Huillier}, {Li}, {Liang}, {Lin}, {Linder}, {Loebman}, {Luki{\'c}}, {Ma}, {MacCrann}, {Magneville}, {Makarem}, {Manera}, {Manser}, {Marshall}, {Martini}, {Massey}, {Matheson}, {McCauley}, {McDonald},
  {McGreer}, {Meisner}, {Metcalfe}, {Miller}, {Miquel}, {Moustakas}, {Myers}, {Naik}, {Newman}, {Nichol}, {Nicola}, {Nicolati da Costa}, {Nie}, {Niz}, {Norberg}, {Nord}, {Norman}, {Nugent}, {O'Brien}, {Oh}, {Olsen}, {Padilla}, {Padmanabhan}, {Padmanabhan}, {Palanque-Delabrouille}, {Palmese}, {Pappalardo}, {P{\^a}ris}, {Park}, {Patej}, {Peacock}, {Peiris}, {Peng}, {Percival}, {Perruchot}, {Pieri}, {Pogge}, {Pollack}, {Poppett}, {Prada}, {Prakash}, {Probst}, {Rabinowitz}, {Raichoor}, {Ree}, {Refregier}, {Regal}, {Reid}, {Reil}, {Rezaie}, {Rockosi}, {Roe}, {Ronayette}, {Roodman}, {Ross}, {Ross}, {Rossi}, {Rozo}, {Ruhlmann-Kleider}, {Rykoff}, {Sabiu}, {Samushia}, {Sanchez}, {Sanchez}, {Schlegel}, {Schneider}, {Schubnell}, {Secroun}, {Seljak}, {Seo}, {Serrano}, {Shafieloo}, {Shan}, {Sharples}, {Sholl}, {Shourt}, {Silber}, {Silva}, {Sirk}, {Slosar}, {Smith}, {Smoot}, {Som}, {Song}, {Sprayberry}, {Staten}, {Stefanik}, {Tarle}, {Sien Tie}, {Tinker}, {Tojeiro}, {Valdes}, {Valenzuela}, {Valluri}, {Vargas-Magana},
  {Verde}, {Walker}, {Wang}, {Wang}, {Weaver}, {Weaverdyck}, {Wechsler}, {Weinberg}, {White}, {Yang}, {Yeche}, {Zhang}, {Zhao}, {Zheng}, {Zhou}, {Zhou}, {Zhu}, {Zou}, \& {Zu}}]{desi}
{DESI Collaboration}, {Aghamousa}, A., {Aguilar}, J., {et~al.} 2016, arXiv e-prints, arXiv:1611.00036, \dodoi{10.48550/arXiv.1611.00036}

\bibitem[{{DESI Collaboration} {et~al.}(2023){DESI Collaboration}, {Adame}, {Aguilar}, {Ahlen}, {Alam}, {Aldering}, {Alexander}, {Alfarsy}, {Allende Prieto}, {Alvarez}, {Alves}, {Anand}, {Andrade-Oliveira}, {Armengaud}, {Asorey}, {Avila}, {Aviles}, {Bailey}, {Balaguera-Antol{\'\i}nez}, {Ballester}, {Baltay}, {Bault}, {Bautista}, {Behera}, {Beltran}, {BenZvi}, {Beraldo e Silva}, {Bermejo-Climent}, {Berti}, {Besuner}, {Beutler}, {Bianchi}, {Blake}, {Blum}, {Bolton}, {Brieden}, {Brodzeller}, {Brooks}, {Brown}, {Buckley-Geer}, {Burtin}, {Cabayol-Garcia}, {Cai}, {Canning}, {Cardiel-Sas}, {Carnero Rosell}, {Castander}, {Cervantes-Cota}, {Chabanier}, {Chaussidon}, {Chaves-Montero}, {Chen}, {Chuang}, {Claybaugh}, {Cole}, {Cooper}, {Cuceu}, {Davis}, {Dawson}, {de Belsunce}, {de la Cruz}, {de la Macorra}, {de Mattia}, {Demina}, {Demirbozan}, {DeRose}, {Dey}, {Dey}, {Dhungana}, {Ding}, {Ding}, {Doel}, {Doshi}, {Douglass}, {Edge}, {Eftekharzadeh}, {Eisenstein}, {Elliott}, {Escoffier}, {Fagrelius}, {Fan}, {Fanning},
  {Fawcett}, {Ferraro}, {Ereza}, {Flaugher}, {Font-Ribera}, {Forero-S{\'a}nchez}, {Forero-Romero}, {Frenk}, {G{\"a}nsicke}, {Garc{\'\i}a}, {Garc{\'\i}a-Bellido}, {Garcia-Quintero}, {Garrison}, {Gil-Mar{\'\i}n}, {Golden-Marx}, {Gontcho}, {Gonzalez-Morales}, {Gonzalez-Perez}, {Gordon}, {Graur}, {Green}, {Gruen}, {Guy}, {Hadzhiyska}, {Hahn}, {Han}, {Hanif}, {Herrera-Alcantar}, {Honscheid}, {Hou}, {Howlett}, {Huterer}, {Ir{\v{s}}i{\v{c}}}, {Ishak}, {Jacques}, {Jana}, {Jiang}, {Jimenez}, {Jing}, {Joudaki}, {Jullo}, {Juneau}, {Kizhuprakkat}, {Kara{\c{c}}ayl{\i}}, {Karim}, {Kehoe}, {Kent}, {Khederlarian}, {Kim}, {Kirkby}, {Kisner}, {Kitaura}, {Kneib}, {Koposov}, {Kov{\'a}cs}, {Kremin}, {Krolewski}, {L'Huillier}, {Lambert}, {Lamman}, {Lan}, {Landriau}, {Lang}, {Lange}, {Lasker}, {Le Guillou}, {Leauthaud}, {Levi}, {Li}, {Linder}, {Lyons}, {Magneville}, {Manera}, {Manser}, {Margala}, {Martini}, {McDonald}, {Medina}, {Medina-Varela}, {Meisner}, {Mena-Fern{\'a}ndez}, {Meneses-Rizo}, {Mezcua}, {Miquel}, {Montero-Camacho},
  {Moon}, {Moore}, {Moustakas}, {Mueller}, {Mundet}, {Mu{\~n}oz-Guti{\'e}rrez}, {Myers}, {Nadathur}, {Napolitano}, {Neveux}, {Newman}, {Nie}, {Nikutta}, {Niz}, {Norberg}, {Noriega}, {Paillas}, {Palanque-Delabrouille}, {Palmese}, {Zhiwei}, {Parkinson}, {Penmetsa}, {Percival}, {P{\'e}rez-Fern{\'a}ndez}, {P{\'e}rez-R{\`a}fols}, {Pieri}, {Poppett}, {Porredon}, {Pothier}, {Prada}, {Pucha}, {Raichoor}, {Ram{\'\i}rez-P{\'e}rez}, {Ramirez-Solano}, {Rashkovetskyi}, {Ravoux}, {Rocher}, {Rockosi}, {Ross}, {Rossi}, {Ruggeri}, {Ruhlmann-Kleider}, {Sabiu}, {Said}, {Saintonge}, {Samushia}, {Sanchez}, {Saulder}, {Schaan}, {Schlafly}, {Schlegel}, {Scholte}, {Schubnell}, {Seo}, {Shafieloo}, {Sharples}, {Sheu}, {Silber}, {Sinigaglia}, {Siudek}, {Slepian}, {Smith}, {Sprayberry}, {Stephey}, {Su{\'a}rez-P{\'e}rez}, {Sun}, {Tan}, {Tarl{\'e}}, {Tojeiro}, {Ure{\~n}a-L{\'o}pez}, {Vaisakh}, {Valcin}, {Valdes}, {Valluri}, {Vargas-Maga{\~n}a}, {Variu}, {Verde}, {Walther}, {Wang}, {Wang}, {Weaver}, {Weaverdyck}, {Wechsler}, {White},
  {Xie}, {Yang}, {Y{\`e}che}, {Yu}, {Yuan}, {Zhang}, {Zhang}, {Zhao}, {Zheng}, {Zhou}, {Zhou}, {Zou}, {Zou}, \& {Zu}}]{desiedr}
{DESI Collaboration}, {Adame}, A.~G., {Aguilar}, J., {et~al.} 2023, arXiv e-prints, arXiv:2306.06308, \dodoi{10.48550/arXiv.2306.06308}

\bibitem[{{Dey} {et~al.}(2019){Dey}, {Schlegel}, {Lang}, {Blum}, {Burleigh}, {Fan}, {Findlay}, {Finkbeiner}, {Herrera}, {Juneau}, {Landriau}, {Levi}, {McGreer}, {Meisner}, {Myers}, {Moustakas}, {Nugent}, {Patej}, {Schlafly}, {Walker}, {Valdes}, {Weaver}, {Y{\`e}che}, {Zou}, {Zhou}, {Abareshi}, {Abbott}, {Abolfathi}, {Aguilera}, {Alam}, {Allen}, {Alvarez}, {Annis}, {Ansarinejad}, {Aubert}, {Beechert}, {Bell}, {BenZvi}, {Beutler}, {Bielby}, {Bolton}, {Brice{\~n}o}, {Buckley-Geer}, {Butler}, {Calamida}, {Carlberg}, {Carter}, {Casas}, {Castander}, {Choi}, {Comparat}, {Cukanovaite}, {Delubac}, {DeVries}, {Dey}, {Dhungana}, {Dickinson}, {Ding}, {Donaldson}, {Duan}, {Duckworth}, {Eftekharzadeh}, {Eisenstein}, {Etourneau}, {Fagrelius}, {Farihi}, {Fitzpatrick}, {Font-Ribera}, {Fulmer}, {G{\"a}nsicke}, {Gaztanaga}, {George}, {Gerdes}, {Gontcho}, {Gorgoni}, {Green}, {Guy}, {Harmer}, {Hernandez}, {Honscheid}, {Huang}, {James}, {Jannuzi}, {Jiang}, {Joyce}, {Karcher}, {Karkar}, {Kehoe}, {Kneib}, {Kueter-Young}, {Lan},
  {Lauer}, {Le Guillou}, {Le Van Suu}, {Lee}, {Lesser}, {Perreault Levasseur}, {Li}, {Mann}, {Marshall}, {Mart{\'\i}nez-V{\'a}zquez}, {Martini}, {du Mas des Bourboux}, {McManus}, {Meier}, {M{\'e}nard}, {Metcalfe}, {Mu{\~n}oz-Guti{\'e}rrez}, {Najita}, {Napier}, {Narayan}, {Newman}, {Nie}, {Nord}, {Norman}, {Olsen}, {Paat}, {Palanque-Delabrouille}, {Peng}, {Poppett}, {Poremba}, {Prakash}, {Rabinowitz}, {Raichoor}, {Rezaie}, {Robertson}, {Roe}, {Ross}, {Ross}, {Rudnick}, {Safonova}, {Saha}, {S{\'a}nchez}, {Savary}, {Schweiker}, {Scott}, {Seo}, {Shan}, {Silva}, {Slepian}, {Soto}, {Sprayberry}, {Staten}, {Stillman}, {Stupak}, {Summers}, {Sien Tie}, {Tirado}, {Vargas-Maga{\~n}a}, {Vivas}, {Wechsler}, {Williams}, {Yang}, {Yang}, {Yapici}, {Zaritsky}, {Zenteno}, {Zhang}, {Zhang}, {Zhou}, \& {Zhou}}]{desi-ls-overview-dey19}
{Dey}, A., {Schlegel}, D.~J., {Lang}, D., {et~al.} 2019, \aj, 157, 168, \dodoi{10.3847/1538-3881/ab089d}

\bibitem[{{Dor{\'e}} {et~al.}(2014){Dor{\'e}}, {Bock}, {Ashby}, {Capak}, {Cooray}, {de Putter}, {Eifler}, {Flagey}, {Gong}, {Habib}, {Heitmann}, {Hirata}, {Jeong}, {Katti}, {Korngut}, {Krause}, {Lee}, {Masters}, {Mauskopf}, {Melnick}, {Mennesson}, {Nguyen}, {{\"O}berg}, {Pullen}, {Raccanelli}, {Smith}, {Song}, {Tolls}, {Unwin}, {Venumadhav}, {Viero}, {Werner}, \& {Zemcov}}]{dore14}
{Dor{\'e}}, O., {Bock}, J., {Ashby}, M., {et~al.} 2014, arXiv e-prints, arXiv:1412.4872, \dodoi{10.48550/arXiv.1412.4872}

\bibitem[{{Dor{\'e}} {et~al.}(2016){Dor{\'e}}, {Werner}, {Ashby}, {Banerjee}, {Battaglia}, {Bauer}, {Benjamin}, {Bleem}, {Bock}, {Boogert}, {Bull}, {Capak}, {Chang}, {Chiar}, {Cohen}, {Cooray}, {Crill}, {Cushing}, {de Putter}, {Driver}, {Eifler}, {Feng}, {Ferraro}, {Finkbeiner}, {Gaudi}, {Greene}, {Hillenbrand}, {H{\"o}flich}, {Hsiao}, {Huffenberger}, {Jansen}, {Jeong}, {Joshi}, {Kim}, {Kim}, {Kirkpatrick}, {Korngut}, {Krause}, {Kriek}, {Leistedt}, {Li}, {Lisse}, {Mauskopf}, {Mechtley}, {Melnick}, {Mohr}, {Murphy}, {Neben}, {Neufeld}, {Nguyen}, {Pierpaoli}, {Pyo}, {Rhodes}, {Sandstrom}, {Schaan}, {Schlaufman}, {Silverman}, {Su}, {Stassun}, {Stevens}, {Strauss}, {Tielens}, {Tsai}, {Tolls}, {Unwin}, {Viero}, {Windhorst}, \& {Zemcov}}]{dore16}
{Dor{\'e}}, O., {Werner}, M.~W., {Ashby}, M., {et~al.} 2016, arXiv e-prints, arXiv:1606.07039.
\newblock \doarXiv{1606.07039}

\bibitem[{{Dor{\'e}} {et~al.}(2018){Dor{\'e}}, {Werner}, {Ashby}, {Bleem}, {Bock}, {Burt}, {Capak}, {Chang}, {Chaves-Montero}, {Chen}, {Civano}, {Cleeves}, {Cooray}, {Crill}, {Crossfield}, {Cushing}, {de la Torre}, {DiMatteo}, {Dvory}, {Dvorkin}, {Espaillat}, {Ferraro}, {Finkbeiner}, {Greene}, {Hewitt}, {Hogg}, {Huffenberger}, {Jun}, {Ilbert}, {Jeong}, {Johnson}, {Kim}, {Kirkpatrick}, {Kowalski}, {Korngut}, {Li}, {Lisse}, {MacGregor}, {Mamajek}, {Mauskopf}, {Melnick}, {M{\'e}nard}, {Neyrinck}, {{\"O}berg}, {Pisani}, {Rocca}, {Salvato}, {Schaan}, {Scoville}, {Song}, {Stevens}, {Tenneti}, {Teplitz}, {Tolls}, {Unwin}, {Urry}, {Wandelt}, {Williams}, {Wilner}, {Windhorst}, {Wolk}, {Yorke}, \& {Zemcov}}]{dore18}
{Dor{\'e}}, O., {Werner}, M.~W., {Ashby}, M. L.~N., {et~al.} 2018, arXiv e-prints, arXiv:1805.05489.
\newblock \doarXiv{1805.05489}

\bibitem[{{Ebeling} {et~al.}(1998){Ebeling}, {Edge}, {Bohringer}, {Allen}, {Crawford}, {Fabian}, {Voges}, \& {Huchra}}]{rasscl}
{Ebeling}, H., {Edge}, A.~C., {Bohringer}, H., {et~al.} 1998, \mnras, 301, 881, \dodoi{10.1046/j.1365-8711.1998.01949.x}

\bibitem[{{Ebeling} {et~al.}(2010){Ebeling}, {Edge}, {Mantz}, {Barrett}, {Henry}, {Ma}, \& {van Speybroeck}}]{macs}
{Ebeling}, H., {Edge}, A.~C., {Mantz}, A., {et~al.} 2010, \mnras, 407, 83, \dodoi{10.1111/j.1365-2966.2010.16920.x}

\bibitem[{{Gardner} {et~al.}(2006){Gardner}, {Mather}, {Clampin}, {Doyon}, {Greenhouse}, {Hammel}, {Hutchings}, {Jakobsen}, {Lilly}, {Long}, {Lunine}, {McCaughrean}, {Mountain}, {Nella}, {Rieke}, {Rieke}, {Rix}, {Smith}, {Sonneborn}, {Stiavelli}, {Stockman}, {Windhorst}, \& {Wright}}]{jwst}
{Gardner}, J.~P., {Mather}, J.~C., {Clampin}, M., {et~al.} 2006, \ssr, 123, 485, \dodoi{10.1007/s11214-006-8315-7}

\bibitem[{{Ginsburg} {et~al.}(2019){Ginsburg}, {Sip{\H o}cz}, {Brasseur}, {Cowperthwaite}, {Craig}, {Deil}, {Guillochon}, {Guzman}, {Liedtke}, {Lian Lim}, {Lockhart}, {Mommert}, {Morris}, {Norman}, {Parikh}, {Persson}, {Robitaille}, {Segovia}, {Singer}, {Tollerud}, {de Val-Borro}, {Valtchanov}, {Woillez}, {The Astroquery collaboration}, \& {a subset of the astropy collaboration}}]{astroquery}
{Ginsburg}, A., {Sip{\H o}cz}, B.~M., {Brasseur}, C.~E., {et~al.} 2019, \aj, 157, 98, \dodoi{10.3847/1538-3881/aafc33}

\bibitem[{{Gladders} {et~al.}(1998){Gladders}, {L{\'o}pez-Cruz}, {Yee}, \& {Kodama}}]{gladders98}
{Gladders}, M.~D., {L{\'o}pez-Cruz}, O., {Yee}, H.~K.~C., \& {Kodama}, T. 1998, \apj, 501, 571, \dodoi{10.1086/305858}

\bibitem[{{Gladders} \& {Yee}(2000)}]{gladders00}
{Gladders}, M.~D., \& {Yee}, H.~K.~C. 2000, \aj, 120, 2148, \dodoi{10.1086/301557}

\bibitem[{{Gonzalez} {et~al.}(2019){Gonzalez}, {Gettings}, {Brodwin}, {Eisenhardt}, {Stanford}, {Wylezalek}, {Decker}, {Marrone}, {Moravec}, {O'Donnell}, {Stalder}, {Stern}, {Abdulla}, {Brown}, {Carlstrom}, {Chambers}, {Hayden}, {Lin}, {Magnier}, {Masci}, {Mantz}, {McDonald}, {Mo}, {Perlmutter}, {Wright}, \& {Zeimann}}]{madcow}
{Gonzalez}, A.~H., {Gettings}, D.~P., {Brodwin}, M., {et~al.} 2019, \apjs, 240, 33, \dodoi{10.3847/1538-4365/aafad2}

\bibitem[{Harris {et~al.}(2020)Harris, Millman, van~der Walt, Gommers, Virtanen, Cournapeau, Wieser, Taylor, Berg, Smith, Kern, Picus, Hoyer, van Kerkwijk, Brett, Haldane, del R{\'{i}}o, Wiebe, Peterson, G{\'{e}}rard-Marchant, Sheppard, Reddy, Weckesser, Abbasi, Gohlke, \& Oliphant}]{numpy}
Harris, C.~R., Millman, K.~J., van~der Walt, S.~J., {et~al.} 2020, Nature, 585, 357, \dodoi{10.1038/s41586-020-2649-2}

\bibitem[{{Hern{\'a}ndez-Lang} {et~al.}(2022){Hern{\'a}ndez-Lang}, {Mohr}, {Klein}, {Grandis}, {Melin}, {Tarr{\'\i}o}, {Arnaud}, {Pratt}, {Abbott}, {Aguena}, {Alves}, {Andrade-Oliveira}, {Bacon}, {Bertin}, {Brooks}, {Burke}, {Carnero Rosell}, {Carrasco Kind}, {Carretero}, {Castander}, {Costanzi}, {da Costa}, {Pereira}, {Desai}, {Diehl}, {Doel}, {Everett}, {Ferrero}, {Flaugher}, {Frieman}, {Garc{\'\i}a-Bellido}, {Gruen}, {Gruendl}, {Gschwend}, {Gutierrez}, {Hinton}, {Hollowood}, {Honscheid}, {James}, {Kuehn}, {Kuropatkin}, {Lahav}, {Lidman}, {Melchior}, {Mena-Fern{\'a}ndez}, {Menanteau}, {Miquel}, {Palmese}, {Paz-Chinch{\'o}n}, {Pieres}, {Plazas Malag{\'o}n}, {Raveri}, {Rodriguez-Monroy}, {Romer}, {Scarpine}, {Sevilla-Noarbe}, {Smith}, {Suchyta}, {Tarle}, {Thomas}, \& {Weaverdyck}}]{madpsz}
{Hern{\'a}ndez-Lang}, D., {Mohr}, J.~J., {Klein}, M., {et~al.} 2022, arXiv e-prints, arXiv:2210.04666, \dodoi{10.48550/arXiv.2210.04666}

\bibitem[{{Hern{\'a}ndez-Lang} {et~al.}(2023){Hern{\'a}ndez-Lang}, {Klein}, {Mohr}, {Grandis}, {Melin}, {Tarr{\'\i}o}, {Arnaud}, {Pratt}, {Abbott}, {Aguena}, {Alves}, {Andrade-Oliveira}, {Bacon}, {Bertin}, {Brooks}, {Burke}, {Carnero Rosell}, {Carrasco Kind}, {Carretero}, {Castander}, {Costanzi}, {da Costa}, {Pereira}, {Desai}, {Diehl}, {Doel}, {Everett}, {Ferrero}, {Flaugher}, {Frieman}, {Garc{\'\i}a-Bellido}, {Gruen}, {Gruendl}, {Gschwend}, {Gutierrez}, {Hinton}, {Hollowood}, {Honscheid}, {James}, {Kuehn}, {Kuropatkin}, {Lahav}, {Lidman}, {Melchior}, {Mena-Fern{\'a}ndez}, {Menanteau}, {Miquel}, {Palmese}, {Paz-Chinch{\'o}n}, {Pieres}, {Plazas Malag{\'o}n}, {Raveri}, {Rodriguez-Monroy}, {Romer}, {Scarpine}, {Sevilla-Noarbe}, {Smith}, {Suchyta}, {Tarle}, {Thomas}, {Weaverdyck}, \& {(DES Collaboration)}}]{psz-mcmf}
{Hern{\'a}ndez-Lang}, D., {Klein}, M., {Mohr}, J.~J., {et~al.} 2023, \mnras, 525, 24, \dodoi{10.1093/mnras/stad2319}

\bibitem[{{Huang} {et~al.}(2020){Huang}, {Bleem}, {Stalder}, {Ade}, {Allen}, {Anderson}, {Austermann}, {Avva}, {Beall}, {Bender}, {Benson}, {Bianchini}, {Bocquet}, {Brodwin}, {Carlstrom}, {Chang}, {Chiang}, {Citron}, {Moran}, {Crawford}, {Crites}, {Haan}, {Dobbs}, {Everett}, {Floyd}, {Gallicchio}, {George}, {Gilbert}, {Gladders}, {Guns}, {Gupta}, {Halverson}, {Harrington}, {Henning}, {Hilton}, {Holder}, {Holzapfel}, {Hrubes}, {Hubmayr}, {Irwin}, {Khullar}, {Knox}, {Lee}, {Li}, {Lowitz}, {McDonald}, {McMahon}, {Meyer}, {Mocanu}, {Montgomery}, {Nadolski}, {Natoli}, {Nibarger}, {Noble}, {Novosad}, {Padin}, {Patil}, {Pryke}, {Reichardt}, {Ruhl}, {Saliwanchik}, {Saro}, {Sayre}, {Schaffer}, {Sharon}, {Sievers}, {Smecher}, {Stark}, {Story}, {Tucker}, {Vanderlinde}, {Veach}, {Vieira}, {Wang}, {Whitehorn}, {Wu}, \& {Yefremenko}}]{sptpol19}
{Huang}, N., {Bleem}, L.~E., {Stalder}, B., {et~al.} 2020, \aj, 159, 110, \dodoi{10.3847/1538-3881/ab6a96}

\bibitem[{Hunter(2007)}]{matplotlib}
Hunter, J.~D. 2007, Computing in Science \& Engineering, 9, 90, \dodoi{10.1109/MCSE.2007.55}

\bibitem[{{Hwang} {et~al.}(2010){Hwang}, {Elbaz}, {Lee}, {Jeong}, {Park}, {Lee}, \& {Lee}}]{hwang10}
{Hwang}, H.~S., {Elbaz}, D., {Lee}, J.~C., {et~al.} 2010, \aap, 522, A33, \dodoi{10.1051/0004-6361/201014807}

\bibitem[{{Hwang} {et~al.}(2014){Hwang}, {Geller}, {Diaferio}, {Rines}, \& {Zahid}}]{hwang14}
{Hwang}, H.~S., {Geller}, M.~J., {Diaferio}, A., {Rines}, K.~J., \& {Zahid}, H.~J. 2014, \apj, 797, 106, \dodoi{10.1088/0004-637X/797/2/106}

\bibitem[{{Ivezi{\'c}} {et~al.}(2019){Ivezi{\'c}}, {Kahn}, {Tyson}, {Abel}, {Acosta}, {Allsman}, {Alonso}, {AlSayyad}, {Anderson}, {Andrew}, {Angel}, {Angeli}, {Ansari}, {Antilogus}, {Araujo}, {Armstrong}, {Arndt}, {Astier}, {Aubourg}, {Auza}, {Axelrod}, {Bard}, {Barr}, {Barrau}, {Bartlett}, {Bauer}, {Bauman}, {Baumont}, {Bechtol}, {Bechtol}, {Becker}, {Becla}, {Beldica}, {Bellavia}, {Bianco}, {Biswas}, {Blanc}, {Blazek}, {Blandford}, {Bloom}, {Bogart}, {Bond}, {Booth}, {Borgland}, {Borne}, {Bosch}, {Boutigny}, {Brackett}, {Bradshaw}, {Brandt}, {Brown}, {Bullock}, {Burchat}, {Burke}, {Cagnoli}, {Calabrese}, {Callahan}, {Callen}, {Carlin}, {Carlson}, {Chandrasekharan}, {Charles-Emerson}, {Chesley}, {Cheu}, {Chiang}, {Chiang}, {Chirino}, {Chow}, {Ciardi}, {Claver}, {Cohen-Tanugi}, {Cockrum}, {Coles}, {Connolly}, {Cook}, {Cooray}, {Covey}, {Cribbs}, {Cui}, {Cutri}, {Daly}, {Daniel}, {Daruich}, {Daubard}, {Daues}, {Dawson}, {Delgado}, {Dellapenna}, {de Peyster}, {de Val-Borro}, {Digel}, {Doherty}, {Dubois},
  {Dubois-Felsmann}, {Durech}, {Economou}, {Eifler}, {Eracleous}, {Emmons}, {Fausti Neto}, {Ferguson}, {Figueroa}, {Fisher-Levine}, {Focke}, {Foss}, {Frank}, {Freemon}, {Gangler}, {Gawiser}, {Geary}, {Gee}, {Geha}, {Gessner}, {Gibson}, {Gilmore}, {Glanzman}, {Glick}, {Goldina}, {Goldstein}, {Goodenow}, {Graham}, {Gressler}, {Gris}, {Guy}, {Guyonnet}, {Haller}, {Harris}, {Hascall}, {Haupt}, {Hernandez}, {Herrmann}, {Hileman}, {Hoblitt}, {Hodgson}, {Hogan}, {Howard}, {Huang}, {Huffer}, {Ingraham}, {Innes}, {Jacoby}, {Jain}, {Jammes}, {Jee}, {Jenness}, {Jernigan}, {Jevremovi{\'c}}, {Johns}, {Johnson}, {Johnson}, {Jones}, {Juramy-Gilles}, {Juri{\'c}}, {Kalirai}, {Kallivayalil}, {Kalmbach}, {Kantor}, {Karst}, {Kasliwal}, {Kelly}, {Kessler}, {Kinnison}, {Kirkby}, {Knox}, {Kotov}, {Krabbendam}, {Krughoff}, {Kub{\'a}nek}, {Kuczewski}, {Kulkarni}, {Ku}, {Kurita}, {Lage}, {Lambert}, {Lange}, {Langton}, {Le Guillou}, {Levine}, {Liang}, {Lim}, {Lintott}, {Long}, {Lopez}, {Lotz}, {Lupton}, {Lust}, {MacArthur}, {Mahabal},
  {Mandelbaum}, {Markiewicz}, {Marsh}, {Marshall}, {Marshall}, {May}, {McKercher}, {McQueen}, {Meyers}, {Migliore}, {Miller}, {Mills}, {Miraval}, {Moeyens}, {Moolekamp}, {Monet}, {Moniez}, {Monkewitz}, {Montgomery}, {Morrison}, {Mueller}, {Muller}, {Mu{\~n}oz Arancibia}, {Neill}, {Newbry}, {Nief}, {Nomerotski}, {Nordby}, {O'Connor}, {Oliver}, {Olivier}, {Olsen}, {O'Mullane}, {Ortiz}, {Osier}, {Owen}, {Pain}, {Palecek}, {Parejko}, {Parsons}, {Pease}, {Peterson}, {Peterson}, {Petravick}, {Libby Petrick}, {Petry}, {Pierfederici}, {Pietrowicz}, {Pike}, {Pinto}, {Plante}, {Plate}, {Plutchak}, {Price}, {Prouza}, {Radeka}, {Rajagopal}, {Rasmussen}, {Regnault}, {Reil}, {Reiss}, {Reuter}, {Ridgway}, {Riot}, {Ritz}, {Robinson}, {Roby}, {Roodman}, {Rosing}, {Roucelle}, {Rumore}, {Russo}, {Saha}, {Sassolas}, {Schalk}, {Schellart}, {Schindler}, {Schmidt}, {Schneider}, {Schneider}, {Schoening}, {Schumacher}, {Schwamb}, {Sebag}, {Selvy}, {Sembroski}, {Seppala}, {Serio}, {Serrano}, {Shaw}, {Shipsey}, {Sick}, {Silvestri},
  {Slater}, {Smith}, {Smith}, {Sobhani}, {Soldahl}, {Storrie-Lombardi}, {Stover}, {Strauss}, {Street}, {Stubbs}, {Sullivan}, {Sweeney}, {Swinbank}, {Szalay}, {Takacs}, {Tether}, {Thaler}, {Thayer}, {Thomas}, {Thornton}, {Thukral}, {Tice}, {Trilling}, {Turri}, {Van Berg}, {Vanden Berk}, {Vetter}, {Virieux}, {Vucina}, {Wahl}, {Walkowicz}, {Walsh}, {Walter}, {Wang}, {Wang}, {Warner}, {Wiecha}, {Willman}, {Winters}, {Wittman}, {Wolff}, {Wood-Vasey}, {Wu}, {Xin}, {Yoachim}, \& {Zhan}}]{lsst}
{Ivezi{\'c}}, {\v{Z}}., {Kahn}, S.~M., {Tyson}, J.~A., {et~al.} 2019, \apj, 873, 111, \dodoi{10.3847/1538-4357/ab042c}

\bibitem[{{Kessler} {et~al.}(1996){Kessler}, {Steinz}, {Anderegg}, {Clavel}, {Drechsel}, {Estaria}, {Faelker}, {Riedinger}, {Robson}, {Taylor}, \& {Xim{\'e}nez de Ferr{\'a}n}}]{iso}
{Kessler}, M.~F., {Steinz}, J.~A., {Anderegg}, M.~E., {et~al.} 1996, \aap, 315, L27

\bibitem[{{Khatri}(2016)}]{khatri16}
{Khatri}, R. 2016, \aap, 592, A48, \dodoi{10.1051/0004-6361/201526479}

\bibitem[{{Klein} {et~al.}(2023){Klein}, {Hern{\'a}ndez-Lang}, {Mohr}, {Bocquet}, \& {Singh}}]{rass-mcmf}
{Klein}, M., {Hern{\'a}ndez-Lang}, D., {Mohr}, J.~J., {Bocquet}, S., \& {Singh}, A. 2023, \mnras, 526, 3757, \dodoi{10.1093/mnras/stad2729}

\bibitem[{{Klein} {et~al.}(2018){Klein}, {Mohr}, {Desai}, {Israel}, {Allam}, {Benoit-L{\'e}vy}, {Brooks}, {Buckley-Geer}, {Carnero Rosell}, {Carrasco Kind}, {Cunha}, {da Costa}, {Dietrich}, {Eifler}, {Evrard}, {Frieman}, {Gruen}, {Gruendl}, {Gutierrez}, {Honscheid}, {James}, {Kuehn}, {Lima}, {Maia}, {March}, {Melchior}, {Menanteau}, {Miquel}, {Plazas}, {Reil}, {Romer}, {Sanchez}, {Santiago}, {Scarpine}, {Schubnell}, {Sevilla-Noarbe}, {Smith}, {Soares-Santos}, {Sobreira}, {Suchyta}, {Swanson}, {Tarle}, \& {DES Collaboration}}]{klein18}
{Klein}, M., {Mohr}, J.~J., {Desai}, S., {et~al.} 2018, \mnras, 474, 3324, \dodoi{10.1093/mnras/stx2929}

\bibitem[{{Klein} {et~al.}(2019){Klein}, {Grandis}, {Mohr}, {Paulus}, {Abbott}, {Annis}, {Avila}, {Bertin}, {Brooks}, {Buckley-Geer}, {Rosell}, {Kind}, {Carretero}, {Castander}, {Cunha}, {D'Andrea}, {da Costa}, {De Vicente}, {Desai}, {Diehl}, {Dietrich}, {Doel}, {Evrard}, {Flaugher}, {Fosalba}, {Frieman}, {Garc{\'\i}a-Bellido}, {Gaztanaga}, {Giles}, {Gruen}, {Gruendl}, {Gschwend}, {Gutierrez}, {Hartley}, {Hollowood}, {Honscheid}, {Hoyle}, {James}, {Jeltema}, {Kuehn}, {Kuropatkin}, {Lima}, {Maia}, {March}, {Marshall}, {Menanteau}, {Miquel}, {Ogando}, {Plazas}, {Romer}, {Roodman}, {Sanchez}, {Scarpine}, {Schindler}, {Serrano}, {Sevilla-Noarbe}, {Smith}, {Smith}, {Soares-Santos}, {Sobreira}, {Suchyta}, {Swanson}, {Tarle}, {Thomas}, {Vikram}, \& {DES Collaboration}}]{klein19}
{Klein}, M., {Grandis}, S., {Mohr}, J.~J., {et~al.} 2019, \mnras, 488, 739, \dodoi{10.1093/mnras/stz1463}

\bibitem[{{Klein} {et~al.}(2022){Klein}, {Oguri}, {Mohr}, {Grandis}, {Ghirardini}, {Liu}, {Liu}, {Bulbul}, {Wolf}, {Comparat}, {Ramos-Ceja}, {Buchner}, {Chiu}, {Clerc}, {Merloni}, {Miyatake}, {Miyazaki}, {Okabe}, {Ota}, {Pacaud}, {Salvato}, \& {Driver}}]{klein22}
{Klein}, M., {Oguri}, M., {Mohr}, J.~J., {et~al.} 2022, \aap, 661, A4, \dodoi{10.1051/0004-6361/202141123}

\bibitem[{{Koulouridis} {et~al.}(2021){Koulouridis}, {Clerc}, {Sadibekova}, {Chira}, {Drigga}, {Faccioli}, {Le F{\`e}vre}, {Garrel}, {Gaynullina}, {Gkini}, {Kosiba}, {Pacaud}, {Pierre}, {Ridl}, {Tazhenova}, {Adami}, {Altieri}, {Baguley}, {Cabanac}, {Cucchetti}, {Khalikova}, {Lieu}, {Melin}, {Molham}, {Ramos-Ceja}, {Soucail}, {Takey}, \& {Valtchanov}}]{xclass}
{Koulouridis}, E., {Clerc}, N., {Sadibekova}, T., {et~al.} 2021, \aap, 652, A12, \dodoi{10.1051/0004-6361/202140566}

\bibitem[{{Kravtsov} \& {Borgani}(2012)}]{kravtsov12}
{Kravtsov}, A.~V., \& {Borgani}, S. 2012, \araa, 50, 353, \dodoi{10.1146/annurev-astro-081811-125502}

\bibitem[{{Lopes}(2007)}]{lopes07}
{Lopes}, P.~A.~A. 2007, \mnras, 380, 1608, \dodoi{10.1111/j.1365-2966.2007.12203.x}

\bibitem[{{Murakami} {et~al.}(2007){Murakami}, {Baba}, {Barthel}, {Clements}, {Cohen}, {Doi}, {Enya}, {Figueredo}, {Fujishiro}, {Fujiwara}, {Fujiwara}, {Garcia-Lario}, {Goto}, {Hasegawa}, {Hibi}, {Hirao}, {Hiromoto}, {Hong}, {Imai}, {Ishigaki}, {Ishiguro}, {Ishihara}, {Ita}, {Jeong}, {Jeong}, {Kaneda}, {Kataza}, {Kawada}, {Kawai}, {Kawamura}, {Kessler}, {Kester}, {Kii}, {Kim}, {Kim}, {Kobayashi}, {Koo}, {Kwon}, {Lee}, {Lorente}, {Makiuti}, {Matsuhara}, {Matsumoto}, {Matsuo}, {Matsuura}, {M{\"U}ller}, {Murakami}, {Nagata}, {Nakagawa}, {Naoi}, {Narita}, {Noda}, {Oh}, {Ohnishi}, {Ohyama}, {Okada}, {Okuda}, {Oliver}, {Onaka}, {Ootsubo}, {Oyabu}, {Pak}, {Park}, {Pearson}, {Rowan-Robinson}, {Saito}, {Sakon}, {Salama}, {Sato}, {Savage}, {Serjeant}, {Shibai}, {Shirahata}, {Sohn}, {Suzuki}, {Takagi}, {Takahashi}, {Tanab{\'E}}, {Takeuchi}, {Takita}, {Thomson}, {Uemizu}, {Ueno}, {Usui}, {Verdugo}, {Wada}, {Wang}, {Watabe}, {Watarai}, {White}, {Yamamura}, {Yamauchi}, \& {Yasuda}}]{akari}
{Murakami}, H., {Baba}, H., {Barthel}, P., {et~al.} 2007, \pasj, 59, S369, \dodoi{10.1093/pasj/59.sp2.S369}

\bibitem[{{Nersesian} {et~al.}(2023){Nersesian}, {van der Wel}, {Gallazzi}, {Leja}, {Bezanson}, {Bell}, {D'Eugenio}, {de Graaff}, {Kaushal}, {Martorano}, {Maseda}, \& {Zibetti}}]{nersesian23}
{Nersesian}, A., {van der Wel}, A., {Gallazzi}, A., {et~al.} 2023, arXiv e-prints, arXiv:2310.18000, \dodoi{10.48550/arXiv.2310.18000}

\bibitem[{{Neugebauer} {et~al.}(1984){Neugebauer}, {Habing}, {van Duinen}, {Aumann}, {Baud}, {Beichman}, {Beintema}, {Boggess}, {Clegg}, {de Jong}, {Emerson}, {Gautier}, {Gillett}, {Harris}, {Hauser}, {Houck}, {Jennings}, {Low}, {Marsden}, {Miley}, {Olnon}, {Pottasch}, {Raimond}, {Rowan-Robinson}, {Soifer}, {Walker}, {Wesselius}, \& {Young}}]{iras}
{Neugebauer}, G., {Habing}, H.~J., {van Duinen}, R., {et~al.} 1984, \apjl, 278, L1, \dodoi{10.1086/184209}

\bibitem[{{Park} \& {Hwang}(2009)}]{park09}
{Park}, C., \& {Hwang}, H.~S. 2009, \apj, 699, 1595, \dodoi{10.1088/0004-637X/699/2/1595}

\bibitem[{{Peebles}(1980)}]{peebles80}
{Peebles}, P.~J.~E. 1980, {The large-scale structure of the universe}

\bibitem[{{Piffaretti} {et~al.}(2011){Piffaretti}, {Arnaud}, {Pratt}, {Pointecouteau}, \& {Melin}}]{mcxc}
{Piffaretti}, R., {Arnaud}, M., {Pratt}, G.~W., {Pointecouteau}, E., \& {Melin}, J.~B. 2011, \aap, 534, A109, \dodoi{10.1051/0004-6361/201015377}

\bibitem[{{Planck Collaboration} {et~al.}(2014){Planck Collaboration}, {Ade}, {Aghanim}, {Armitage-Caplan}, {Arnaud}, {Ashdown}, {Atrio-Barandela}, {Aumont}, {Aussel}, {Baccigalupi}, {Banday}, {Barreiro}, {Barrena}, {Bartelmann}, {Bartlett}, {Battaner}, {Benabed}, {Beno{\^\i}t}, {Benoit-L{\'e}vy}, {Bernard}, {Bersanelli}, {Bielewicz}, {Bikmaev}, {Bobin}, {Bock}, {B{\"o}hringer}, {Bonaldi}, {Bond}, {Borrill}, {Bouchet}, {Bridges}, {Bucher}, {Burenin}, {Burigana}, {Butler}, {Cardoso}, {Carvalho}, {Catalano}, {Challinor}, {Chamballu}, {Chary}, {Chen}, {Chiang}, {Chiang}, {Chon}, {Christensen}, {Churazov}, {Church}, {Clements}, {Colombi}, {Colombo}, {Comis}, {Couchot}, {Coulais}, {Crill}, {Curto}, {Cuttaia}, {Da Silva}, {Dahle}, {Danese}, {Davies}, {Davis}, {de Bernardis}, {de Rosa}, {de Zotti}, {Delabrouille}, {Delouis}, {D{\'e}mocl{\`e}s}, {D{\'e}sert}, {Dickinson}, {Diego}, {Dolag}, {Dole}, {Donzelli}, {Dor{\'e}}, {Douspis}, {Dupac}, {Efstathiou}, {Eisenhardt}, {En{\ss}lin}, {Eriksen}, {Feroz}, {Finelli},
  {Flores-Cacho}, {Forni}, {Frailis}, {Franceschi}, {Fromenteau}, {Galeotta}, {Ganga}, {G{\'e}nova-Santos}, {Giard}, {Giardino}, {Gilfanov}, {Giraud-H{\'e}raud}, {Gonz{\'a}lez-Nuevo}, {G{\'o}rski}, {Grainge}, {Gratton}, {Gregorio}, {Groeneboom}, {Gruppuso}, {Hansen}, {Hanson}, {Harrison}, {Hempel}, {Henrot-Versill{\'e}}, {Hern{\'a}ndez-Monteagudo}, {Herranz}, {Hildebrandt}, {Hivon}, {Hobson}, {Holmes}, {Hornstrup}, {Hovest}, {Huffenberger}, {Hurier}, {Hurley-Walker}, {Jaffe}, {Jaffe}, {Jones}, {Juvela}, {Keih{\"a}nen}, {Keskitalo}, {Khamitov}, {Kisner}, {Kneissl}, {Knoche}, {Knox}, {Kunz}, {Kurki-Suonio}, {Lagache}, {L{\"a}hteenm{\"a}ki}, {Lamarre}, {Lasenby}, {Laureijs}, {Lawrence}, {Leahy}, {Leonardi}, {Le{\'o}n-Tavares}, {Lesgourgues}, {Li}, {Liddle}, {Liguori}, {Lilje}, {Linden-V{\o}rnle}, {L{\'o}pez-Caniego}, {Lubin}, {Mac{\'\i}as-P{\'e}rez}, {MacTavish}, {Maffei}, {Maino}, {Mandolesi}, {Maris}, {Marshall}, {Martin}, {Mart{\'\i}nez-Gonz{\'a}lez}, {Masi}, {Massardi}, {Matarrese}, {Matthai}, {Mazzotta},
  {Mei}, {Meinhold}, {Melchiorri}, {Melin}, {Mendes}, {Mennella}, {Migliaccio}, {Mikkelsen}, {Mitra}, {Miville-Desch{\^e}nes}, {Moneti}, {Montier}, {Morgante}, {Mortlock}, {Munshi}, {Murphy}, {Naselsky}, {Nati}, {Natoli}, {Nesvadba}, {Netterfield}, {N{\o}rgaard-Nielsen}, {Noviello}, {Novikov}, {Novikov}, {O'Dwyer}, {Olamaie}, {Osborne}, {Oxborrow}, {Paci}, {Pagano}, {Pajot}, {Paoletti}, {Pasian}, {Patanchon}, {Pearson}, {Perdereau}, {Perotto}, {Perrott}, {Perrotta}, {Piacentini}, {Piat}, {Pierpaoli}, {Pietrobon}, {Plaszczynski}, {Pointecouteau}, {Polenta}, {Ponthieu}, {Popa}, {Poutanen}, {Pratt}, {Pr{\'e}zeau}, {Prunet}, {Puget}, {Rachen}, {Reach}, {Rebolo}, {Reinecke}, {Remazeilles}, {Renault}, {Ricciardi}, {Riller}, {Ristorcelli}, {Rocha}, {Rosset}, {Roudier}, {Rowan-Robinson}, {Rubi{\~n}o-Mart{\'\i}n}, {Rumsey}, {Rusholme}, {Sandri}, {Santos}, {Saunders}, {Savini}, {Schammel}, {Scott}, {Seiffert}, {Shellard}, {Shimwell}, {Spencer}, {Stanford}, {Starck}, {Stolyarov}, {Stompor}, {Sudiwala}, {Sunyaev},
  {Sureau}, {Sutton}, {Suur-Uski}, {Sygnet}, {Tauber}, {Tavagnacco}, {Terenzi}, {Toffolatti}, {Tomasi}, {Tristram}, {Tucci}, {Tuovinen}, {T{\"u}rler}, {Umana}, {Valenziano}, {Valiviita}, {Van Tent}, {Vibert}, {Vielva}, {Villa}, {Vittorio}, {Wade}, {Wandelt}, {White}, {White}, {Yvon}, {Zacchei}, \& {Zonca}}]{psz1}
{Planck Collaboration}, {Ade}, P.~A.~R., {Aghanim}, N., {et~al.} 2014, \aap, 571, A29, \dodoi{10.1051/0004-6361/201321523}

\bibitem[{{Planck Collaboration} {et~al.}(2015{\natexlab{a}}){Planck Collaboration}, {Ade}, {Aghanim}, {Armitage-Caplan}, {Arnaud}, {Ashdown}, {Atrio-Barandela}, {Aumont}, {Aussel}, {Baccigalupi}, {Banday}, {Barreiro}, {Barrena}, {Bartelmann}, {Bartlett}, {Battaner}, {Benabed}, {Beno{\^\i}t}, {Benoit-L{\'e}vy}, {Bernard}, {Bersanelli}, {Bielewicz}, {Bikmaev}, {Bobin}, {Bock}, {B{\"o}hringer}, {Bonaldi}, {Bond}, {Borrill}, {Bouchet}, {Bridges}, {Bucher}, {Burenin}, {Burigana}, {Butler}, {Cardoso}, {Carvalho}, {Catalano}, {Challinor}, {Chamballu}, {Chary}, {Chen}, {Chiang}, {Chiang}, {Chon}, {Christensen}, {Churazov}, {Church}, {Clements}, {Colombi}, {Colombo}, {Comis}, {Couchot}, {Coulais}, {Crill}, {Curto}, {Cuttaia}, {Da Silva}, {Dahle}, {Danese}, {Davies}, {Davis}, {de Bernardis}, {de Rosa}, {de Zotti}, {Delabrouille}, {Delouis}, {D{\'e}mocl{\`e}s}, {D{\'e}sert}, {Dickinson}, {Diego}, {Dolag}, {Dole}, {Donzelli}, {Dor{\'e}}, {Douspis}, {Dupac}, {Efstathiou}, {En{\ss}lin}, {Eriksen}, {Feroz}, {Ferragamo},
  {Finelli}, {Flores-Cacho}, {Forni}, {Frailis}, {Franceschi}, {Fromenteau}, {Galeotta}, {Ganga}, {G{\'e}nova-Santos}, {Giard}, {Giardino}, {Gilfanov}, {Giraud-H{\'e}raud}, {Gonz{\'a}lez-Nuevo}, {G{\'o}rski}, {Grainge}, {Gratton}, {Gregorio}, {Groeneboom}, {Gruppuso}, {Hansen}, {Hanson}, {Harrison}, {Hempel}, {Henrot-Versill{\'e}}, {Hern{\'a}ndez-Monteagudo}, {Herranz}, {Hildebrandt}, {Hivon}, {Hobson}, {Holmes}, {Hornstrup}, {Hovest}, {Huffenberger}, {Hurier}, {Hurley-Walker}, {Jaffe}, {Jaffe}, {Jones}, {Juvela}, {Keih{\"a}nen}, {Keskitalo}, {Khamitov}, {Kisner}, {Kneissl}, {Knoche}, {Knox}, {Kunz}, {Kurki-Suonio}, {Lagache}, {L{\"a}hteenm{\"a}ki}, {Lamarre}, {Lasenby}, {Laureijs}, {Lawrence}, {Leahy}, {Leonardi}, {Le{\'o}n-Tavares}, {Lesgourgues}, {Li}, {Liddle}, {Liguori}, {Lilje}, {Linden-V{\o}rnle}, {L{\'o}pez-Caniego}, {Lubin}, {Mac{\'\i}as-P{\'e}rez}, {MacTavish}, {Maffei}, {Maino}, {Mandolesi}, {Maris}, {Marshall}, {Martin}, {Mart{\'\i}nez-Gonz{\'a}lez}, {Masi}, {Massardi}, {Matarrese}, {Matthai},
  {Mazzotta}, {Mei}, {Meinhold}, {Melchiorri}, {Melin}, {Mendes}, {Mennella}, {Migliaccio}, {Mikkelsen}, {Mitra}, {Miville-Desch{\^e}nes}, {Moneti}, {Montier}, {Morgante}, {Mortlock}, {Munshi}, {Murphy}, {Naselsky}, {Nastasi}, {Nati}, {Natoli}, {Nesvadba}, {Netterfield}, {N{\o}rgaard-Nielsen}, {Noviello}, {Novikov}, {Novikov}, {O'Dwyer}, {Olamaie}, {Osborne}, {Oxborrow}, {Paci}, {Pagano}, {Pajot}, {Paoletti}, {Pasian}, {Patanchon}, {Pearson}, {Perdereau}, {Perotto}, {Perrott}, {Perrotta}, {Piacentini}, {Piat}, {Pierpaoli}, {Pietrobon}, {Plaszczynski}, {Pointecouteau}, {Polenta}, {Ponthieu}, {Popa}, {Poutanen}, {Pratt}, {Pr{\'e}zeau}, {Prunet}, {Puget}, {Rachen}, {Reach}, {Rebolo}, {Reinecke}, {Remazeilles}, {Renault}, {Ricciardi}, {Riller}, {Ristorcelli}, {Rocha}, {Rosset}, {Roudier}, {Rowan-Robinson}, {Rubi{\~n}o-Mart{\'\i}n}, {Rumsey}, {Rusholme}, {Sandri}, {Santos}, {Saunders}, {Savini}, {Schammel}, {Scott}, {Seiffert}, {Shellard}, {Shimwell}, {Spencer}, {Starck}, {Stolyarov}, {Stompor}, {Streblyanska},
  {Sudiwala}, {Sunyaev}, {Sureau}, {Sutton}, {Suur-Uski}, {Sygnet}, {Tauber}, {Tavagnacco}, {Terenzi}, {Toffolatti}, {Tomasi}, {Tramonte}, {Tristram}, {Tucci}, {Tuovinen}, {T{\"u}rler}, {Umana}, {Valenziano}, {Valiviita}, {Van Tent}, {Vibert}, {Vielva}, {Villa}, {Vittorio}, {Wade}, {Wandelt}, {White}, {White}, {Yvon}, {Zacchei}, \& {Zonca}}]{psz1v2}
---. 2015{\natexlab{a}}, \aap, 581, A14, \dodoi{10.1051/0004-6361/201525787}

\bibitem[{{Planck Collaboration} {et~al.}(2015{\natexlab{b}}){Planck Collaboration}, {Ade}, {Aghanim}, {Arnaud}, {Ashdown}, {Aumont}, {Baccigalupi}, {Banday}, {Barreiro}, {Barrena}, {Bartolo}, {Battaner}, {Benabed}, {Benoit-L{\'e}vy}, {Bernard}, {Bersanelli}, {Bielewicz}, {Bikmaev}, {B{\"o}hringer}, {Bonaldi}, {Bonavera}, {Bond}, {Borrill}, {Bouchet}, {Burenin}, {Burigana}, {Butler}, {Calabrese}, {Carvalho}, {Catalano}, {Chamballu}, {Chiang}, {Chon}, {Christensen}, {Churazov}, {Clements}, {Colombo}, {Comis}, {Couchot}, {Curto}, {Cuttaia}, {Dahle}, {Danese}, {Davies}, {Davis}, {de Bernardis}, {de Rosa}, {de Zotti}, {Delabrouille}, {Diego}, {Dole}, {Dor{\'e}}, {Douspis}, {Ducout}, {Dupac}, {Efstathiou}, {Elsner}, {En{\ss}lin}, {Eriksen}, {Finelli}, {Flores-Cacho}, {Forni}, {Frailis}, {Fraisse}, {Franceschi}, {Frejsel}, {Fromenteau}, {Galeotta}, {Ganga}, {G{\'e}nova-Santos}, {Giard}, {Gilfanov}, {Giraud-H{\'e}raud}, {Gjerl{\o}w}, {Gonz{\'a}lez-Nuevo}, {G{\'o}rski}, {Gruppuso}, {Hansen}, {Hanson}, {Harrison},
  {Hempel}, {Henrot-Versill{\'e}}, {Hern{\'a}ndez-Monteagudo}, {Herranz}, {Hildebrandt}, {Hivon}, {Hobson}, {Holmes}, {Hornstrup}, {Hovest}, {Huffenberger}, {Hurier}, {Jaffe}, {Jones}, {Juvela}, {Keih{\"a}nen}, {Keskitalo}, {Khamitov}, {Kisner}, {Kneissl}, {Knoche}, {Kunz}, {Kurki-Suonio}, {Lagache}, {Lamarre}, {Lasenby}, {Lattanzi}, {Lawrence}, {Leonardi}, {Levrier}, {Liguori}, {Lilje}, {Linden-V{\o}rnle}, {L{\'o}pez-Caniego}, {Lubin}, {Mac{\'\i}as-P{\'e}rez}, {Maino}, {Mandolesi}, {Maris}, {Martin}, {Mart{\'\i}nez-Gonz{\'a}lez}, {Masi}, {Matarrese}, {Mazzotta}, {Melin}, {Mendes}, {Mennella}, {Migliaccio}, {Miville-Desch{\^e}nes}, {Moneti}, {Montier}, {Morgante}, {Mortlock}, {Munshi}, {Murphy}, {Naselsky}, {Nati}, {Natoli}, {N{\o}rgaard-Nielsen}, {Novikov}, {Novikov}, {Oxborrow}, {Pagano}, {Pajot}, {Paoletti}, {Pasian}, {Perdereau}, {Perotto}, {Perrotta}, {Pettorino}, {Piacentini}, {Piat}, {Pietrobon}, {Plaszczynski}, {Pointecouteau}, {Polenta}, {Popa}, {Pratt}, {Prunet}, {Puget}, {Rachen}, {Reinecke},
  {Remazeilles}, {Renault}, {Ricciardi}, {Ristorcelli}, {Rocha}, {Roman}, {Rosset}, {Rossetti}, {Roudier}, {Rubi{\~n}o-Mart{\'\i}n}, {Rusholme}, {Sandri}, {Scott}, {Spencer}, {Stolyarov}, {Sudiwala}, {Sunyaev}, {Sutton}, {Suur-Uski}, {Sygnet}, {Tauber}, {Terenzi}, {Toffolatti}, {Tomasi}, {Tristram}, {Tucci}, {Valenziano}, {Valiviita}, {Van Tent}, {Vibert}, {Vielva}, {Villa}, {Wade}, {Wandelt}, {Wehus}, {Yvon}, {Zacchei}, \& {Zonca}}]{rttfollowup}
---. 2015{\natexlab{b}}, \aap, 582, A29, \dodoi{10.1051/0004-6361/201424674}

\bibitem[{{Planck Collaboration} {et~al.}(2016{\natexlab{a}}){Planck Collaboration}, {Ade}, {Aghanim}, {Arnaud}, {Ashdown}, {Aumont}, {Baccigalupi}, {Banday}, {Barreiro}, {Barrena}, {Bartlett}, {Bartolo}, {Battaner}, {Battye}, {Benabed}, {Beno{\^\i}t}, {Benoit-L{\'e}vy}, {Bernard}, {Bersanelli}, {Bielewicz}, {Bikmaev}, {B{\"o}hringer}, {Bonaldi}, {Bonavera}, {Bond}, {Borrill}, {Bouchet}, {Bucher}, {Burenin}, {Burigana}, {Butler}, {Calabrese}, {Cardoso}, {Carvalho}, {Catalano}, {Challinor}, {Chamballu}, {Chary}, {Chiang}, {Chon}, {Christensen}, {Clements}, {Colombi}, {Colombo}, {Combet}, {Comis}, {Couchot}, {Coulais}, {Crill}, {Curto}, {Cuttaia}, {Dahle}, {Danese}, {Davies}, {Davis}, {de Bernardis}, {de Rosa}, {de Zotti}, {Delabrouille}, {D{\'e}sert}, {Dickinson}, {Diego}, {Dolag}, {Dole}, {Donzelli}, {Dor{\'e}}, {Douspis}, {Ducout}, {Dupac}, {Efstathiou}, {Eisenhardt}, {Elsner}, {En{\ss}lin}, {Eriksen}, {Falgarone}, {Fergusson}, {Feroz}, {Ferragamo}, {Finelli}, {Forni}, {Frailis}, {Fraisse}, {Franceschi},
  {Frejsel}, {Galeotta}, {Galli}, {Ganga}, {G{\'e}nova-Santos}, {Giard}, {Giraud-H{\'e}raud}, {Gjerl{\o}w}, {Gonz{\'a}lez-Nuevo}, {G{\'o}rski}, {Grainge}, {Gratton}, {Gregorio}, {Gruppuso}, {Gudmundsson}, {Hansen}, {Hanson}, {Harrison}, {Hempel}, {Henrot-Versill{\'e}}, {Hern{\'a}ndez-Monteagudo}, {Herranz}, {Hildebrandt}, {Hivon}, {Hobson}, {Holmes}, {Hornstrup}, {Hovest}, {Huffenberger}, {Hurier}, {Jaffe}, {Jaffe}, {Jin}, {Jones}, {Juvela}, {Keih{\"a}nen}, {Keskitalo}, {Khamitov}, {Kisner}, {Kneissl}, {Knoche}, {Kunz}, {Kurki-Suonio}, {Lagache}, {Lamarre}, {Lasenby}, {Lattanzi}, {Lawrence}, {Leonardi}, {Lesgourgues}, {Levrier}, {Liguori}, {Lilje}, {Linden-V{\o}rnle}, {L{\'o}pez-Caniego}, {Lubin}, {Mac{\'\i}as-P{\'e}rez}, {Maggio}, {Maino}, {Mak}, {Mandolesi}, {Mangilli}, {Martin}, {Mart{\'\i}nez-Gonz{\'a}lez}, {Masi}, {Matarrese}, {Mazzotta}, {McGehee}, {Mei}, {Melchiorri}, {Melin}, {Mendes}, {Mennella}, {Migliaccio}, {Mitra}, {Miville-Desch{\^e}nes}, {Moneti}, {Montier}, {Morgante}, {Mortlock}, {Moss},
  {Munshi}, {Murphy}, {Naselsky}, {Nastasi}, {Nati}, {Natoli}, {Netterfield}, {N{\o}rgaard-Nielsen}, {Noviello}, {Novikov}, {Novikov}, {Olamaie}, {Oxborrow}, {Paci}, {Pagano}, {Pajot}, {Paoletti}, {Pasian}, {Patanchon}, {Pearson}, {Perdereau}, {Perotto}, {Perrott}, {Perrotta}, {Pettorino}, {Piacentini}, {Piat}, {Pierpaoli}, {Pietrobon}, {Plaszczynski}, {Pointecouteau}, {Polenta}, {Pratt}, {Pr{\'e}zeau}, {Prunet}, {Puget}, {Rachen}, {Reach}, {Rebolo}, {Reinecke}, {Remazeilles}, {Renault}, {Renzi}, {Ristorcelli}, {Rocha}, {Rosset}, {Rossetti}, {Roudier}, {Rozo}, {Rubi{\~n}o-Mart{\'\i}n}, {Rumsey}, {Rusholme}, {Rykoff}, {Sandri}, {Santos}, {Saunders}, {Savelainen}, {Savini}, {Schammel}, {Scott}, {Seiffert}, {Shellard}, {Shimwell}, {Spencer}, {Stanford}, {Stern}, {Stolyarov}, {Stompor}, {Streblyanska}, {Sudiwala}, {Sunyaev}, {Sutton}, {Suur-Uski}, {Sygnet}, {Tauber}, {Terenzi}, {Toffolatti}, {Tomasi}, {Tramonte}, {Tristram}, {Tucci}, {Tuovinen}, {Umana}, {Valenziano}, {Valiviita}, {Van Tent}, {Vielva}, {Villa},
  {Wade}, {Wandelt}, {Wehus}, {White}, {Wright}, {Yvon}, {Zacchei}, \& {Zonca}}]{psz2}
---. 2016{\natexlab{a}}, \aap, 594, A27, \dodoi{10.1051/0004-6361/201525823}

\bibitem[{{Planck Collaboration} {et~al.}(2016{\natexlab{b}}){Planck Collaboration}, {Ade}, {Aghanim}, {Arnaud}, {Ashdown}, {Aumont}, {Baccigalupi}, {Banday}, {Barreiro}, {Barrena}, {Bartolo}, {Battaner}, {Benabed}, {Benoit-L{\'e}vy}, {Bernard}, {Bersanelli}, {Bielewicz}, {Bikmaev}, {B{\"o}hringer}, {Bonaldi}, {Bonavera}, {Bond}, {Borrill}, {Bouchet}, {Burenin}, {Burigana}, {Calabrese}, {Cardoso}, {Catalano}, {Chamballu}, {Chary}, {Chiang}, {Chon}, {Christensen}, {Clements}, {Colombo}, {Combet}, {Comis}, {Crill}, {Curto}, {Cuttaia}, {Dahle}, {Danese}, {Davies}, {Davis}, {de Bernardis}, {de Rosa}, {de Zotti}, {Delabrouille}, {Diego}, {Dole}, {Donzelli}, {Dor{\'e}}, {Douspis}, {Dupac}, {Efstathiou}, {Elsner}, {En{\ss}lin}, {Eriksen}, {Ferragamo}, {Finelli}, {Forni}, {Frailis}, {Fraisse}, {Franceschi}, {Fromenteau}, {Galeotta}, {Galli}, {Ganga}, {G{\'e}nova-Santos}, {Giard}, {Gjerl{\o}w}, {Gonz{\'a}lez-Nuevo}, {G{\'o}rski}, {Gruppuso}, {Hansen}, {Harrison}, {Hempel}, {Hern{\'a}ndez-Monteagudo}, {Herranz},
  {Hildebrandt}, {Hivon}, {Hornstrup}, {Hovest}, {Huffenberger}, {Hurier}, {Jaffe}, {Keih{\"a}nen}, {Keskitalo}, {Khamitov}, {Kisner}, {Kneissl}, {Knoche}, {Kunz}, {Kurki-Suonio}, {Lamarre}, {Lasenby}, {Lattanzi}, {Lawrence}, {Leonardi}, {Le{\'o}n-Tavares}, {Levrier}, {Lietzen}, {Liguori}, {Lilje}, {Linden-V{\o}rnle}, {L{\'o}pez-Caniego}, {Lubin}, {Mac{\'\i}as-P{\'e}rez}, {Maffei}, {Maino}, {Mandolesi}, {Maris}, {Martin}, {Mart{\'\i}nez-Gonz{\'a}lez}, {Masi}, {Matarrese}, {McGehee}, {Melchiorri}, {Mennella}, {Migliaccio}, {Miville-Desch{\^e}nes}, {Moneti}, {Montier}, {Morgante}, {Mortlock}, {Munshi}, {Murphy}, {Naselsky}, {Nati}, {Natoli}, {Novikov}, {Novikov}, {Oxborrow}, {Pagano}, {Pajot}, {Paoletti}, {Pasian}, {Perdereau}, {Pettorino}, {Piacentini}, {Piat}, {Pierpaoli}, {Plaszczynski}, {Pointecouteau}, {Polenta}, {Pratt}, {Prunet}, {Puget}, {Rachen}, {Rebolo}, {Reinecke}, {Remazeilles}, {Renault}, {Renzi}, {Ristorcelli}, {Rocha}, {Rosset}, {Rossetti}, {Roudier}, {Rubi{\~n}o-Mart{\'\i}n}, {Rusholme},
  {Sandri}, {Santos}, {Savelainen}, {Savini}, {Scott}, {Stolyarov}, {Streblyanska}, {Sudiwala}, {Sunyaev}, {Suur-Uski}, {Sygnet}, {Tauber}, {Terenzi}, {Toffolatti}, {Tomasi}, {Tramonte}, {Tristram}, {Tucci}, {Valenziano}, {Valiviita}, {Van Tent}, {Vielva}, {Villa}, {Wade}, {Wandelt}, {Wehus}, {Yvon}, {Zacchei}, \& {Zonca}}]{eno}
---. 2016{\natexlab{b}}, \aap, 586, A139, \dodoi{10.1051/0004-6361/201526345}

\bibitem[{{Planck Collaboration} {et~al.}(2016{\natexlab{c}}){Planck Collaboration}, {Aghanim}, {Arnaud}, {Ashdown}, {Aumont}, {Baccigalupi}, {Banday}, {Barreiro}, {Bartlett}, {Bartolo}, {Battaner}, {Battye}, {Benabed}, {Beno{\^\i}t}, {Benoit-L{\'e}vy}, {Bernard}, {Bersanelli}, {Bielewicz}, {Bock}, {Bonaldi}, {Bonavera}, {Bond}, {Borrill}, {Bouchet}, {Burigana}, {Butler}, {Calabrese}, {Cardoso}, {Catalano}, {Challinor}, {Chiang}, {Christensen}, {Churazov}, {Clements}, {Colombo}, {Combet}, {Comis}, {Coulais}, {Crill}, {Curto}, {Cuttaia}, {Danese}, {Davies}, {Davis}, {de Bernardis}, {de Rosa}, {de Zotti}, {Delabrouille}, {D{\'e}sert}, {Dickinson}, {Diego}, {Dolag}, {Dole}, {Donzelli}, {Dor{\'e}}, {Douspis}, {Ducout}, {Dupac}, {Efstathiou}, {Elsner}, {En{\ss}lin}, {Eriksen}, {Fergusson}, {Finelli}, {Forni}, {Frailis}, {Fraisse}, {Franceschi}, {Frejsel}, {Galeotta}, {Galli}, {Ganga}, {G{\'e}nova-Santos}, {Giard}, {Gonz{\'a}lez-Nuevo}, {G{\'o}rski}, {Gregorio}, {Gruppuso}, {Gudmundsson}, {Hansen}, {Harrison},
  {Henrot-Versill{\'e}}, {Hern{\'a}ndez-Monteagudo}, {Herranz}, {Hildebrandt}, {Hivon}, {Holmes}, {Hornstrup}, {Huffenberger}, {Hurier}, {Jaffe}, {Jones}, {Juvela}, {Keih{\"a}nen}, {Keskitalo}, {Kneissl}, {Knoche}, {Kunz}, {Kurki-Suonio}, {Lacasa}, {Lagache}, {L{\"a}hteenm{\"a}ki}, {Lamarre}, {Lasenby}, {Lattanzi}, {Leonardi}, {Lesgourgues}, {Levrier}, {Liguori}, {Lilje}, {Linden-V{\o}rnle}, {L{\'o}pez-Caniego}, {Mac{\'\i}as-P{\'e}rez}, {Maffei}, {Maggio}, {Maino}, {Mandolesi}, {Mangilli}, {Maris}, {Martin}, {Mart{\'\i}nez-Gonz{\'a}lez}, {Masi}, {Matarrese}, {Melchiorri}, {Melin}, {Migliaccio}, {Miville-Desch{\^e}nes}, {Moneti}, {Montier}, {Morgante}, {Mortlock}, {Munshi}, {Murphy}, {Naselsky}, {Nati}, {Natoli}, {Noviello}, {Novikov}, {Novikov}, {Paci}, {Pagano}, {Pajot}, {Paoletti}, {Pasian}, {Patanchon}, {Perdereau}, {Perotto}, {Pettorino}, {Piacentini}, {Piat}, {Pierpaoli}, {Pietrobon}, {Plaszczynski}, {Pointecouteau}, {Polenta}, {Ponthieu}, {Pratt}, {Prunet}, {Puget}, {Rachen}, {Reinecke}, {Remazeilles},
  {Renault}, {Renzi}, {Ristorcelli}, {Rocha}, {Rossetti}, {Roudier}, {Rubi{\~n}o-Mart{\'\i}n}, {Rusholme}, {Sandri}, {Santos}, {Sauv{\'e}}, {Savelainen}, {Savini}, {Scott}, {Spencer}, {Stolyarov}, {Stompor}, {Sunyaev}, {Sutton}, {Suur-Uski}, {Sygnet}, {Tauber}, {Terenzi}, {Toffolatti}, {Tomasi}, {Tramonte}, {Tristram}, {Tucci}, {Tuovinen}, {Valenziano}, {Valiviita}, {Van Tent}, {Vielva}, {Villa}, {Wade}, {Wandelt}, {Wehus}, {Yvon}, {Zacchei}, \& {Zonca}}]{tszmap}
{Planck Collaboration}, {Aghanim}, N., {Arnaud}, M., {et~al.} 2016{\natexlab{c}}, \aap, 594, A22, \dodoi{10.1051/0004-6361/201525826}

\bibitem[{{Rines} {et~al.}(2013){Rines}, {Geller}, {Diaferio}, \& {Kurtz}}]{hecs13}
{Rines}, K., {Geller}, M.~J., {Diaferio}, A., \& {Kurtz}, M.~J. 2013, \apj, 767, 15, \dodoi{10.1088/0004-637X/767/1/15}

\bibitem[{{Rines} {et~al.}(2016){Rines}, {Geller}, {Diaferio}, \& {Hwang}}]{hecs-sz}
{Rines}, K.~J., {Geller}, M.~J., {Diaferio}, A., \& {Hwang}, H.~S. 2016, \apj, 819, 63, \dodoi{10.3847/0004-637X/819/1/63}

\bibitem[{{Rines} {et~al.}(2018){Rines}, {Geller}, {Diaferio}, {Hwang}, \& {Sohn}}]{hecs-red}
{Rines}, K.~J., {Geller}, M.~J., {Diaferio}, A., {Hwang}, H.~S., \& {Sohn}, J. 2018, \apj, 862, 172, \dodoi{10.3847/1538-4357/aacd49}

\bibitem[{{Rykoff} {et~al.}(2014){Rykoff}, {Rozo}, {Busha}, {Cunha}, {Finoguenov}, {Evrard}, {Hao}, {Koester}, {Leauthaud}, {Nord}, {Pierre}, {Reddick}, {Sadibekova}, {Sheldon}, \& {Wechsler}}]{redmapper}
{Rykoff}, E.~S., {Rozo}, E., {Busha}, M.~T., {et~al.} 2014, \apj, 785, 104, \dodoi{10.1088/0004-637X/785/2/104}

\bibitem[{{Schlegel} {et~al.}(1998){Schlegel}, {Finkbeiner}, \& {Davis}}]{sfd98}
{Schlegel}, D.~J., {Finkbeiner}, D.~P., \& {Davis}, M. 1998, \apj, 500, 525, \dodoi{10.1086/305772}

\bibitem[{{Skrutskie} {et~al.}(2006){Skrutskie}, {Cutri}, {Stiening}, {Weinberg}, {Schneider}, {Carpenter}, {Beichman}, {Capps}, {Chester}, {Elias}, {Huchra}, {Liebert}, {Lonsdale}, {Monet}, {Price}, {Seitzer}, {Jarrett}, {Kirkpatrick}, {Gizis}, {Howard}, {Evans}, {Fowler}, {Fullmer}, {Hurt}, {Light}, {Kopan}, {Marsh}, {McCallon}, {Tam}, {Van Dyk}, \& {Wheelock}}]{2mass}
{Skrutskie}, M.~F., {Cutri}, R.~M., {Stiening}, R., {et~al.} 2006, \aj, 131, 1163, \dodoi{10.1086/498708}

\bibitem[{{Sohn} {et~al.}(2023){Sohn}, {Geller}, {Hwang}, {Fabricant}, {Utsumi}, \& {Damjanov}}]{hectomap_dr2}
{Sohn}, J., {Geller}, M.~J., {Hwang}, H.~S., {et~al.} 2023, \apj, 945, 94, \dodoi{10.3847/1538-4357/acb925}

\bibitem[{{Streblyanska} {et~al.}(2019){Streblyanska}, {Aguado-Barahona}, {Ferragamo}, {Barrena}, {Rubi{\~n}o-Mart{\'\i}n}, {Tramonte}, {Genova-Santos}, \& {Lietzen}}]{lp15a}
{Streblyanska}, A., {Aguado-Barahona}, A., {Ferragamo}, A., {et~al.} 2019, \aap, 628, A13, \dodoi{10.1051/0004-6361/201935674}

\bibitem[{{Streblyanska} {et~al.}(2018){Streblyanska}, {Barrena}, {Rubi{\~n}o-Mart{\'\i}n}, {van der Burg}, {Aghanim}, {Aguado-Barahona}, {Ferragamo}, \& {Lietzen}}]{streblyanska18}
{Streblyanska}, A., {Barrena}, R., {Rubi{\~n}o-Mart{\'\i}n}, J.~A., {et~al.} 2018, \aap, 617, A71, \dodoi{10.1051/0004-6361/201732306}

\bibitem[{{Sunyaev} \& {Zeldovich}(1972)}]{sz}
{Sunyaev}, R.~A., \& {Zeldovich}, Y.~B. 1972, Comments on Astrophysics and Space Physics, 4, 173

\bibitem[{{van der Burg} {et~al.}(2016){van der Burg}, {Aussel}, {Pratt}, {Arnaud}, {Melin}, {Aghanim}, {Barrena}, {Dahle}, {Douspis}, {Ferragamo}, {Fromenteau}, {Herbonnet}, {Hurier}, {Pointecouteau}, {Rubi{\~n}o-Mart{\'\i}n}, \& {Streblyanska}}]{vanderburg16}
{van der Burg}, R.~F.~J., {Aussel}, H., {Pratt}, G.~W., {et~al.} 2016, \aap, 587, A23, \dodoi{10.1051/0004-6361/201527299}

\bibitem[{Virtanen {et~al.}(2020)Virtanen, Gommers, Oliphant, Haberland, Reddy, Cournapeau, Burovski, Peterson, Weckesser, Bright, {van der Walt}, Brett, Wilson, Millman, Mayorov, Nelson, Jones, Kern, Larson, Carey, Polat, Feng, Moore, {VanderPlas}, Laxalde, Perktold, Cimrman, Henriksen, Quintero, Harris, Archibald, Ribeiro, Pedregosa, {van Mulbregt}, \& {SciPy 1.0 Contributors}}]{2020SciPy-NMeth}
Virtanen, P., Gommers, R., Oliphant, T.~E., {et~al.} 2020, Nature Methods, 17, 261, \dodoi{10.1038/s41592-019-0686-2}

\bibitem[{{Voges} {et~al.}(1999){Voges}, {Aschenbach}, {Boller}, {Br{\"a}uninger}, {Briel}, {Burkert}, {Dennerl}, {Englhauser}, {Gruber}, {Haberl}, {Hartner}, {Hasinger}, {K{\"u}rster}, {Pfeffermann}, {Pietsch}, {Predehl}, {Rosso}, {Schmitt}, {Tr{\"u}mper}, \& {Zimmermann}}]{rasscat_all}
{Voges}, W., {Aschenbach}, B., {Boller}, T., {et~al.} 1999, \aap, 349, 389.
\newblock \doarXiv{astro-ph/9909315}

\bibitem[{{Werner} {et~al.}(2004){Werner}, {Roellig}, {Low}, {Rieke}, {Rieke}, {Hoffmann}, {Young}, {Houck}, {Brandl}, {Fazio}, {Hora}, {Gehrz}, {Helou}, {Soifer}, {Stauffer}, {Keene}, {Eisenhardt}, {Gallagher}, {Gautier}, {Irace}, {Lawrence}, {Simmons}, {Van Cleve}, {Jura}, {Wright}, \& {Cruikshank}}]{spitzer}
{Werner}, M.~W., {Roellig}, T.~L., {Low}, F.~J., {et~al.} 2004, \apjs, 154, 1, \dodoi{10.1086/422992}

\bibitem[{{Wright} {et~al.}(2010){Wright}, {Eisenhardt}, {Mainzer}, {Ressler}, {Cutri}, {Jarrett}, {Kirkpatrick}, {Padgett}, {McMillan}, {Skrutskie}, {Stanford}, {Cohen}, {Walker}, {Mather}, {Leisawitz}, {Gautier}, {McLean}, {Benford}, {Lonsdale}, {Blain}, {Mendez}, {Irace}, {Duval}, {Liu}, {Royer}, {Heinrichsen}, {Howard}, {Shannon}, {Kendall}, {Walsh}, {Larsen}, {Cardon}, {Schick}, {Schwalm}, {Abid}, {Fabinsky}, {Naes}, \& {Tsai}}]{wise}
{Wright}, E.~L., {Eisenhardt}, P. R.~M., {Mainzer}, A.~K., {et~al.} 2010, \aj, 140, 1868, \dodoi{10.1088/0004-6256/140/6/1868}

\bibitem[{{Zaznobin} {et~al.}(2019){Zaznobin}, {Burenin}, {Bikmaev}, {Khamitov}, {Khorunzhev}, {Konoplev}, {Eselevich}, {Afanasiev}, {Dodonov}, {Rubi{\~n}o-Mart{\'\i}n}, {Aghanim}, \& {Sunyaev}}]{zaznobin19}
{Zaznobin}, I.~A., {Burenin}, R.~A., {Bikmaev}, I.~F., {et~al.} 2019, Astronomy Letters, 45, 49, \dodoi{10.1134/S1063773719020063}

\bibitem[{{Zaznobin} {et~al.}(2020){Zaznobin}, {Burenin}, {Bikmaev}, {Khamitov}, {Khorunzhev}, {Lyapin}, {Eselevich}, {Afanasiev}, {Dodonov}, \& {Sunyaev}}]{zaznobin20}
---. 2020, Astronomy Letters, 46, 79, \dodoi{10.1134/S1063773720020048}

\bibitem[{{Zaznobin} {et~al.}(2021){Zaznobin}, {Burenin}, {Bikmaev}, {Khamitov}, {Khorunzhev}, {Lyapin}, {Eselevich}, {Lyskova}, {Medvedev}, {Gilfanov}, \& {Sunyaev}}]{zaznobin21}
---. 2021, Astronomy Letters, 47, 61, \dodoi{10.1134/S1063773721020055}

\bibitem[{{Zohren} {et~al.}(2019){Zohren}, {Schrabback}, {van der Burg}, {Arnaud}, {Melin}, {van den Busch}, {Hoekstra}, \& {Klein}}]{zohren19}
{Zohren}, H., {Schrabback}, T., {van der Burg}, R. F.~J., {et~al.} 2019, \mnras, 488, 2523, \dodoi{10.1093/mnras/stz1838}

\end{thebibliography}

\end{document}